%% file: main.tex
\documentclass[sigconf,usenames,dvipsnames]{acmart}

\copyrightyear{2024}
\acmYear{2024}
\setcopyright{acmlicensed}\acmConference[arXiv]{}{May, 2024}{}

\AtBeginDocument{%
  \providecommand\BibTeX{{%
    \normalfont B\kern-0.5em{\scshape i\kern-0.25em b}\kern-0.8em\TeX}}}
    
\begin{document}

\newcommand{\system}{Video2MR}
\title{\system{}: Automatically Generating Mixed Reality 3D Instructions by Augmenting Extracted Motion from 2D Videos}

\author{Keiichi Ihara}
\affiliation{%
  \institution{University of Tsukuba}
  \city{Tsukuba}
  \country{Japan}}
\affiliation{%
  \institution{University of Calgary}
  \city{Calgary}
  \country{Canada}}  
\email{kihara@iplab.cs.tsukuba.ac.jp}

\author{Kyzyl Monteiro}
\affiliation{%
  \institution{Carnegie Mellon University}
  \city{Pittsburgh}
  \country{United States}}
\email{kyzyl@cmu.edu}

\author{Mehrad Faridan}
\affiliation{%
  \institution{University of Calgary}
  \city{Calgary}
  \country{Canada}}
\email{mehrad.faridan1@ucalgary.ca}

\author{Rubaiat Habib Kazi}
\affiliation{%
  \institution{Adobe Research}
  \city{Seattle}
  \country{United States}}
\email{rhabib@adobe.com}

\author{Ryo Suzuki}
\affiliation{%
  \institution{University of Calgary}
  \city{Calgary}
  \country{Canada}}
\email{ryo.suzuki@ucalgary.ca}

\renewcommand{\shortauthors}{Ihara, et al.}

\input{0-abstract}

\begin{teaserfigure}
\centering
\resizebox{\columnwidth}{!}{
\includegraphics[height=\linewidth]{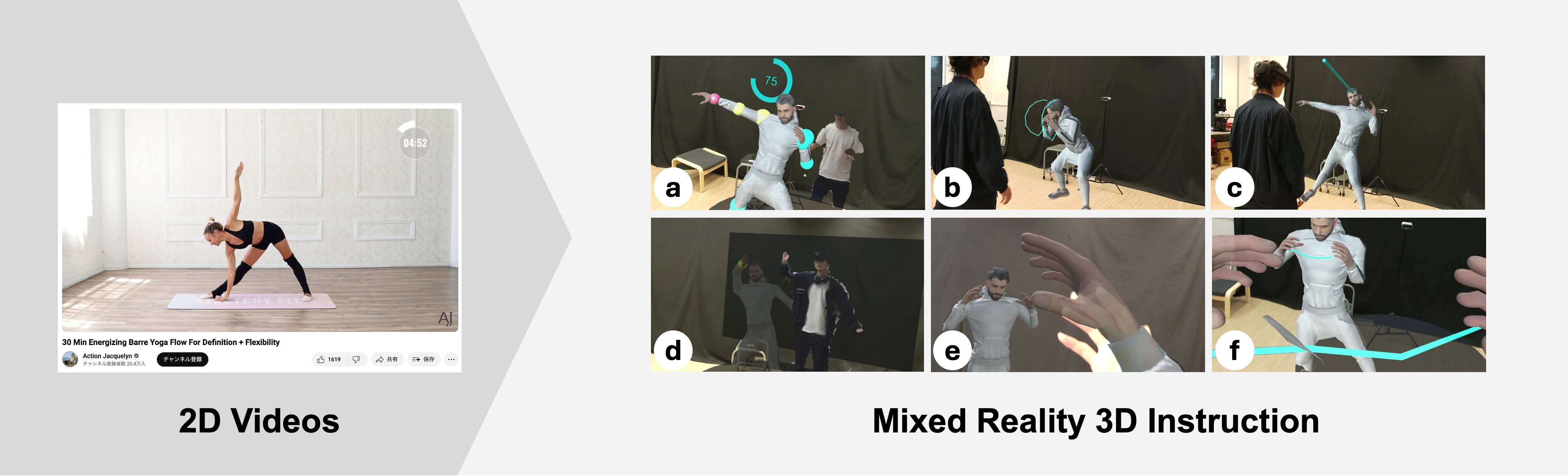}}
\caption{We introduce \system{}, an enhanced mixed reality instruction that extract body motion from 2D videos. \system{} augments the instructions by (a) comparing the user's movements with the instructor's movement, (b,c) visualizing the instructor's avatar by showing trajectories and highlighting gaze, (d) navigating the avatar motion based on the user's movement, (e, f) repositioning the avatar in first-person and highlighting them.}
\label{fig:teaser}
\end{teaserfigure}

\maketitle

\input{1-introduction}

\input{2-related-work}
\input{3-formative-study}

\input{4-system}

\input{5-user-study}

\input{6-future-work}

\input{7-conclusion}

\balance
\bibliographystyle{ACM-Reference-Format}
\bibliography{references}

\end{document}

%% file: 0-abstract.tex
\begin{abstract}
This paper introduces Video2MR, a mixed reality system that automatically generates 3D sports and exercise instructions from 2D videos. Mixed reality instructions have great potential for physical training, but existing works require substantial time and cost to create these 3D experiences. Video2MR overcomes this limitation by transforming arbitrary instructional videos available online into MR 3D avatars with AI-enabled motion capture (\textit{DeepMotion}). Then, it automatically enhances the avatar motion through the following augmentation techniques: 1) contrasting and highlighting differences between the user and avatar postures, 2) visualizing key trajectories and movements of specific body parts, 3) manipulation of time and speed using body motion, and 4) spatially repositioning avatars for different perspectives. Developed on Hololens 2 and Azure Kinect, we showcase various use cases, including yoga, dancing, soccer, tennis, and other physical exercises. The study results confirm that Video2MR provides more engaging and playful learning experiences, compared to existing 2D video instructions. 
\end{abstract}

\ccsdesc[500]{Human-centered computing~Mixed / augmented reality}

\keywords{Mixed Reality; Sports and Exercises; Videos; Motion Capture; Avatar; Automated Generation}

%% file: 1-introduction.tex
\section{Introduction}
Mixed reality instructions have great potential for sports and exercise training. 
They offer an immersive and interactive learning experience that are not possible with traditional 2D video instructions.
For example, prior works have shown that the ability to see a 3D avatar can improve their understanding of the postures \cite{ikeda2018ar, hamanishi2019assisting} and interactive MR visualizations are effective guidance to learn and notice differences in contrast to merely imitating actions from a 2D screen \cite{tharatipyakul2020pose}.

However, creating high-quality mixed reality instructions often requires substantial time and cost. The process typically requires 3D motion capture and programming, which presents a significant technical challenge for creating 3D instructions for professional instructors. 
In addition, as the manual creation process is time-consuming and tedious, these challenges significantly limit the scalability, availability, and diversity of mixed reality 3D instructions.

In this paper, we explore the idea of \textit{automatically generating mixed reality 3D instructions} by transforming existing online 2D videos into immersive sports and exercise training. 
Our idea is driven by advances in computer vision and generative AI \cite{fang2017rmpe, xiu2018pose}, as well as the recent democratisation of these techniques \cite{noauthor_deepmotion_nodate}, which enables us to extract 3D human motions from arbitrary 2D videos. With this, we can leverage a vast variety of professional videos already available online (eg. YouTube) to create immersive instructional experiences for various physical activities, such as yoga, dancing, exercise, and many other sports. 
However, several important questions remain: 1) What are the design challenges and limitations of automatically generated 3D instructions from the user's perspective? 2) How could we design and improve an automated 3D avatar instructor for a better training experiences? 3) What is the effectiveness of automatically generated Mixed Reality 3D instructions, compared to 2D videos? 4) Can this automated approach generate effective instructions across different kinds of sports and physical activities?

This paper investigates these research questions by adapting \textit{Buchenau’s 2-step experience prototyping approach}~\cite{buchenau2000experience}. 
To gain insights from the user's experiences, we first develop a simple working prototype that extracts human motions from 2D videos \textit{DeepMotion}~\cite{noauthor_deepmotion_nodate}, translates them to a 3D avatar and presents it in a MR spatial experience via a Hololens 2.
We evaluated this minimal prototype through a formative study with 8 participants to identify benefits and challenges of automated 3D instructions. Participants found that mixed reality instructions have significant benefits as compared to their previous experiences with 2D online tutorial videos, even when automatically generated. For example, the participants expressed that they felt the increased co-presence of the instructor and appreciated the ability to see the instructor from different angles. On the other hand, areas for improvement included - difficulty in comparing their movements to the instructor's, tracking specific body parts, navigating time and controlling speed, and switching between different viewing perspectives. 


To address these challenges, we present Video2MR, a system that leverages extracted human motion from existing videos, to generate a 3D avatar. Video2MR then further augments and enhances this avatar to create a mixed reality instructional experience. Our concept builds upon the previous research in MR instructional visualisations\cite{zhou2021syncup,hoang2016onebody,iannucci2023arrow} and body-based experiences which utilize 3D avatars~\cite{han2017my, ikeda2018ar, hamanishi2019assisting}, but we make two key contributions beyond them.

First, we explore a \textbf{broader design space} of the enhancements and 3D augmentations of a 3D avatar in an immersive AR instructional experience. Based on the formative study and informed by previous literature, we identify our design space which includes: 1) \textbf{Posture Comparison:} contrasting and highlighting differences between the user's and instructor's avatar postures, 2) \textbf{Motion Visualization:} visualizing key trajectories and movements of specific body parts, 3) \textbf{Embodied Temporal Navigation:} manipulation of time and speed using body motion, and 4) \textbf{Avatar Repositioning:} spatially repositioning avatars to view the avatar from different perspectives. These four features explore augmentation techniques like the color indicator and scoring for comparison, footprints and trajectories for visualization, body motion-driven temporal navigation, and switching between the first-person view and third-person view. These design space elements have been inspired by prior work, for example 
\textit{LightGuide}~\cite{sodhi2012lightguide} for indicators, \textit{RealitySketch}~\cite{suzuki2020realitysketch} for pose match, \textit{ARrow}~\cite{iannucci2023arrow} for trajectories, 
\textit{Projection based AR}~\cite{sekhavat2018projection} for footprints, 
\textit{ReactiveVideo}~\cite{clarke2020reactive} for body-based temporal navigation, \textit{OneBody}~\cite{hoang2016onebody} for first-person, and \textit{OutsideMe}~\cite{yan2015outsideme} for third-person view. While these have been introduced individually in previous work, in our design space, we aggregate and combine these to apply them to a new application scenario, an immersive 3D AR instructional experience.



Second, we contribute to a \textbf{holistic user evaluation} of the system to better understand the usability and the versatility of the system. To this end, we design and conduct a study with three parts which include: 1) a usability study with 12 participants that compares Video, Avatar, and Video+Avatar 
2) a versatility evaluation which checks the accuracy and feasibility of the system across six physical activities and 
3) expert interviews with five domain experts to gain in-depth qualitative feedback about Video2MR. The usability study results confirm that our approach helps automatically generate MR instructional experiences which are generally more engaging, fun, and easy-to-follow as compared to video tutorials. 
The expert review confirms the value of our system and our features.
Also, they provided valuable feedback on each feature's practical uses, unique qualitative insights of potential use cases, educational benefits, and direction for future feature improvements. 
Finally, we discuss the limitations of our approach and explore future opportunities for automated MR instructional experiences.

Finally, this paper contributes:
\begin{enumerate}
\item Insights from eight users through an experience prototyping protocol that elicits potential benefits and challenges with automatically generated mixed reality instructions and a design space of features that cater to these needs. 
\item Video2MR, a system that automatically generates 3D avatar animations and augmentations from 2D videos in a mixed reality immersive setting.
\item Results and insights gained from an evaluation study with three parts which evaluates the Video2MR system and design space quantitatively and qualitatively.



\end{enumerate}







%% file: 2-related-work.tex
\section{Related Work}
In this research, we developed a system that automatically generates instructional experiences in MR by creating 3D avatars from videos and automatically augmenting them with several features. This section presents prior research on the use of 3D avatars and the variety of features proposed in the previous instructional systems.

\subsection{Usage of 3D Avatars}
Systems employing 3D avatars have been proposed in the context of remote instruction and collaboration. These avatars can be categorized into three types: avatars that reflect real-time movements of the remote users, pre-designed avatars, and automatically generated avatars. 

\textbf{Remote Manipulated Avatars:}
There are several studies utilizing motion capture of experts in remote locations. For instance, \textit{OneBody}~\cite{hoang2016onebody} presents the posture of a remote person from a first-person perspective. In the context of remote collaboration, there are studies displaying the full body of a remote person~\cite{thoravi2019loki, ihara2023holobots}, parts of the body (e.g., hands~\cite{amores2015showme, faridan2023chameleoncontrol}, gaze~\cite{bai2020user}), avatars of different sizes~\cite{piumsomboon2018mini, piumsomboon2019shoulder}, and multiple avatars~\cite{thanyadit2019observar}. These enable detailed feedback in real time. However, these systems require the presence of an expert, making it difficult for many users to utilize them easily.

\textbf{Pre-designed Movement Avatars:}
There is also research that utilizes avatars with pre-designed movements~\cite{wu2023ar, jan2021augmented, mostajeran2020augmented, han2017my, ikeda2018ar}. For example, \textit{ARenhanced Workout}~\cite{wu2023ar} combines 3D avatars with visualizations to facilitate users' understanding of correct posture. \textit{My Chi Coaches}~\cite{han2017my} use multiple 3D avatars to offer views from various angles. Furthermore, in the context of tutorial creation, there are studies capturing the movements of an expert to be replayed later as avatar movements~\cite{huang2021adaptutar}. These approaches allow for the setting of various movements and the creation of accurate avatars, but creating these movements can be time-consuming. Additionally, users are limited to using pre-made movements, unable to obtain specific instructions they want immediately.

\textbf{Automatically Generated Avatars:}
Research also exists on the automatic generation of avatars from videos for purposes such as tutorial generation~\cite{eckhoff2018tutar}, motion-based browsing~\cite{hamanishi2021motion}, and animation authoring~\cite{wang2022videoposevr}. These studies enable the creation of 3D interactive content distinct from 2D content. However, previous research has not utilized this approach for creating MR instructional experiences, nor has it investigated what functionalities can be added to the generated avatars and which of those functionalities are useful. Therefore, we employed AI-based tools for 3D avatar generation to create MR instructional experiences in 3D, proposed various functionalities, conducted user studies, and investigated the effectiveness of these functionalities.

\subsection{Features in Instructional Systems}
Various instructional experiences have been proposed in past research, not only in MR environments but also using 2D screens, projections, and VR devices. Here, we show the types of instructional experiences proposed in other media and contexts and how we have applied them in our 3D MR experience.

\textbf{Visualization Techniques:}
A wide range of visualization techniques have been proposed for instruction systems. For example, in the use of projection, several visualizations have been proposed, such as visualizing footprints~\cite{sekhavat2018projection, kosmalla2021virtualladder}, indicating correct directions~\cite{sodhi2012lightguide}, and visualizing users’ posture~\cite{turmo2020bodylights}. Additionally, using the 2D screens, displaying trajectories on 2D screens~\cite{iannucci2023arrow}, and presenting synchronization accuracy~\cite{zhou2021syncup} are also explored. Moreover, various 3D visualization techniques have been proposed, such as visualizing the movement of a badminton shuttle in VR~\cite{ye2020shuttlespace} and visualizations for ski training in VR~\cite{matsumoto2022skiing}. In AR, several visualization techniques have been proposed, such as visualization techniques for workouts with 3D avatars~\cite{wu2023ar} and for improving free throws in basketball practice~\cite{lin2021towards}. In contrast, our focus is on visualization applied to generated avatars. By automatically generating visualized instructional experiences from videos, we support a wide range of sports and exercises.

\textbf{Comparison Methods:}
Many studies have proposed using stick figures on 2D screens for comparison~\cite{velloso2013motionma, tharatipyakul2020pose, zhao20223d, conner2016correcting, dittakavi2022pose, marquardt2012super, saenz2016kinect}. These studies compare the user's 2D posture with the instructor's posture~\cite{velloso2013motionma, tharatipyakul2020pose}. For example, \textit{MotionMA}~\cite{velloso2013motionma} compares the user's input video with the instructor's movements. Tharatipyakul et al.~\cite{tharatipyakul2020pose} conducted a study comparing feedback using videos and stick figures. Other studies compare using 3D avatars on 2D screens~\cite{fieraru2021aifit, liu2022posecoach}. \textit{AIFit}~\cite{fieraru2021aifit} converts input videos into 3D postures and provides feedback on differences from correct postures. \textit{PoseCoach}~\cite{liu2022posecoach} is a study that utilizes 3D avatars in 3D space. Our system compares the user's 3D posture captured by an RGBD camera with the automatically generated instructor's 3D posture.

\textbf{Temporal Navigation Techniques:}
To change the timing in videos, mainly 2D UI elements like scroll bars are used, but different methods, such as those using avatars and body movements, have also been proposed. For example, Hamanishi et al.~\cite{hamanishi2021motion} arrange avatars in a timeline and manipulate time by interacting with them. There are also studies that compare the user's 2D posture with the 2D posture in videos to navigate to specific times~\cite{clarke2020reactive, hamanishi2020poseasquery}. In contrast, our system proposes changing the generated avatar's 3D posture based on the user's 3D posture.

\textbf{Utilization of First-Person Cues:}
Several studies use first-person cues to instruct on how to move their bodies~\cite{han2016ar, hoang2016onebody, yu2020perspective, elsayed2022understanding}. There are also studies specialized in specific applications, such as Aikido~\cite{suzuki2023gino}, musical instrument performance~\cite{liu2023pianosyncar, skreinig2022ar}, agility training~\cite{kosmalla2021virtualladder}, and juggling~\cite{adolf2019juggling}. However, these first-person cues are manually created, and their production can be time-consuming. Therefore, we aim to automatically generate avatars and then convert them into first-person cues.

\textbf{Multimodal Feedback:}
Instructional methods utilizing not only visual but also auditory and haptic feedback have been proposed~\cite{xia2022volearn, schonauer2012multimodal, katzakis2017stylo}. For example, \textit{VoLearn}~\cite{xia2022volearn} provides feedback on the user's movements using wearable devices and auditory cues. \textit{Multimodal motion guidance}~\cite{schonauer2012multimodal} uses vibrations to give instructions on speed and direction. Furthermore, \textit{Stylo and Handifact}~\cite{katzakis2017stylo} offer the sensation of the hand being pressed to assist in improving movements. In contrast, our system focuses on generating visual guidance and feedback.

%% file: 3-formative-study.tex
\section{Experience Prototyping}
To identify the challenges and potential benefits of automatically generated 3D instructions, we adapted \textit{Buchenau's experience prototyping approach} ~\cite{buchenau2000experience} by developing an initial prototype and then conducting a formative study. This is because MR experiences are often difficult to imagine before experiencing them, thus it is difficult to get an appropriate insight or design guidelines without a functional prototype. Therefore, we developed a simple prototype to test our concept and gain user feedback through the formative user evaluation. 

\subsection{Initial Prototype}
\begin{figure}[ht]
\centering
\includegraphics[width=0.6\linewidth]{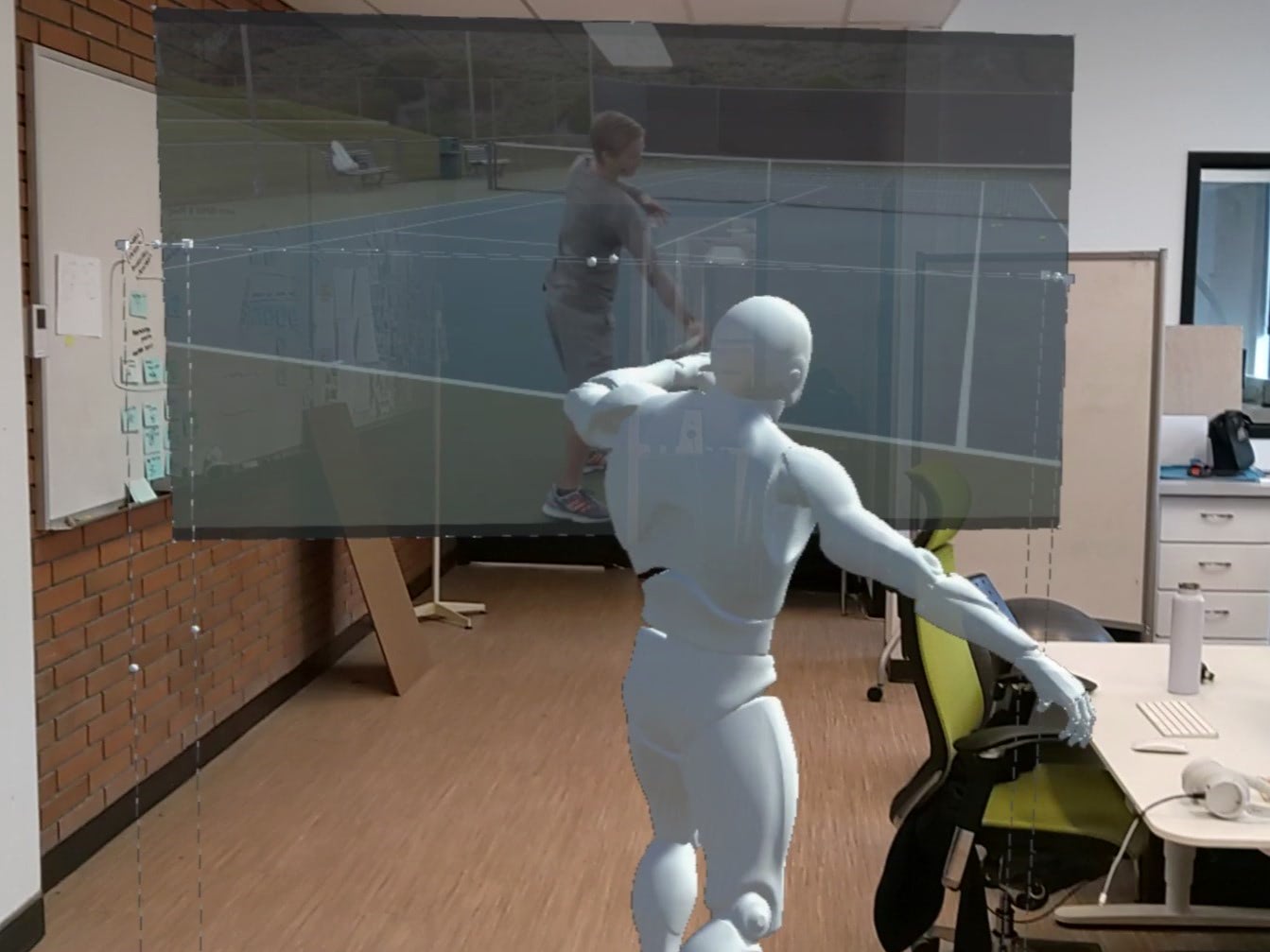}
\caption{Formative Study}
\label{fig:FormativeStudy}
\end{figure}

For the initial prototype, we use off-the-shelf software that automatically extracts human motion from online 2D videos. We initially tested four different software, including \textit{DeepMotion}~\footnote{\url{https://www.deepmotion.com/}}, \textit{Plask Motion}~\footnote{\url{https://plask.ai/}}, \textit{Kinetix} ~\footnote{\url{https://www.kinetix.tech/}}, and \textit{Rokoko} ~\footnote{\url{https://www.rokoko.com/}}. 
After the initial investigation, we decide to use \textit{DeepMotion} as it can produce the highest quality and most accurate results for our purpose. 
We converted six 2D video tutorials into 3D avatars using \textit{DeepMotion}. 
These videos include tennis, dance, baseball, yoga, taichi, and exercise. We generated 30-second instructions for each video. 
Based on the extracted human motion, we developed a simple Unity application that shows a 3D avatar animated based on extracted body motion and the associated video on the background in the mixed reality scene through Hololens 2 (Fig~\ref{fig:FormativeStudy}). 

\begin{figure*}[h]
\centering
\includegraphics[width=\textwidth]{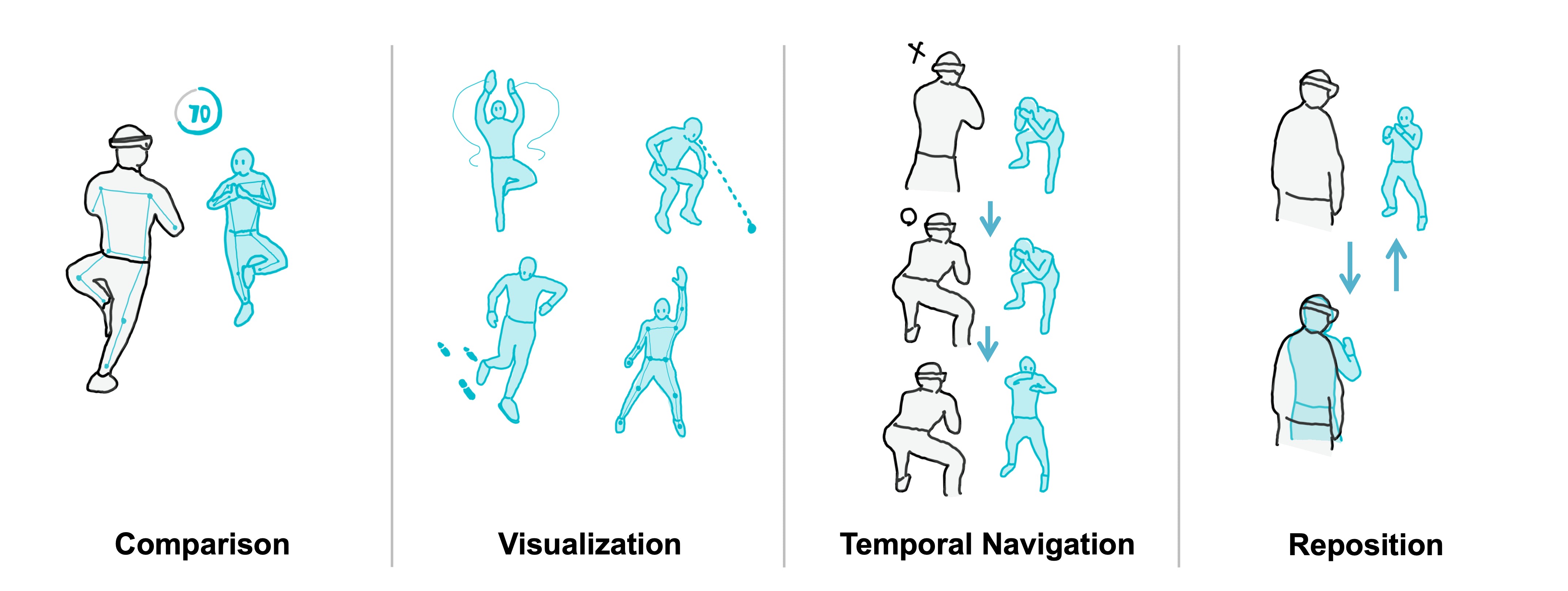}
\caption{Design Space of \system{}}
\label{fig:design-space}
\end{figure*}

\subsection{Formative Study Protocol}
To better understand the benefits and challenges of using 3D avatars generated from 2D videos, we conducted a formative study with eight participants (5 males, 3 females, ages 21 - 38). All participants had previously learned physical sports like dance, soccer, yoga, martial arts, and table tennis through online videos. 

Each interview for the formative study was 50-70 minutes long. All interviews were recorded and later transcribed with the consent of the participants. The purpose of the formative study was to understand the challenges users faced in 3D Mixed Reality tutorials and elicit improvements. 
First, participants were given a brief description of the system they would be experiencing and were then given a demo of six instructional MR experiences. 
The demos were aimed at giving participants an understanding of MR instructions. Following the demos, participants shared their experience of using the prototype and compared it with their previous experience in online learning. Their transcribed responses were thematically coded by two authors, the sections below highlight the themes that emerged in the form of benefits and challenges experienced and expressed by participants.


\subsection{Benefits}
Overall, participants appreciated the 3D avatar instruction. 
A significant advantage highlighted by the participants was that it was easy to understand the spatial orientation of the instructor's poses, a challenge often faced with traditional 2D formats.
\textit{``I think the advantage is quite obvious that if I just watched the 2D video, I may not know so clearly about some 3D things, like, that this leg is in the front and the other is at the back''} (P1). The ability to adjust their viewpoint around the 3D avatar was particularly appreciated (P3-4,P6, P8). \textit{``I guess something that's useful - being able to walk around it. Because it is 3D, its useful, as sometimes from certain camera positions, you lose sight of things.''} (P4) \textit{``It's nice to see the backside because sometimes that's easier to follow [...] I like that as an option and that's something that 2D videos lack''} (P3). 

Another benefit that participants mentioned is the presence of the instructor in the MR environment. One participant mentioned that \textit{``the biggest difference that I can feel is that the model is really here I feel myself more immersed''} (P6). Another participant echoed this aspect by saying \textit{``it's kind of closer to the presence of another person explaining than rather than 2D video.''} (P7). 

Conversely, participants also appreciated that the instructor wasn't physically present, allowing for pressure-free learning (P5, P7). \textit{``One thing I like about learning through visual media is that I won’t be judged by another person [...] if I dance poorly I don't want others to see me dance but if I dance in front of a virtual instructor I won't be judged ... it mitigates social awkwardness''} (P5)

\subsection{Challenges and Needs}
On the other hand, participants shared several challenges and needs they experienced and discuss how the MR experience can be improved by resolving these problems. 
Through the interview, we identify the following four main challenges of the current simple prototype.

\subsubsection{\textbf{Posture Comparison: Needs to Easily Compare between User's and Instructor's Motion}}
One of the benefits of mixed reality instructions highlighted by participants is that the user can see the instructors in the real-world scale (P3, P6-8). This helps participants to mimic the instructor's movements, but they felt that the current system could do more to further enhance this benefit.  \textit{``Whenever you are practicing, the avatar can only show you what to do, but it cannot tell you what you are doing wrong.''} (P6) \textit{``It would be great if you could compare your position to the 3D avatar model and if there is some way that it can detect something like - your arm isn't high enough because at least for me it's hard to know if I am doing the same thing as the video''} (P3). 
To address this gap, participants suggested enhancements to the MR instructions by providing real-time feedback through comparison visualizations.
\textit{``if you had some intelligence that could criticize you and like it could actually see your motion and go you're doing this wrong, you know, why don't you modify this? This is what yours looks like here is us and adapt in real-time that would be super cool.''} (P2) 

\subsubsection{\textbf{Focusing and Highlighting: Difficulty in Tracking Specific Body Parts}}
Participants also express the need to focus on the motion of specific body parts (P1-2, P4-7). \textit{``Like when he's swinging, you want to see the hand and when he is waiting, you want to see the leg.''} (P1) \textit{``And in table tennis the ball is important but even the hand motion is important.''} (P2) Participants highlighted the importance of focusing on the motion of a body part which is the point-of-interest in the instructional experience (P4, P6-7). \textit{``I have noticed that some videos are focused on a particular body part, like for moonwalk we focus on the feet''}(P4) They also further shared visualizations which could be potentially useful. \textit{``I was thinking of showing trails if I wanted to track one of his hands showing that it moved from here to here - showing the path of hands or feet, like in animation, you have onion skinning''} (P4)

\subsubsection{\textbf{Temporal Navigation: Challenges in Controlling Speed and Time of the Avatar Instruction}}
Participants mentioned that they find it challenging to navigate through the instructions, especially when they want to simultaneously follow the instructions and navigate through the video content (P1, P3, P5-7). They shared a strong need for an intuitive way to navigate through the instructional content (P1, P3, P5, P8). \textit{``I usually have to stop the video when I miss a certain part of the video, sometimes rewind the part I want to mimic''} (P3). Participants also suggested techniques to make it more useful. \textit{``If the video could adapt to my progress that would be cool because I feel sometimes I couldn't follow the videos if they are too fast (...) So if they could adapt like if I am still doing the last [step] it could slow down a little bit''} (P5). \textit{``If I want to skip to the other pose I know, but want to re-watch, I can just do that pose. I want to control the progress of the video in a natural way, like moving your body.''} (P1).


\subsubsection{\textbf{Spatial Reposition: Needs of Seamlessly Switching between First- and Third-Person Perspectives}}
Participants also mentioned they often would want to teleport to the instructor's first-person perspective, as they believe it would facilitate a more accurate replication of the instructor's actions (P2, P6).
\textit{``Sometimes with these types of instructions, we get left and right mixed up so being able to switch to a first-person view could be helpful.''} (P4)
Participants also mentioned that they need to alternate between watching a video and monitoring their own actions to ensure they are following instructions accurately (P2, P8). The first-person perspective could also solve this juggling problem.
``that's always the challenge, you're doing a move and they're saying now do this and you're like, well, I can't look at you because you're over here and my face is supposed to be turned the other way.'' (P8) \textit{``...then you can't see the video, but it doesn't have to be because now that it's in [first-person] AR you can always look at it.''} (P2)

%% file: 4-system.tex
\section{\system{}: Augmenting Auto Generated MR Instructions}

Based on the insights, we designed \system{}, a system that augments the automatically generated mixed reality instruction. 
Similar to the initial prototype, we used \textit{DeepMotion} to convert 2D video into 3D animation, and show the avatar model chosen from Mixamo~\footnote{\url{https://www.mixamo.com/}}. 
We utilized Unity (version 2020.3.35f1) and Mixed Reality Toolkit (version 2.8.3) to create our system.
The scene is rendered in HoloLens2~\footnote{\url{https://www.microsoft.com/en-us/hololens/}} that is connected to a laptop PC (G-Tune, Intel Core
i7-11800H 2.30GHz CPU, NVIDIA GeForce RTX 3060 GPU, 64GB RAM) to show the MR scene through the Holographic Remoting Player~\footnote{\url{https://learn.microsoft.com/en-us/windows/mixed-reality/develop/native/holographic-remoting-player}}.
We also capture the user's body using one Azure Kinect RGB-D camera~\footnote{\url{https://azure.microsoft.com/en-ca/products/kinect-dk}} placed in front of the user with a tripod.
The Kinect camera is connected to the same laptop through a USB-C cable. 
We use both skeleton position data and 3D mesh data for real-time capturing. 
Additionally, we utilized the Azure Kinect Examples for Unity package~\footnote{\url{https://assetstore.unity.com/packages/tools/149700}} to generate 3D meshes from RGBD data, enabling users to view their 3D posture. Azure Kinect's Body Tracking feature allowed for the estimation of joint angles, capturing 32 joint angle data from users~\footnote{\url{https://learn.microsoft.com/en-us/azure/kinect-dk/body-joints@:}}. Similarly, DeepMotion could acquire rotational data for 20 joints within a video~\footnote{\url{https://blog.deepmotion.com/2020/11/19/animate-3d-custom-characters/}}.

Based on the formative study, we design the following four features: 1) \textbf{\textit{Posture Comparison:}} contrasting and highlighting differences between user and avatar postures, 2) \textbf{\textit{Motion Visualization:}} visualizing key trajectories and movements of specific body parts, 3) \textbf{\textit{Embodied Temporal Navigation:}} embodied manipulation for time and speed through body motion, and 4) \textbf{\textit{Avatar Repositioning:}} spatially repositioning avatars for different perspectives. 
All of these features are automatically generated based on the extracted body motion and user posture captured by Azure Kinect. 

\begin{figure}[h]
\centering
\includegraphics[width=\linewidth]{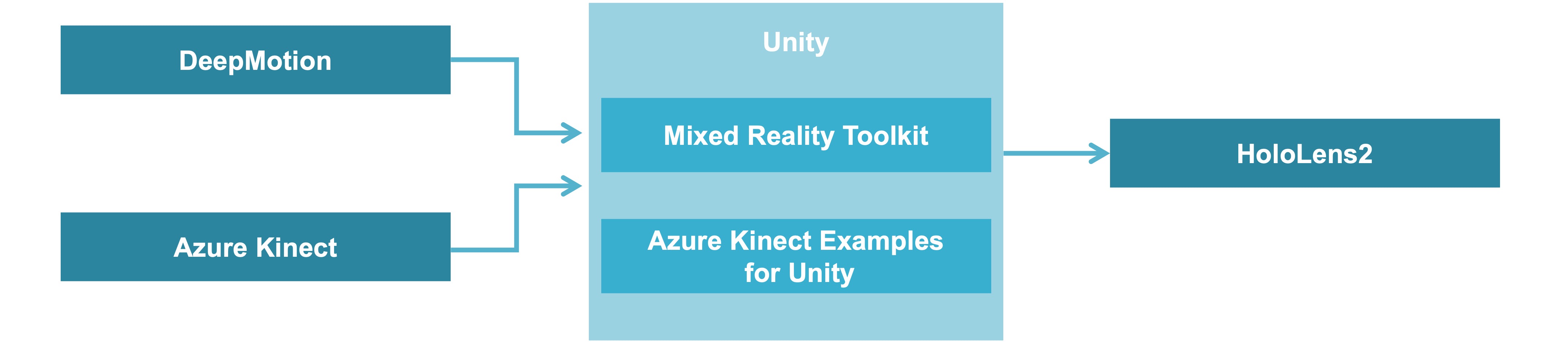}
\caption{System Overview}
\label{fig:design-space}
\end{figure}

\subsection{Posture Comparison}
In the formative study, users sometimes wanted to learn by mimicking the instructor's movement. 
To support this, the system shows the user’s real-time mesh on the side or top of the instructor's avatar. 
We not only show their mesh but also calculate the difference between the user's and instructor's posture and provide feedback by changing the color or providing the synchronization score. 

\subsubsection{Pose Match Indicator}
\begin{figure}[h]
\centering
\includegraphics[width=0.32\linewidth]{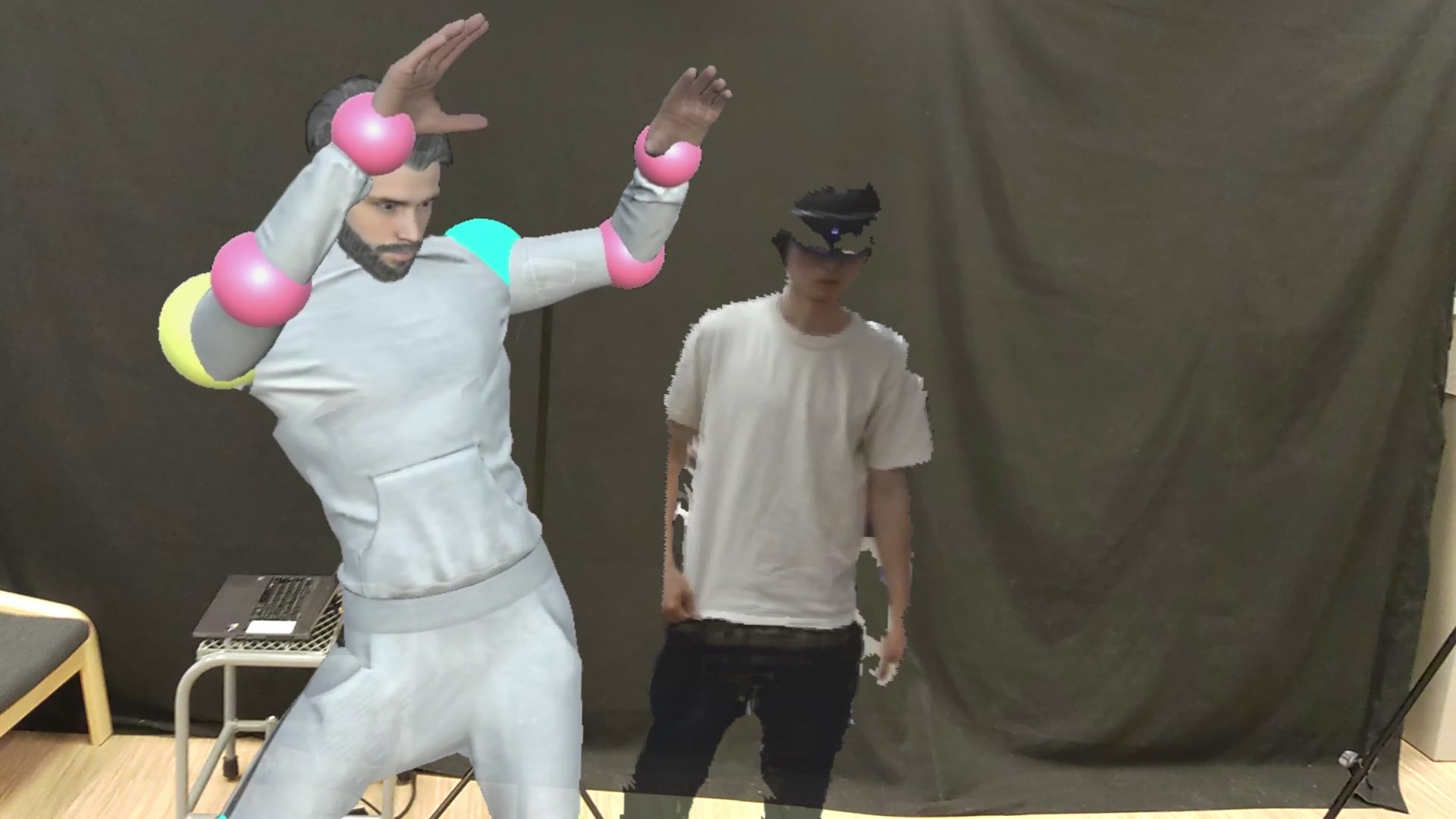}
\includegraphics[width=0.32\linewidth]{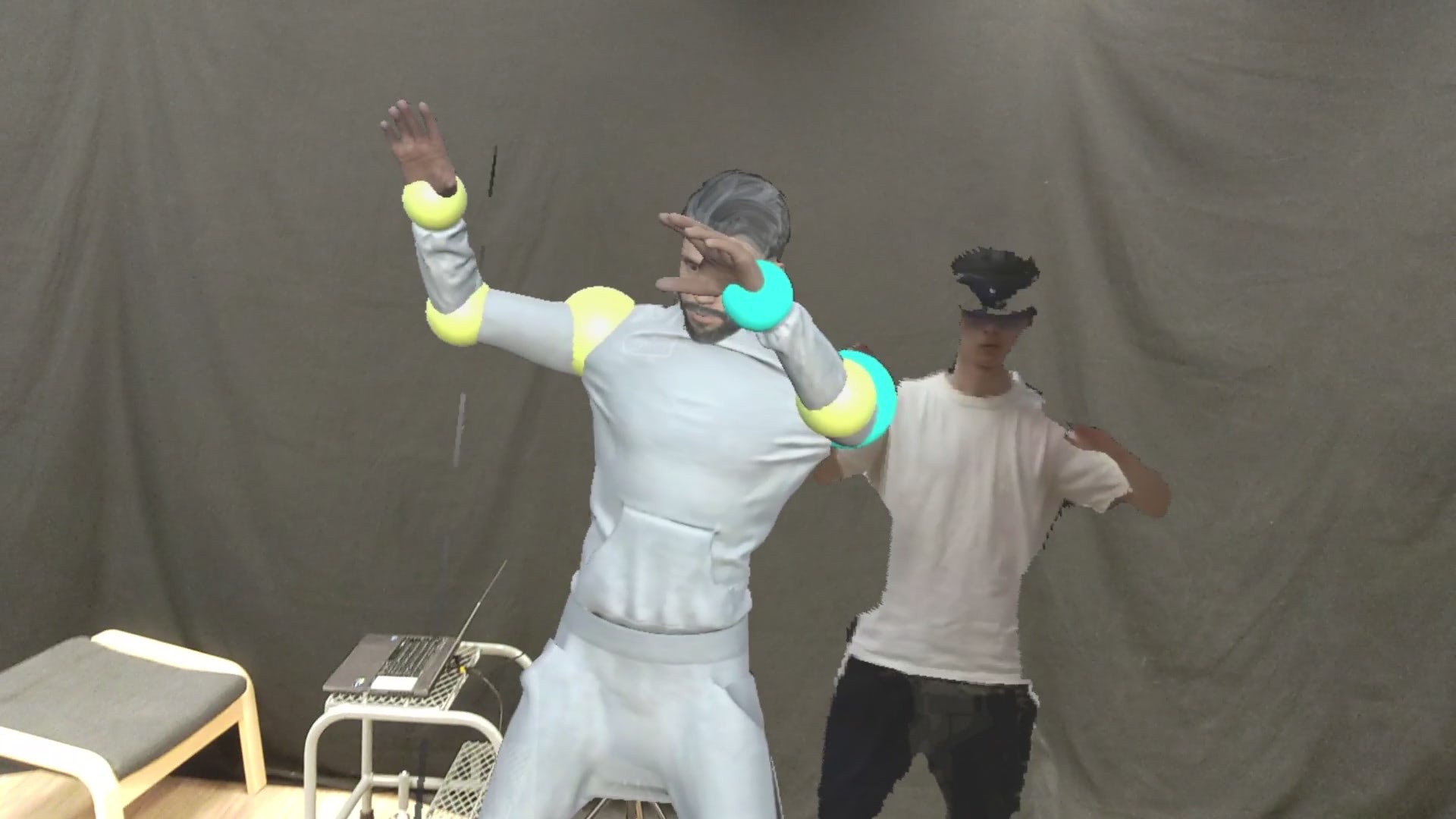}
\includegraphics[width=0.32\linewidth]{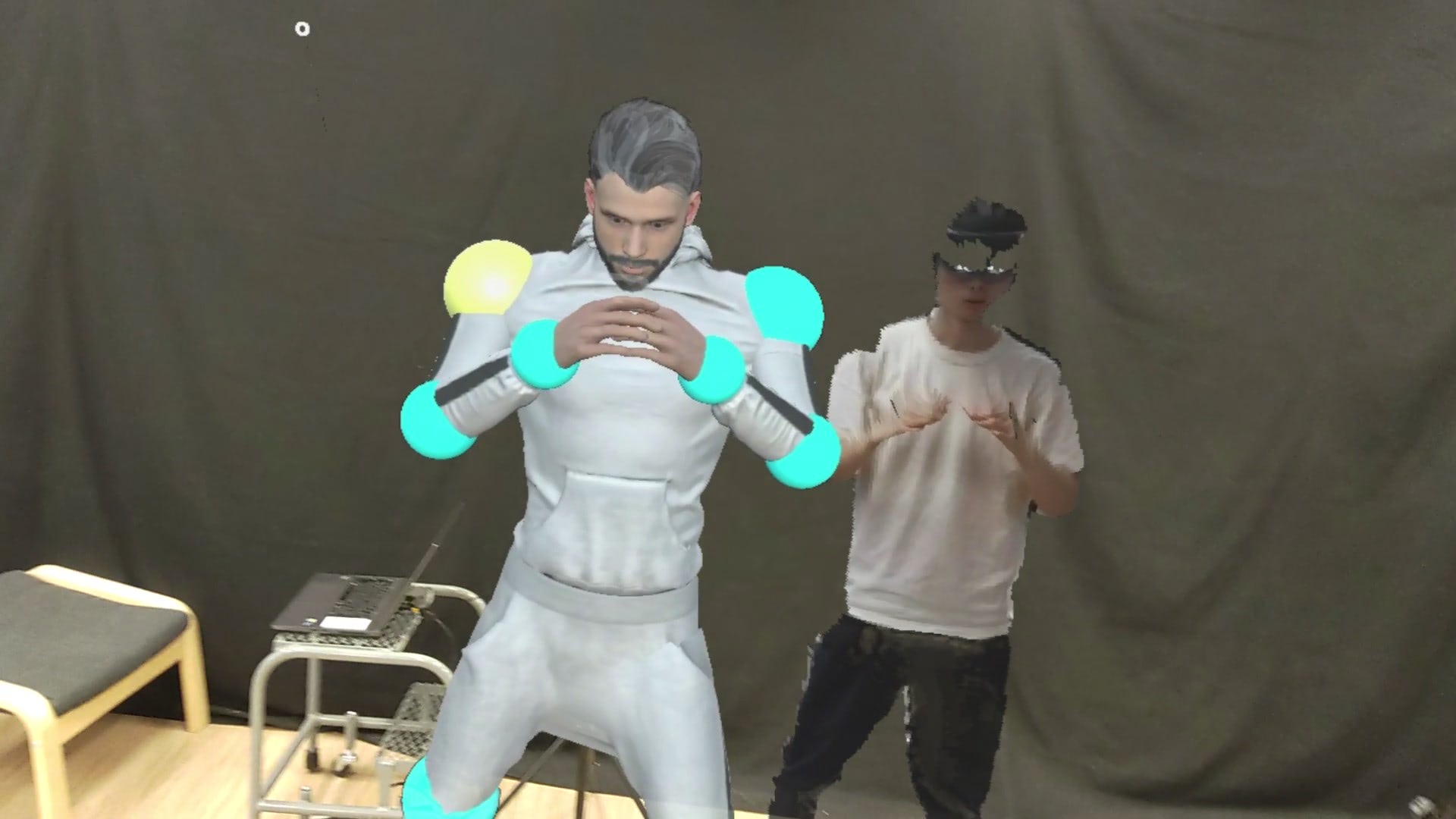}
\caption{Indicator}
\label{fig:posematchindicator}
\end{figure}

The accuracy of user postures relative to the instructor's avatar is crucial for ensuring effective practice and understanding of movements. 
This indicator provides real-time feedback to users by visualizing the alignment discrepancies between the user's limbs and those of the avatar.
Specifically, colored spheres are dynamically displayed on the avatar's limbs, including left and right arms, and left and right legs. The color of each sphere indicates the degree of alignment. 
Blue spheres signify close alignment, yellow indicates minor deviations, and red marks significant misalignments (Fig.~\ref{fig:posematchindicator}). 
For example, during a complex dance sequence, users may struggle to synchronize their footwork with the instructor's. 
The Pose Match Indicator visually alerts them of their inaccuracies, allowing for immediate correction and improvement.

\subsubsection{Pose Match Score}\label{posematchscore}
\begin{figure}[h]
\centering
\includegraphics[width=0.32\linewidth]{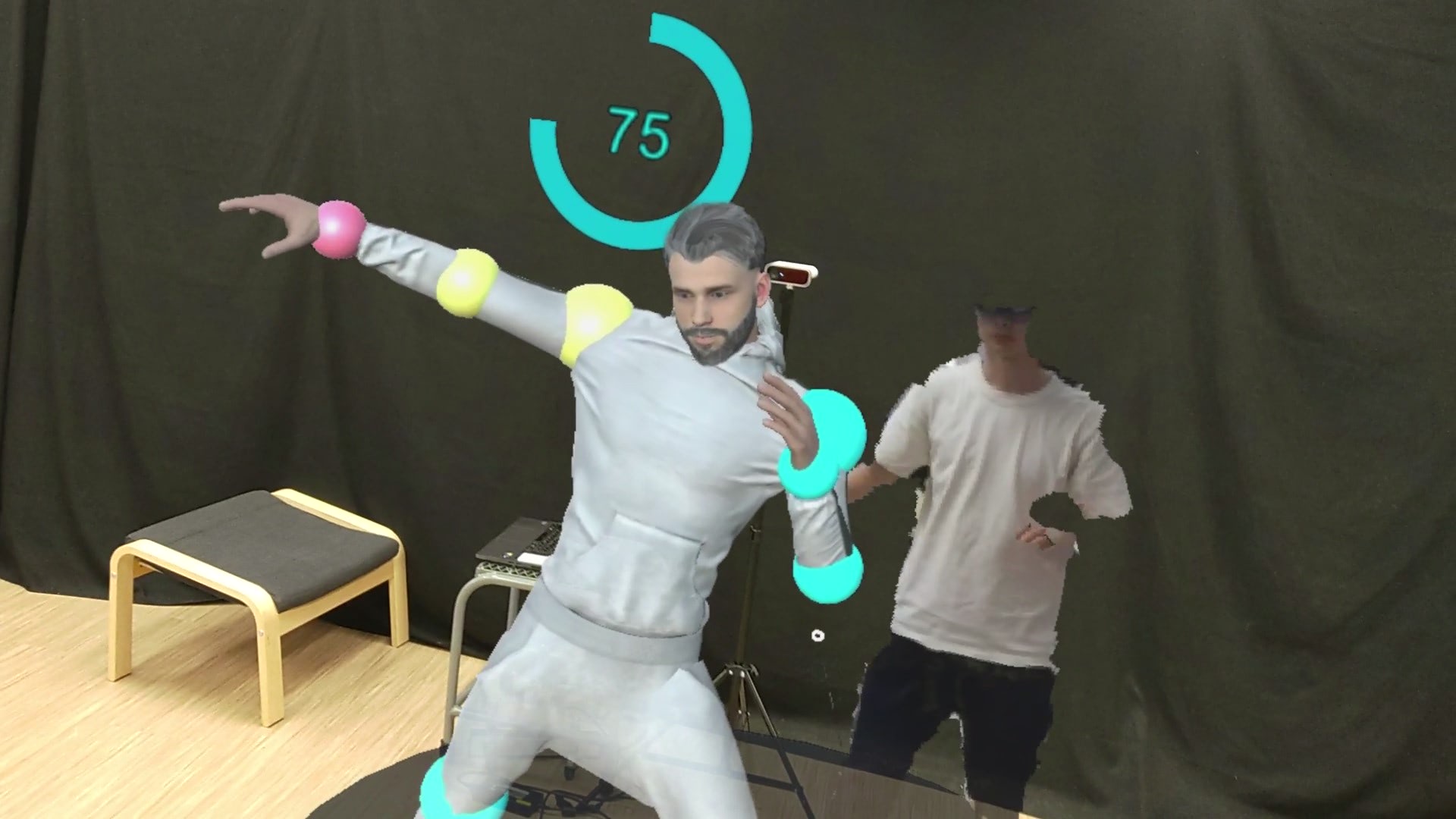}
\includegraphics[width=0.32\linewidth]{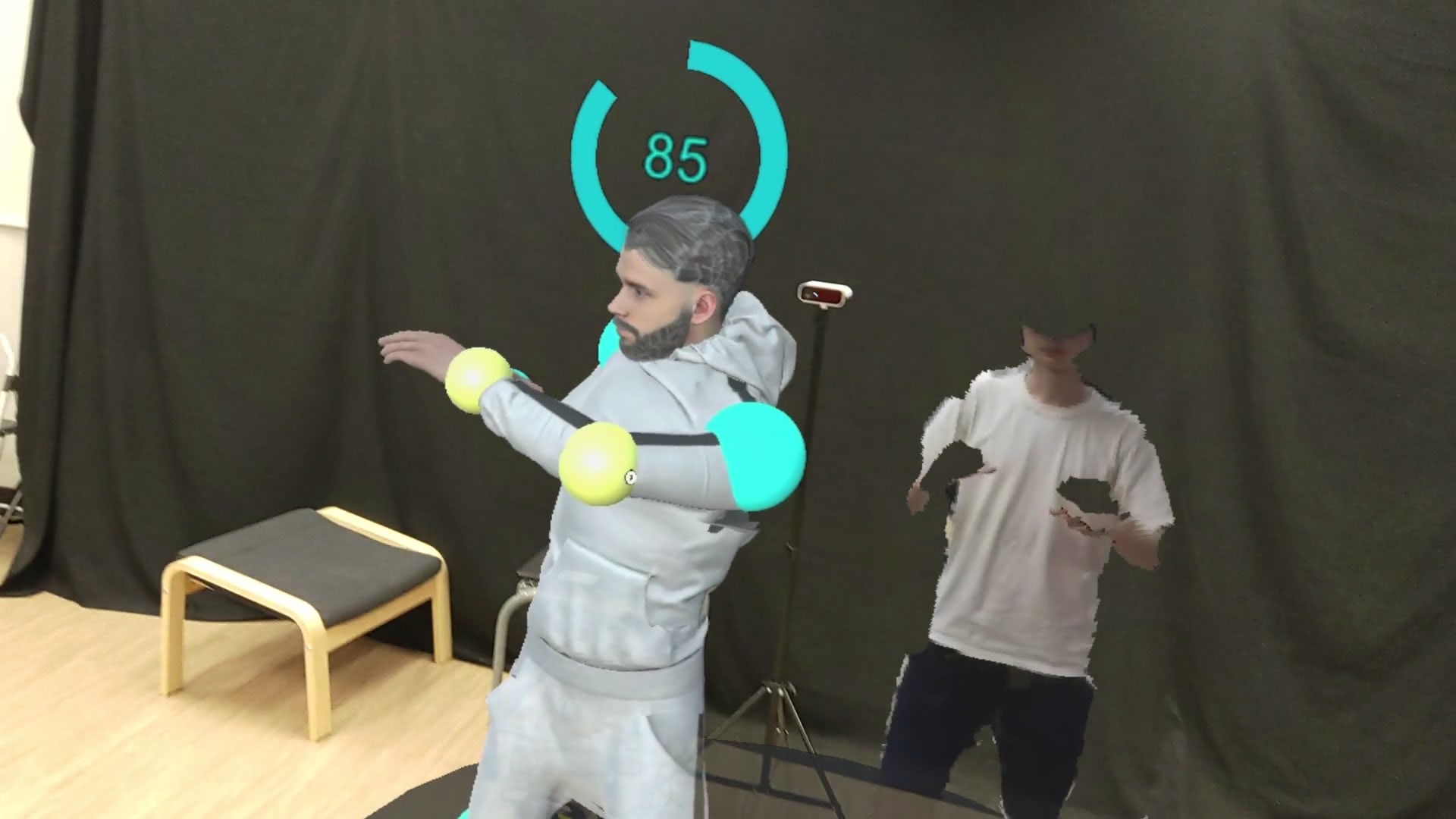}
\includegraphics[width=0.32\linewidth]{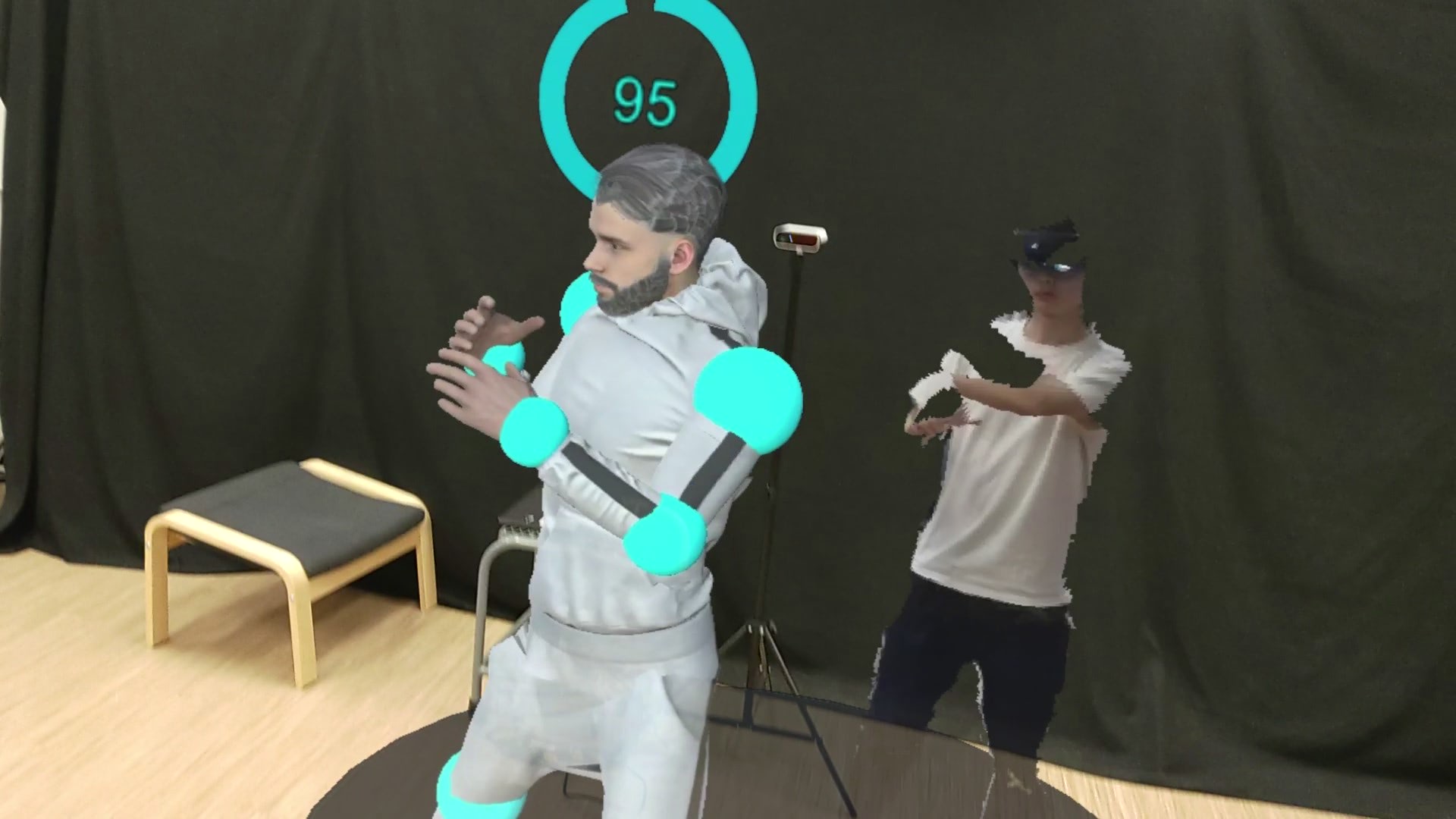}
\caption{Pose Match Score}
\label{fig:PoseMatchScore}
\end{figure}

The pose match score quantifies the alignment between the user's posture and the instructor's ideal posture in an abstract way (Fig.~\ref{fig:PoseMatchScore}).
The score is shown as a numeric value and a circle graph.
The score is calculated based on the difference between the instructor's and the user's body positions.
We first calculate the score for each joint, add all the scores for each joint, and then show the total score.
In this system, we used ten joints, each with a maximum of 10 points, and the maximum total score is 100.
This could be used when overall body alignment is more important than the pinpoint accuracy of individual joints.
For example, in a boxing workout, we don't have to match the posture with the instructor accurately, but we have to make the whole body posture similar to the instructor's for an extended period.

\subsection{Motion Visualization}
The user also sometimes wants to focus on the motion of specific body motions.
For example, for dancing, the user wanted to focus on the foot motion.
To address this, we highlight the specific motions including head gaze, footprints, and trajectory.

\subsubsection{Head Gaze}
\begin{figure}[h]
\centering
\includegraphics[width=0.32\linewidth]{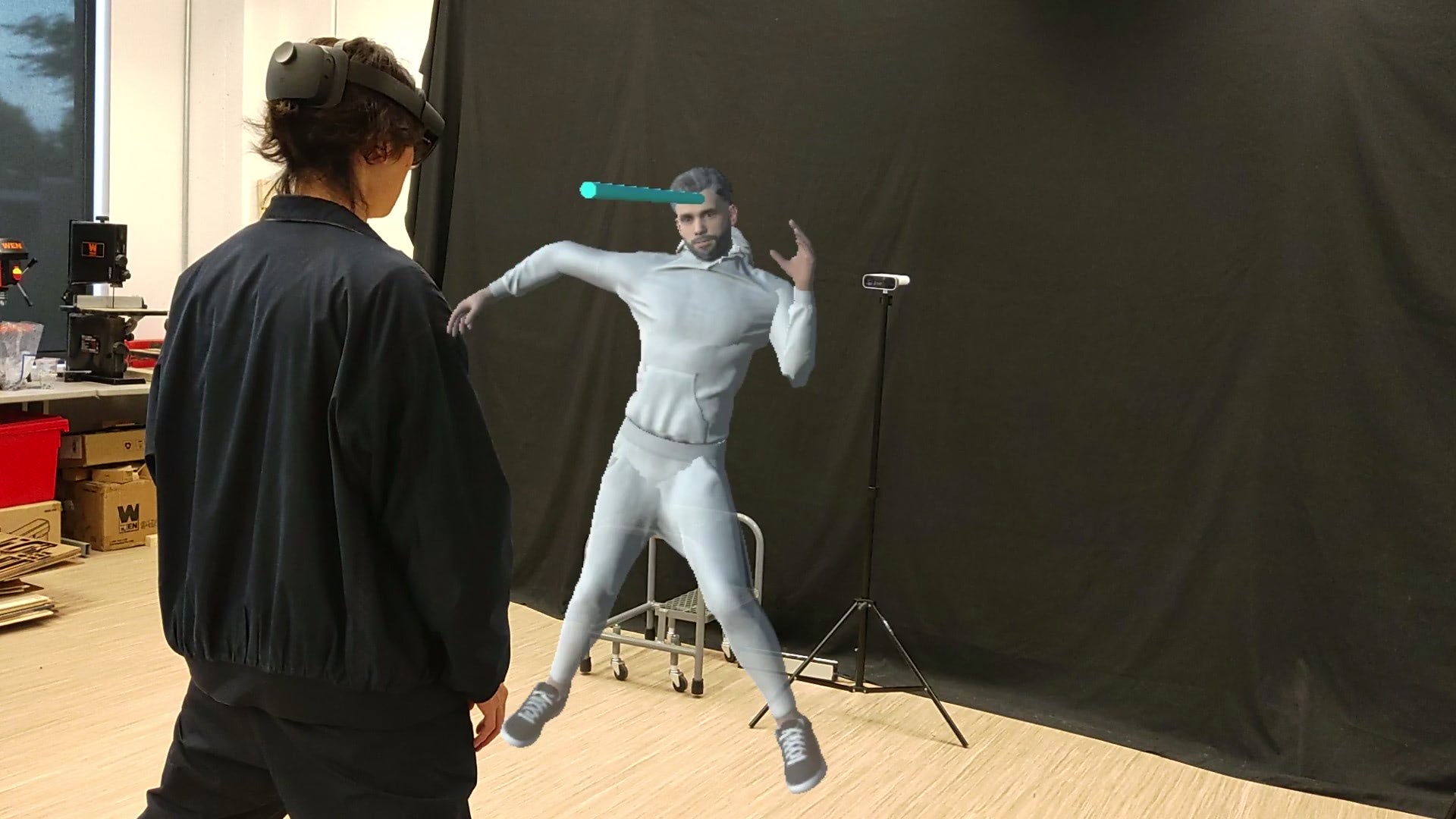}
\includegraphics[width=0.32\linewidth]{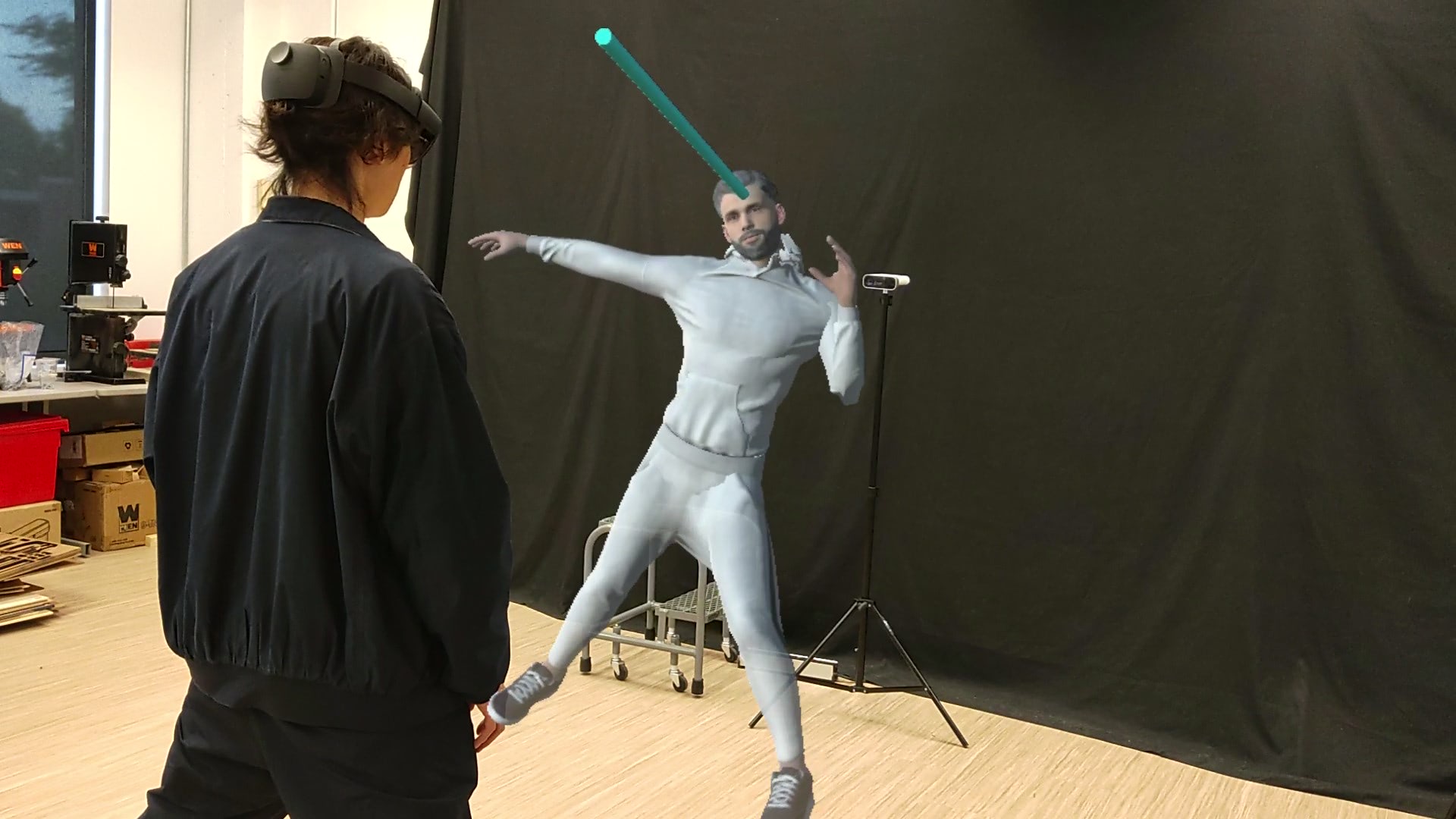}
\includegraphics[width=0.32\linewidth]{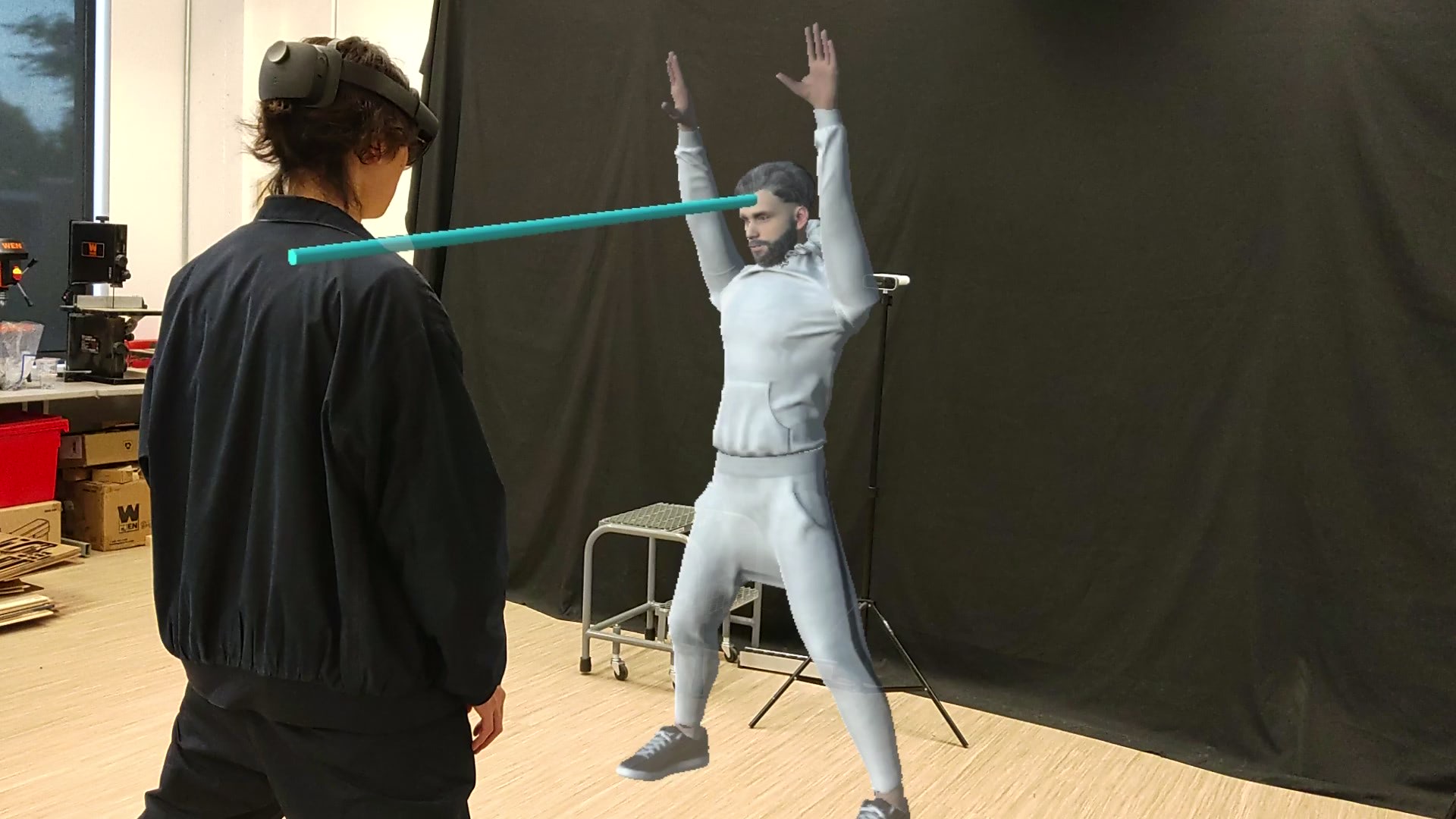}
\caption{Head Gaze}
\label{fig:headgaze}
\end{figure}

Some previous work visualized the body direction by using lights~\cite{turmo2020bodylights, turmo2019enlightened}.
Inspired by these works, we visualize the head gaze of the instructor.
The instructor's head gaze can provide valuable insights into where their focus is directed during the activity.
Showing this can improve the users' understanding of the instructor's intent or the adequate head direction.
The gaze is shown using the ray, which starts from the instructor's avatar's head (Fig.~\ref{fig:headgaze}).
For instance, in dance instruction, although it is difficult to understand the head direction while overlaying the body to the instructor's avatar, by using this, the user can understand the direction by looking at the gaze ray. 

\subsubsection{Trajectory}
\begin{figure}[h]
\centering
\includegraphics[width=0.32\linewidth]{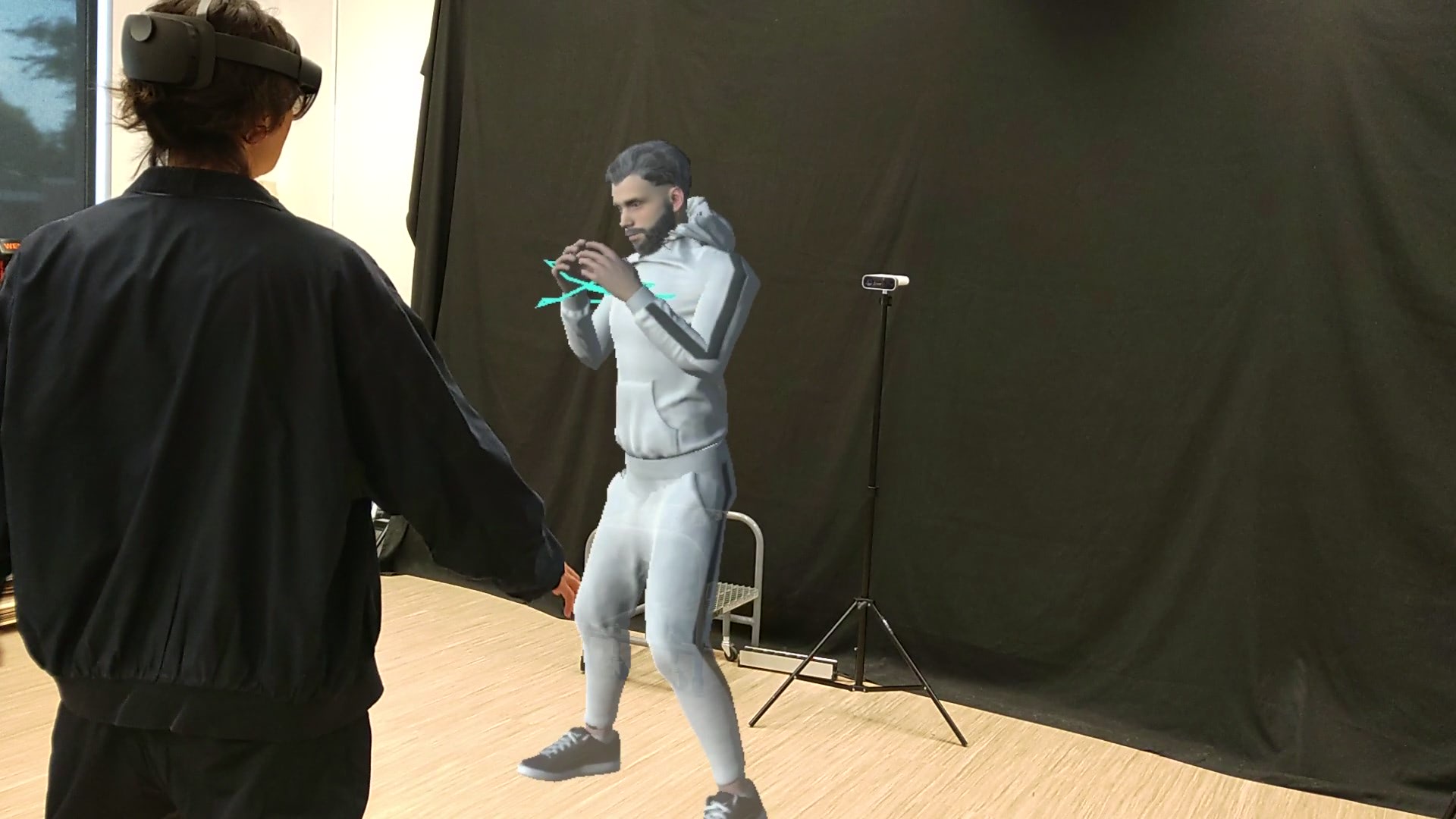}
\includegraphics[width=0.32\linewidth]{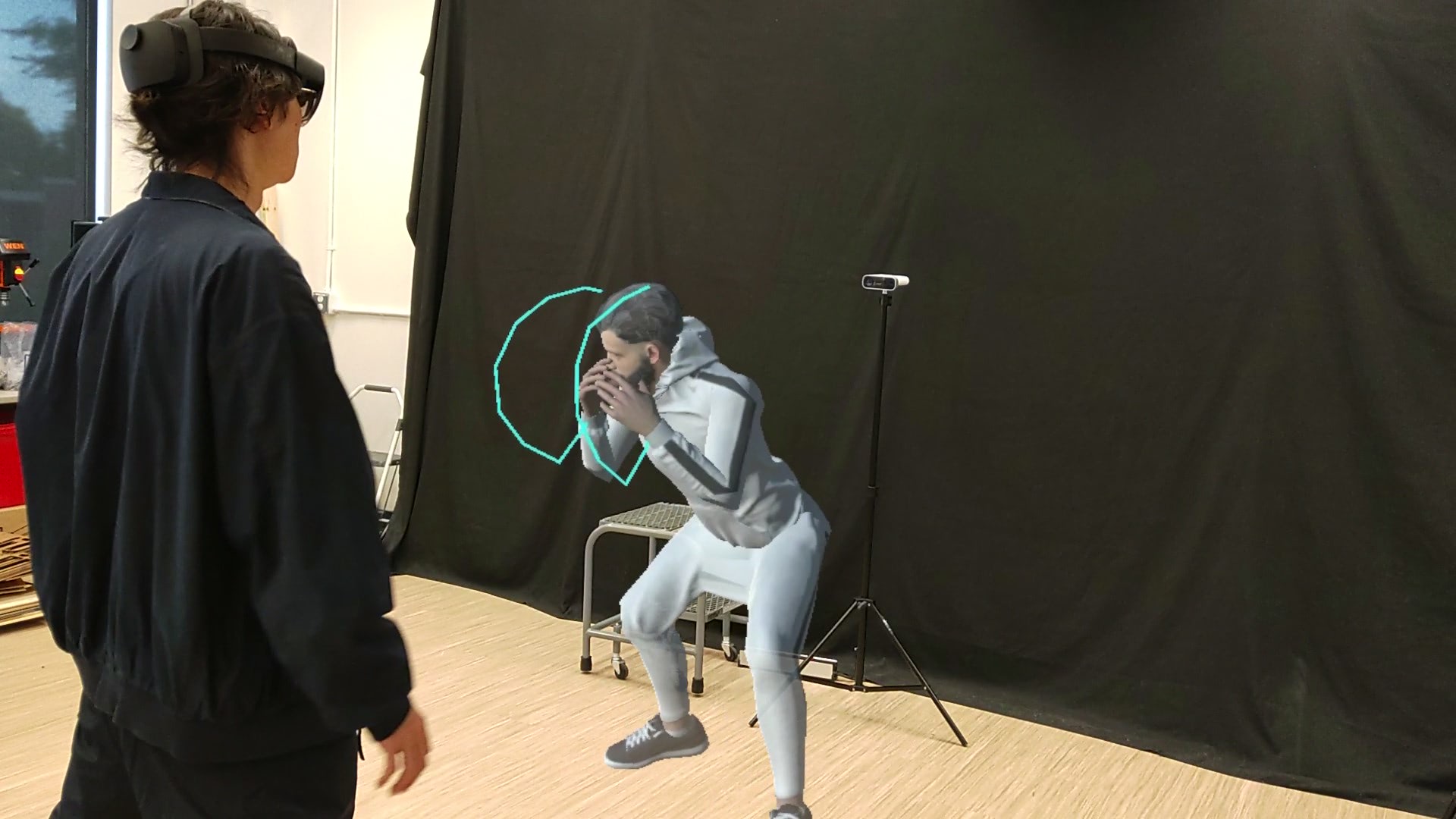}
\includegraphics[width=0.32\linewidth]{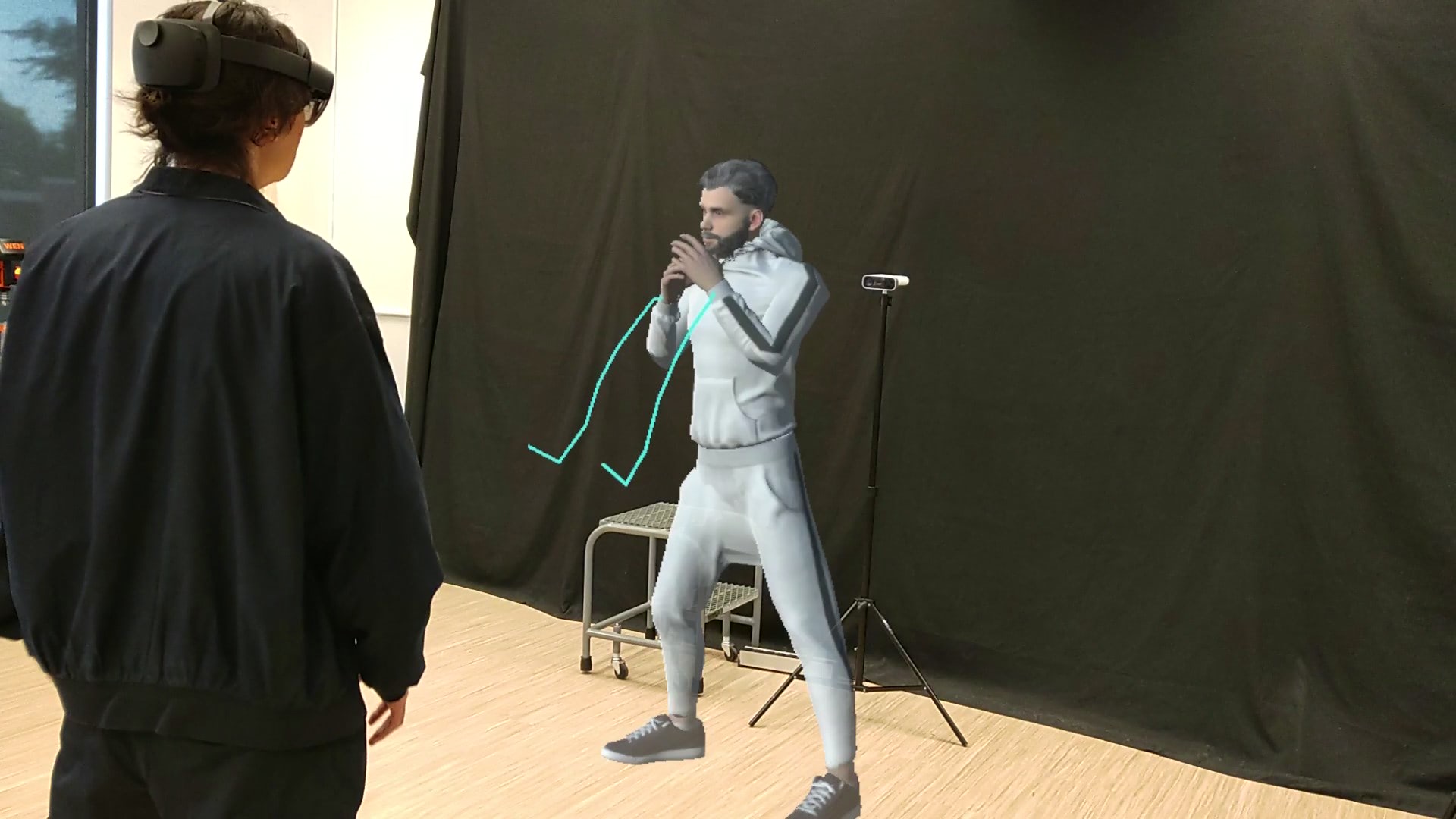}
\caption{Trajectory}
\label{fig:motionpath}
\end{figure}

This feature displays a motion path, providing a visual guide for users to follow.
In Figure.~\ref{fig:motionpath}, the instructor's hand trajectory is shown, which shows the hand's position in the past. 
This can be used for users who struggle to keep up with the instructor's pace due to the complexity or speed of the movements.
For instance, in tennis instruction, users can use this mode to understand and replicate fast and complex swings that might otherwise be challenging to follow.

\subsubsection{Footprints}
\begin{figure}[h]
\centering
\includegraphics[width=0.32\linewidth]{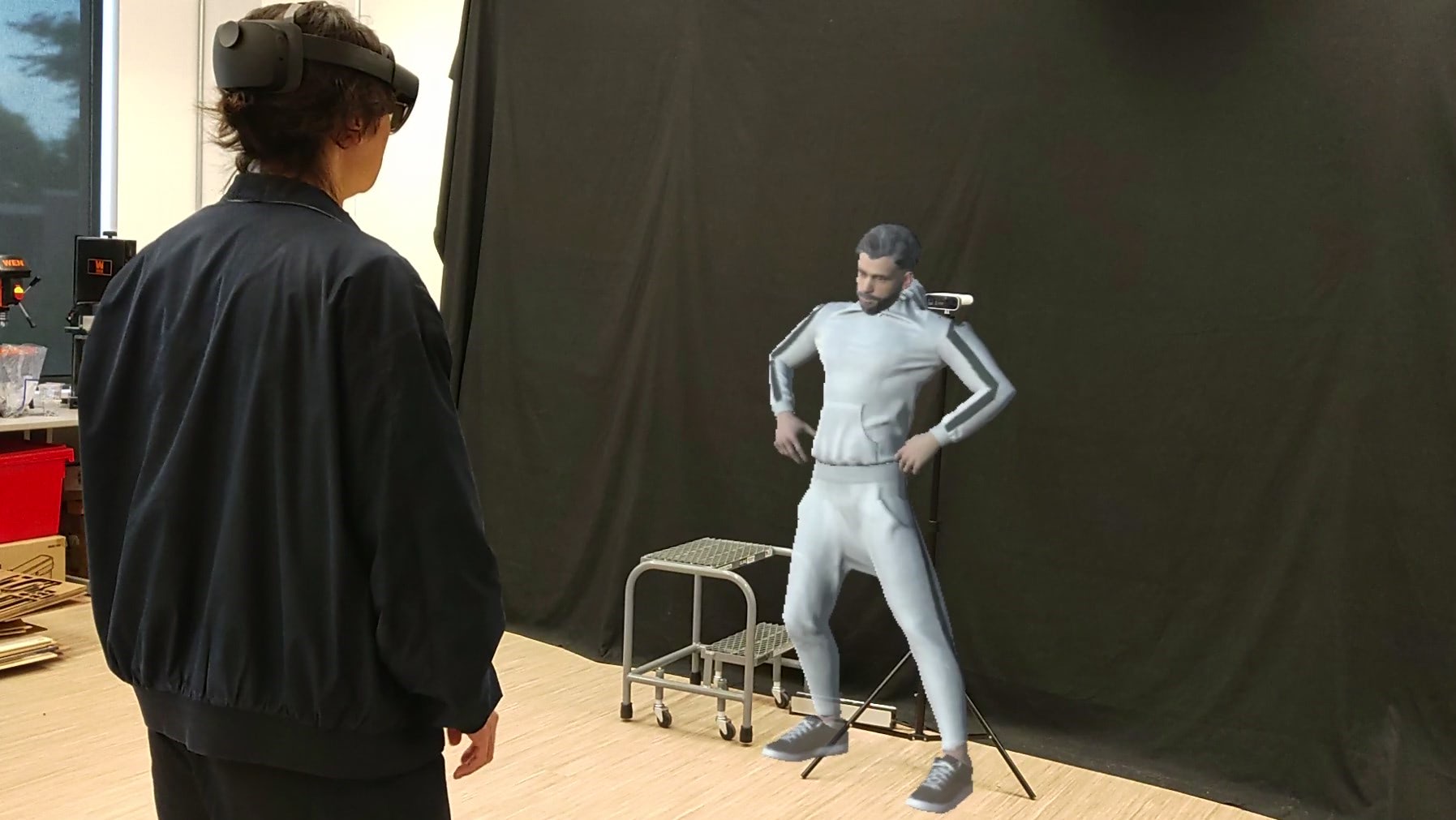}
\includegraphics[width=0.32\linewidth]{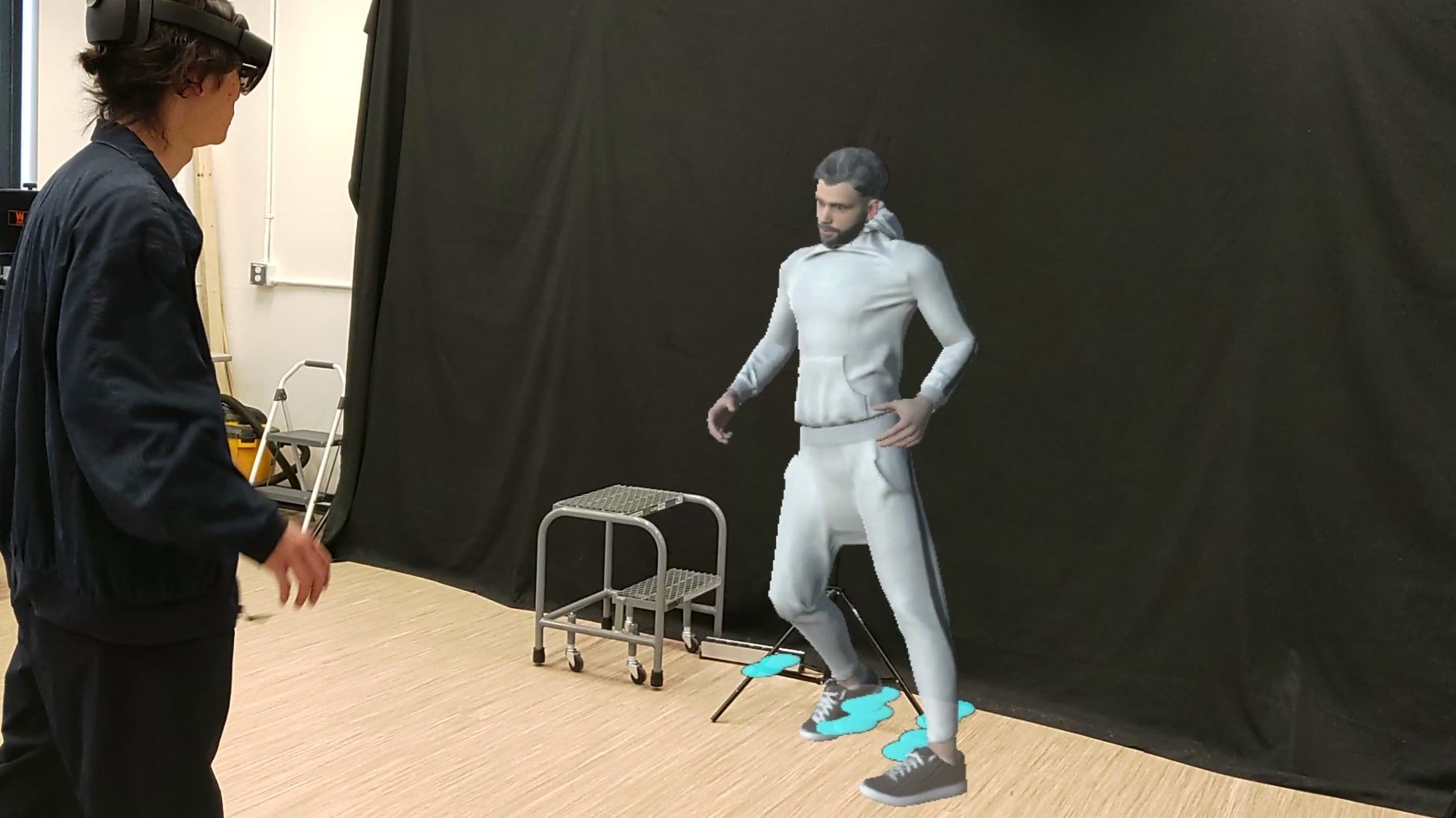}
\includegraphics[width=0.32\linewidth]{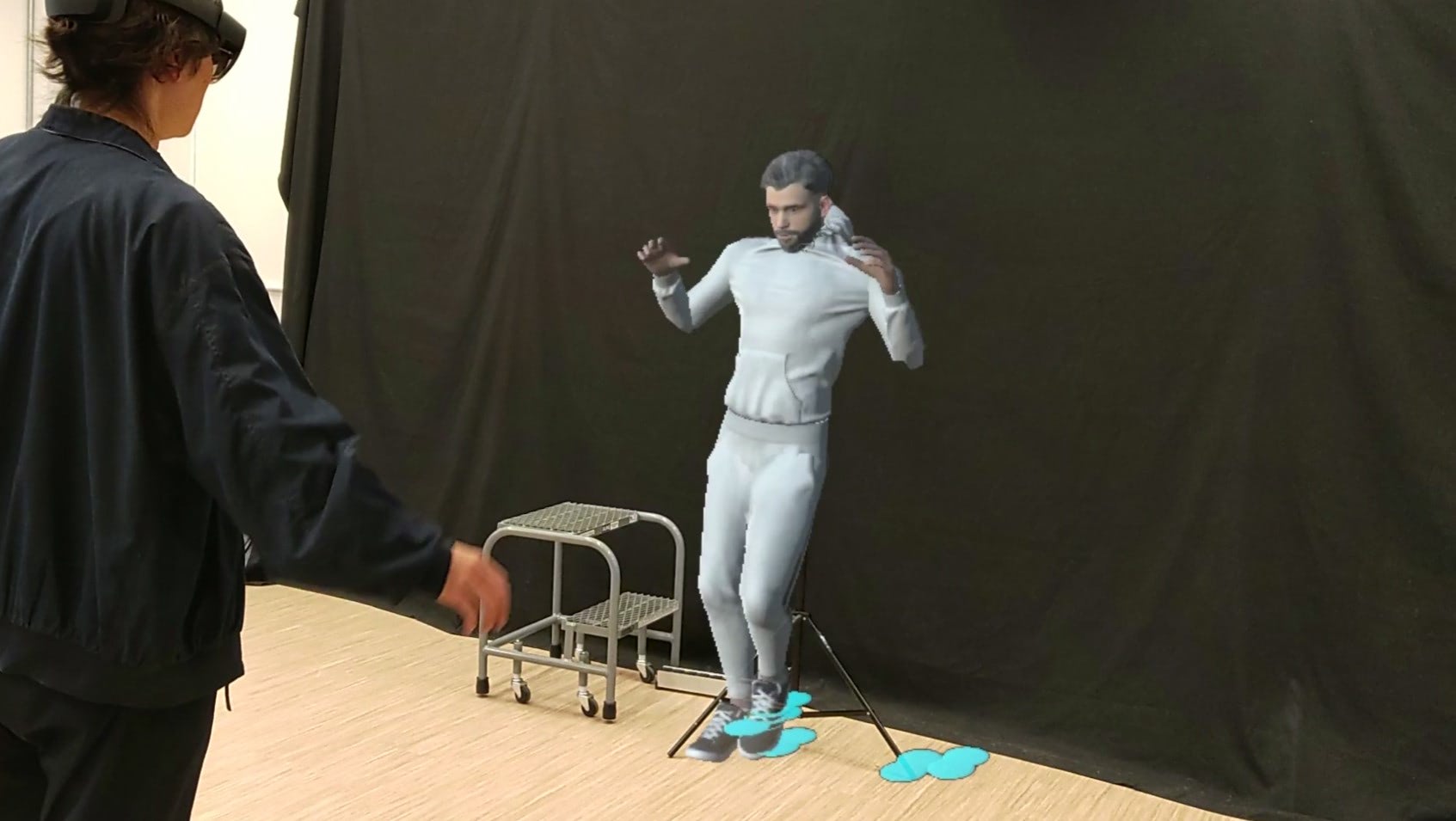}
\caption{FootPrint}
\label{fig:footprint}
\end{figure}

Inspired by Projection-Based AR~\cite{sekhavat2018projection} that projects footprints using a projector, we utilized footprints to highlight the 3D avatar's feet positions.
Footprints provide important information about foot placement, which is often occluded by the instructor's body.
The foot positions of the instructor's avatar are acquired and marked with blue on the ground to indicate the footprints (Fig.~\ref{fig:footprint}).
A new footprint appears every few seconds and subsequently fades by making it gradually transparent.
This feature is beneficial in learning activities such as baseball instruction, where understanding the correct foot positioning is important for mastering the technique of throwing a ball. 

\subsection{Embodied Temporal Navigation}
Inspired by the previous work, such as \textit{Reactive Video}~\cite{clarke2020reactive} and \textit{PoseAsQuery}~\cite{hamanishi2020poseasquery}, that implements body-based navigation for 2D videos, we extended this concept to navigate 3D avatar motions.
Our system allows the user to stop the avatar motion until the user synchronizes the exact same posture. 
Then, the avatar motion starts moving as if the avatar moves step-by-step, based on the user's motion. 
We achieve this by calculating using the scoring system mentioned in \ref{posematchscore}.
We set a threshold score, and if the user's score is better than the score, the avatar will jump the animation timeline to move to the next step.
We set the threshold score high if the high accuracy is important, and we set it low if it is not.
Our system changes how much we jump based on whether the user wants to understand a long duration of instruction quickly or understand the detailed movement of a short duration.

\begin{figure}[h]
\centering
\includegraphics[width=0.32\linewidth]{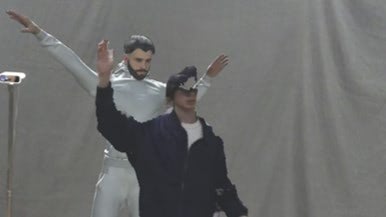}
\includegraphics[width=0.32\linewidth]{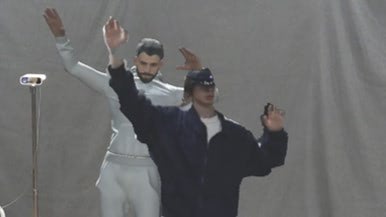}
\includegraphics[width=0.32\linewidth]{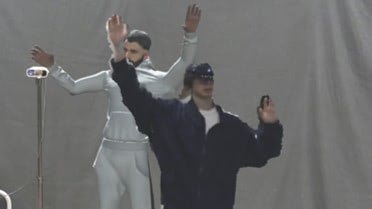}
\caption{Body-based Navigation using User's Avatar}
\label{fig:Body-based}
\end{figure}


\subsection{Avatar Repositioning}
\begin{figure}[h]
\centering
\includegraphics[width=0.32\linewidth]{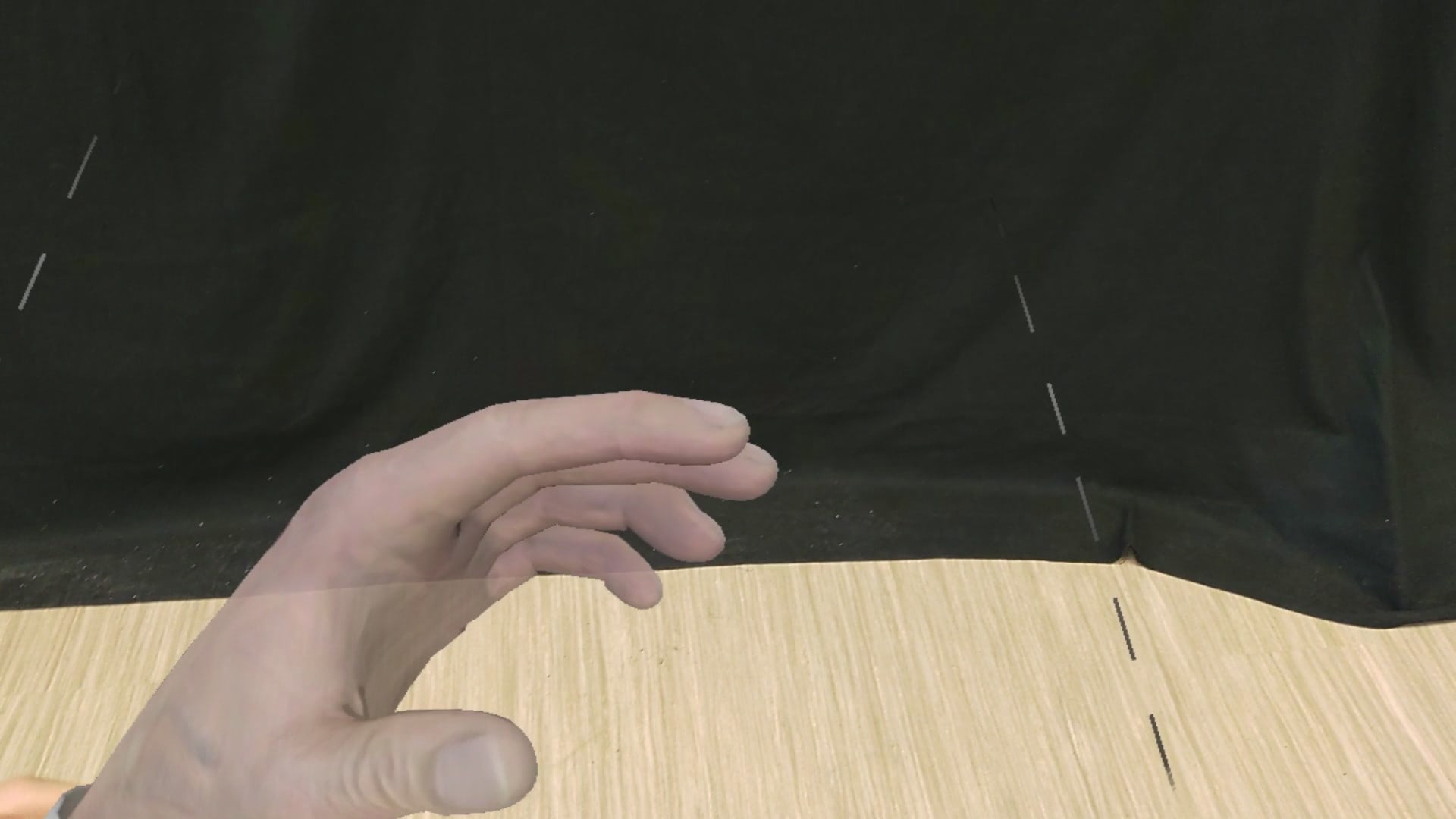}
\includegraphics[width=0.32\linewidth]{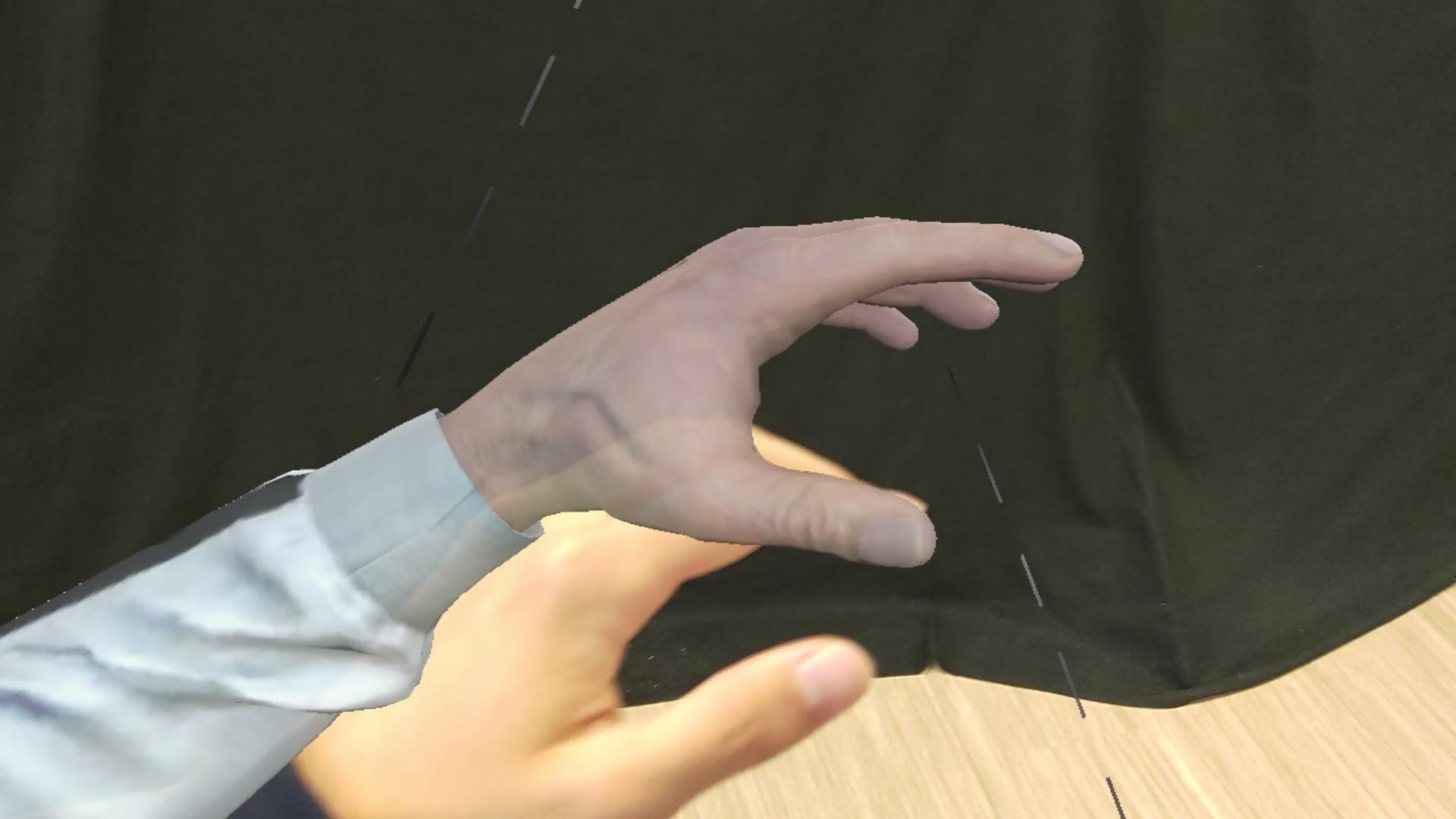}
\includegraphics[width=0.32\linewidth]{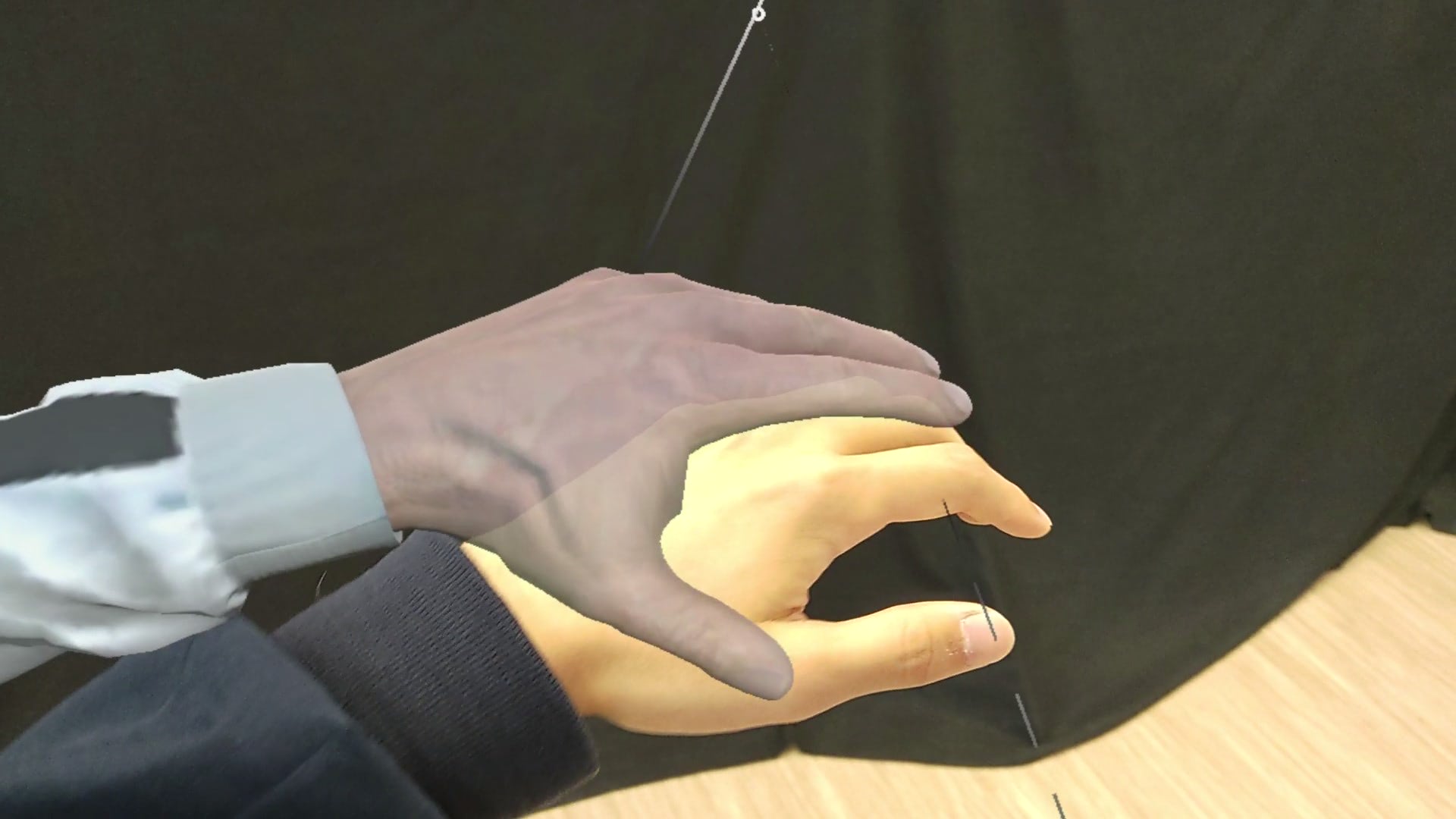}
\caption{First-Person View}
\label{fig:First-personView}
\end{figure}

\begin{figure}[h]
\centering
\includegraphics[width=0.32\linewidth]{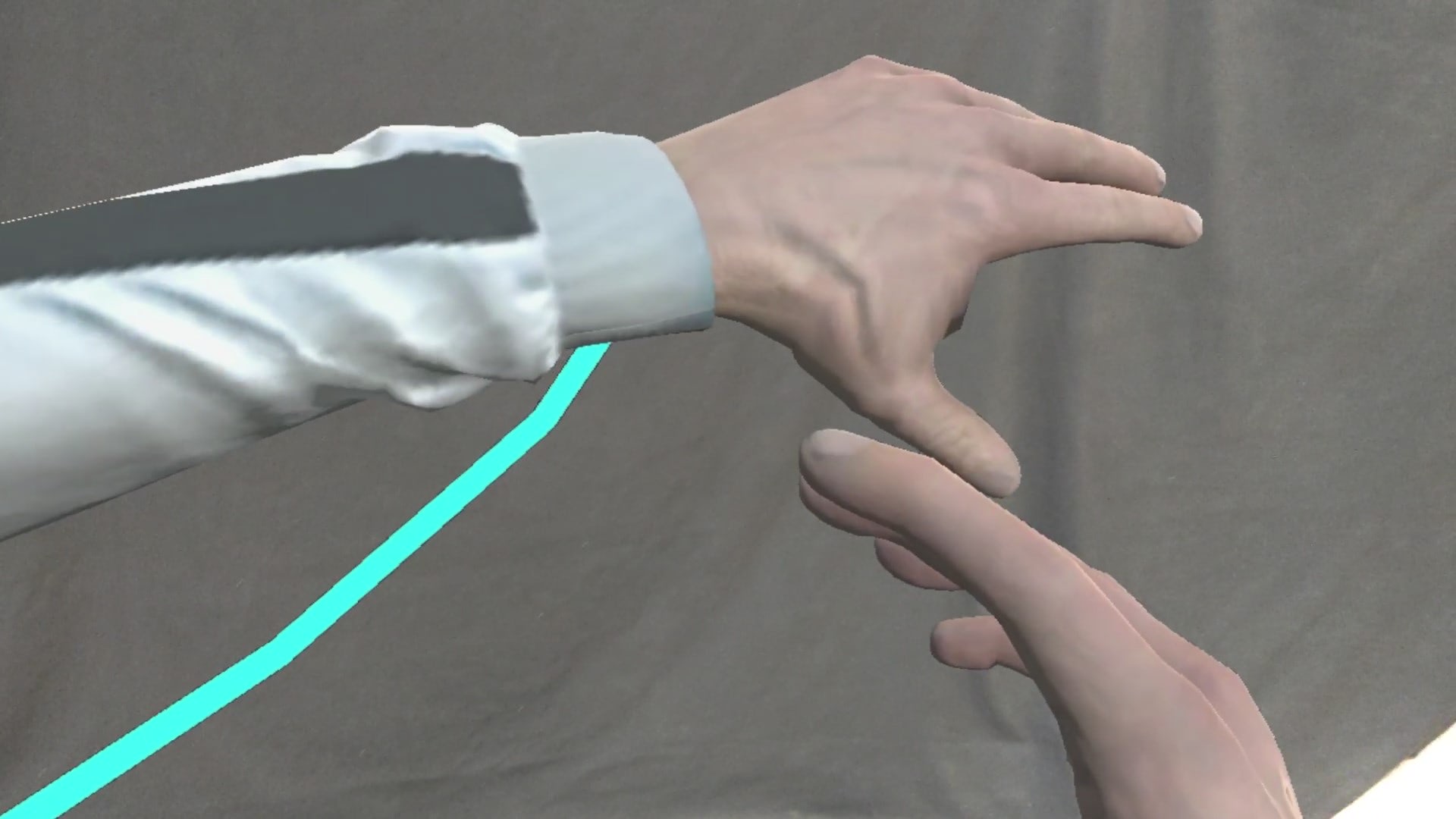}
\includegraphics[width=0.32\linewidth]{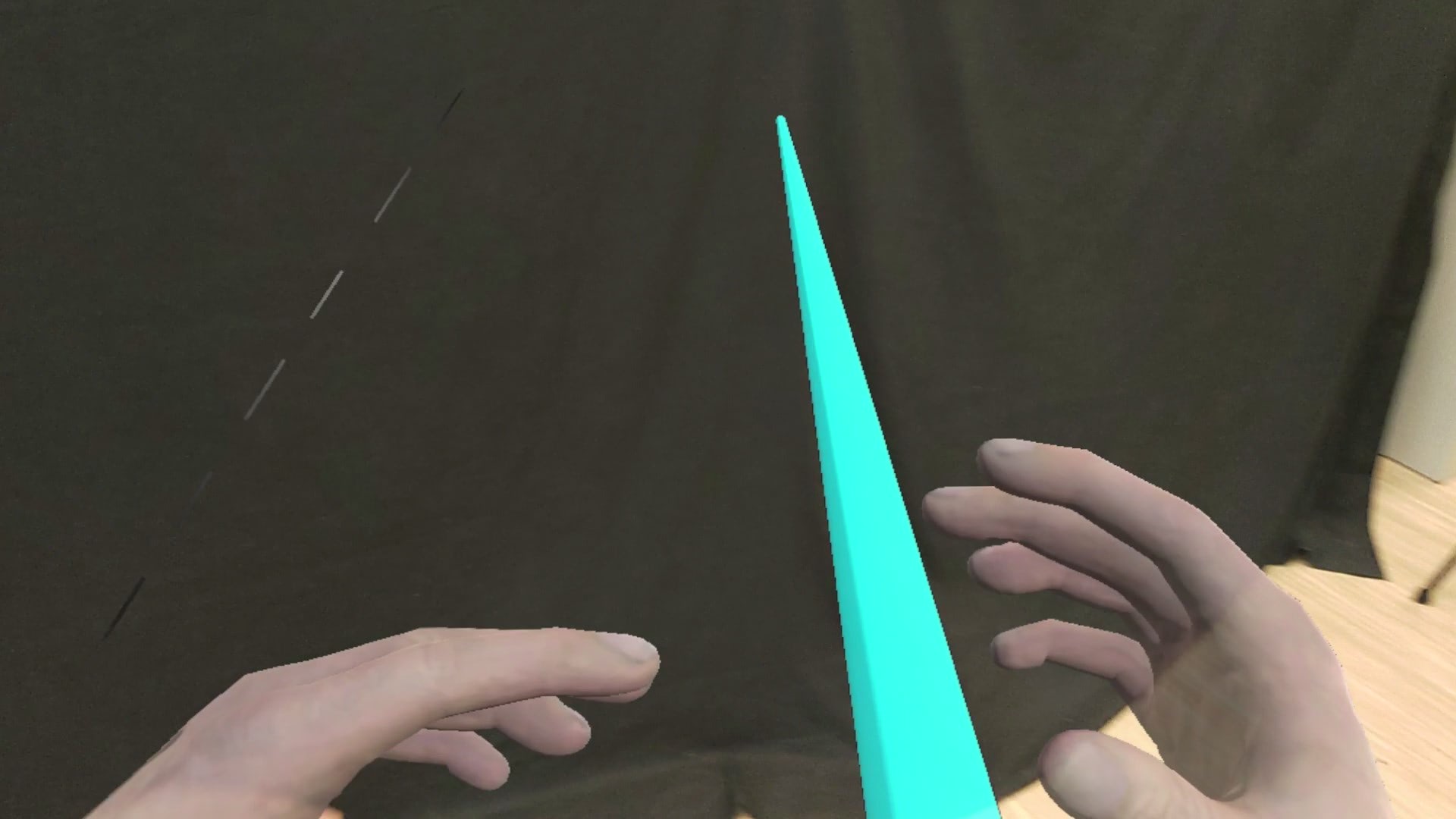}
\includegraphics[width=0.32\linewidth]{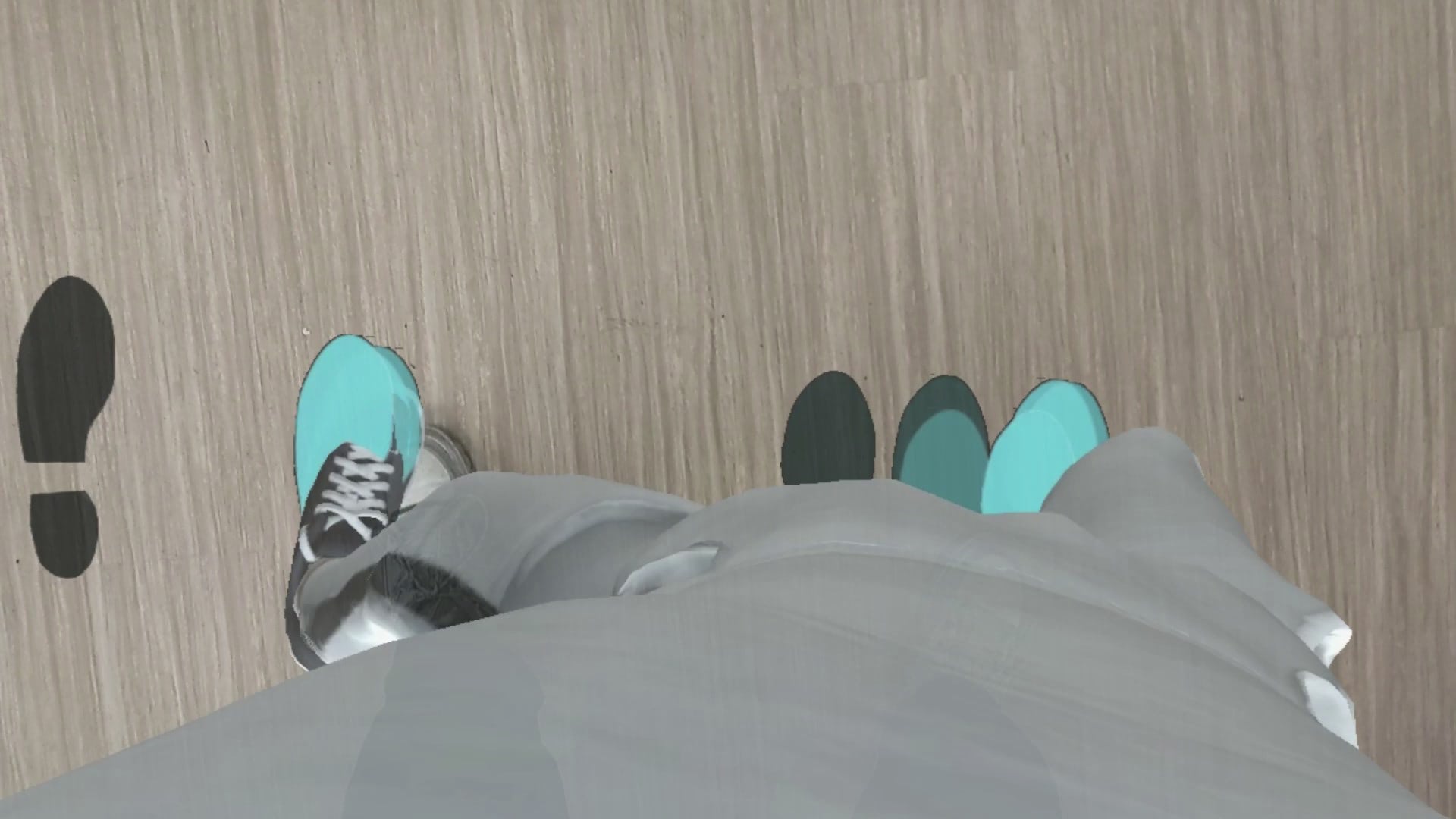}
\caption{First-Person Visualization. (a) Trajectory, (b) Head Gaze, (c) Footprint.}
\label{fig:FP-Visualization}
\end{figure}

In the formative study, the users wanted to mimic their behaviors from different perspectives. 
Inspired by OneBody\cite{hoang2016onebody} and AR-Arm~\cite{han2016ar}, our system allows the user to transport from the third-person view to the first-person view, and vice versa (Fig.~\ref{fig:First-personView}). 
We enabled this by automatically synchronizing the user’s head position to the avatar’s so that the first person will be shown correctly whether the user moves around.
By using the first-person view, users can just move their body parts to the first-person instructor's body part to understand the proper position of those.
Also, this could be combined with visualization, including trajectory, head gaze, and footprint (Fig.~\ref{fig:FP-Visualization}).
For example, hand trajectory in first-person view could help users follow the hand movement when the movement is rapid and expansive.

\subsection{Versatility Evaluation}

To determine what kinds of online sports videos are and are not suitable for our system, we conducted a versatility evaluation. 

\subsubsection{Dataset}
We selected 10 online sports videos from diverse sports categories: 
Yoga~\footnote{\url{https://www.youtube.com/watch?v=j7rKKpwdXNE}},
Dancing~\footnote{\url{https://www.youtube.com/watch?v=cUJRn-WfTbw}}, 
Martial arts~\footnote{\url{https://www.youtube.com/watch?v=e64AtWekQVo&t=395s}},
Gym workout~\footnote{\url{https://www.youtube.com/watch?v=sthD8ziGP1c}},
At-home workout~\footnote{\url{https://www.youtube.com/watch?v=cbKkB3POqaY}},
Swimming~\footnote{\url{https://www.youtube.com/watch?v=LijdyVaaDnY&t=2s}}, 
Baseball~\footnote{\url{https://www.youtube.com/watch?v=YY9tErIBVQw}},
Tennis~\footnote{\url{https://www.youtube.com/watch?v=CXgfNBnetzQ&t=341s}}, 
Boxing~\footnote{\url{https://www.youtube.com/watch?v=kKDHdsVN0b8&t=170s}}
Fencing~\footnote{\url{https://www.youtube.com/watch?v=s-bXSFtXqQk&t=1020s}}.
To avoid cherrypicking videos that would work well for our system, we selected videos popular on YouTube, at least within the top 5 videos. Furthermore, to conduct our evaluation in a standardized manner, we searched for the online videos by using the same search prompt: "Category Name" + "Tutorial" or "Lesson" on YouTube. 
We chose videos between 3 and 5 minutes long, and for longer videos, we trimmed to 5 minutes that featured the instruction's main performance. 




\begin{figure}[h]
\centering
\includegraphics[width=\linewidth]{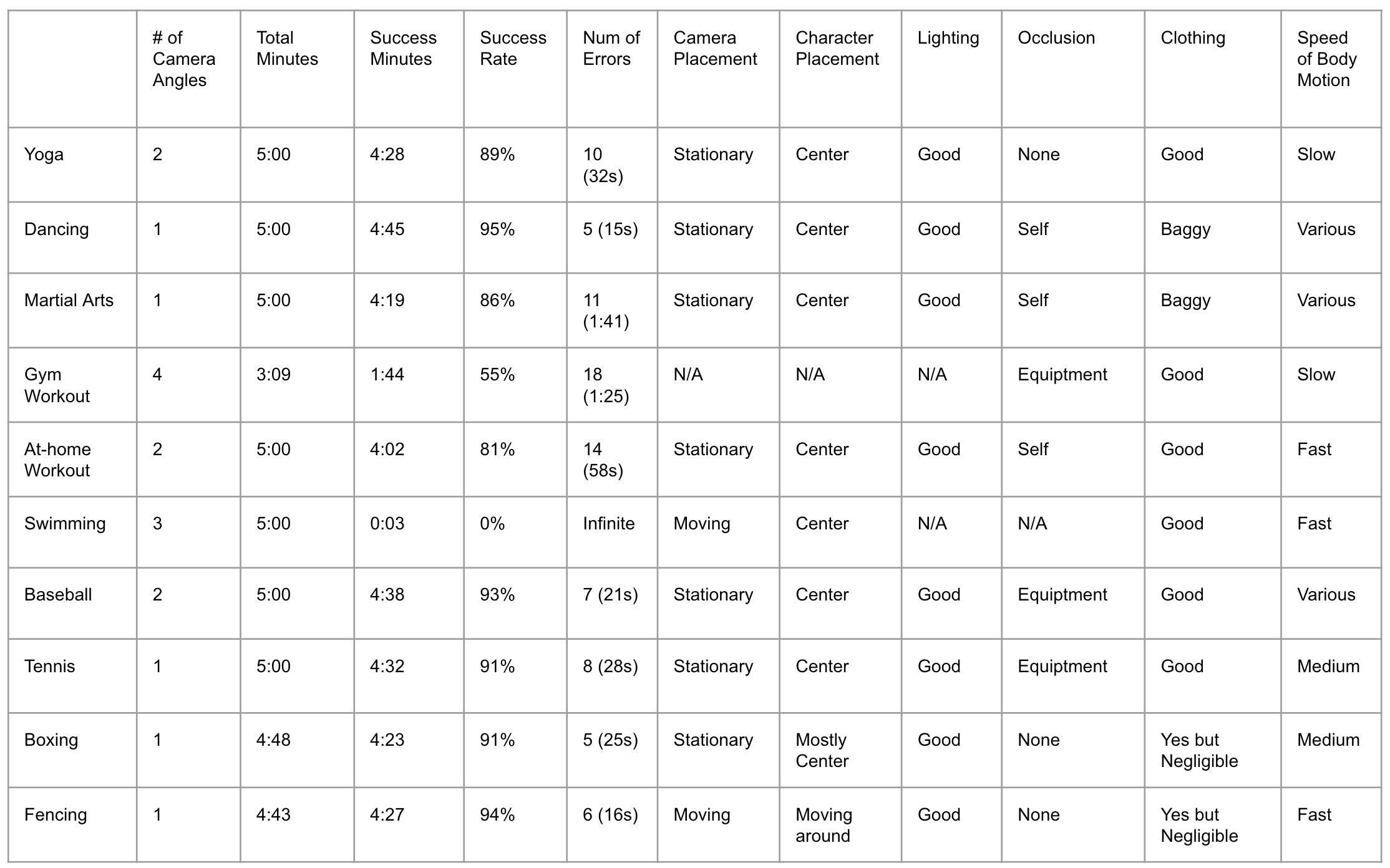}
\caption{Versatility Evaluation Results}
\label{fig:table}
\end{figure}

\subsubsection{Results and Findings}
The table above shows the result of our evaluation. Generally, most videos performed quite well, above 85\% accuracy, though all videos had visual glitches to some level. Swimming performed the worst by far, glitching out completely 3 seconds into the video. For most videos, the fingers were incorrect. For example, in tennis, boxing, tennis and baseball the fingers of the avatar were outstretched. However, the input video shows an instructor gripping some equipment, making a fist or otherwise not what the avatar's fingers look like. Additionally, if the input video had the instructors hands close together, this made it much more likely that the avatar's arms would clip through the other arm or the torso. For yoga, at-home workout and especially martial arts, tracking suffered when the instructor was on the floor. In particular for martial arts, the baggy and mono-colored martial arts clothing made tracking accuracy very poor at times. The gym workout's lower accuracy was primarily due to occlusion issues as well as numerous camera angles.

%% file: 5-user-study.tex
\section{User Study}
To evaluate the effectiveness of \system{}, we conducted a user study comparing \system{} with 2D videos.
Also, we evaluated the usefulness for each feature for six videos.

\subsection{Method}
\subsubsection{Participants}
We recruited 12 participants (7 male, 5 female), aged between 21-38 years  ($M = 25.17, SD = 4.73$).
To evaluate their familiarity with video instructions, we surveyed it using a 7-point Likert scale, ranging from 1 (Not familiar at all) to 7 (Extremely Familiar).
The mean score of the familiarity was 6.42 ($SD = 0.79$).

\subsubsection{Conditions}
We evaluated the system under the following three conditions:
\begin{description}
\item[Video]: Participants viewed the videos on an iPad, using the default video player.
\item[Avatar]: Participants viewed a MR scene using Hololens2, showing avatars both the instructor's avatar and the user's mesh positioned in front of the user. 
\item[Avatar + Video] This condition combined both avatars and video. The MR scene displayed avatars with the video running in the background through Hololens2. 
\end{description}
We conducted the study in an with-in-subjects design.
Therefore, each participant used three conditions.

\subsubsection{Study Setup}
\begin{figure}[h]
\centering
\includegraphics[width=\linewidth]{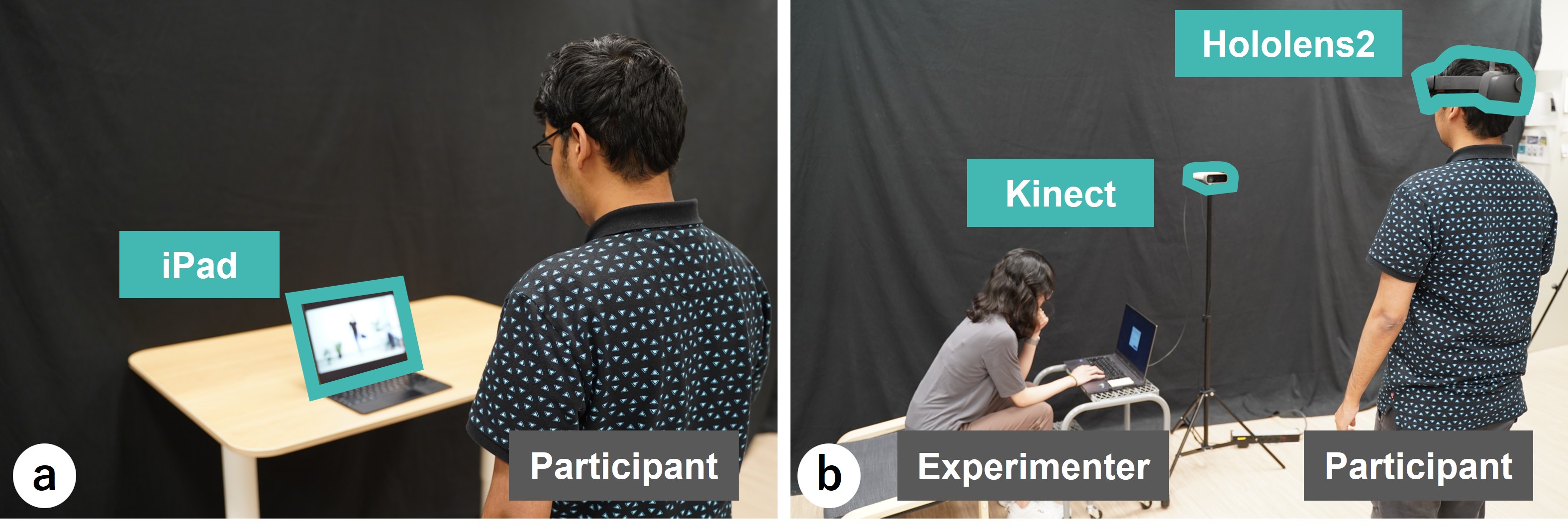}
\caption{Study Setup. (a) \textit{Video} condition (b) \textit{Avatar} and \textit{Avatar+Video} condition.  }
\label{fig:studysetup}
\end{figure}

\begin{figure*}[h]
\centering
\includegraphics[width = \linewidth]{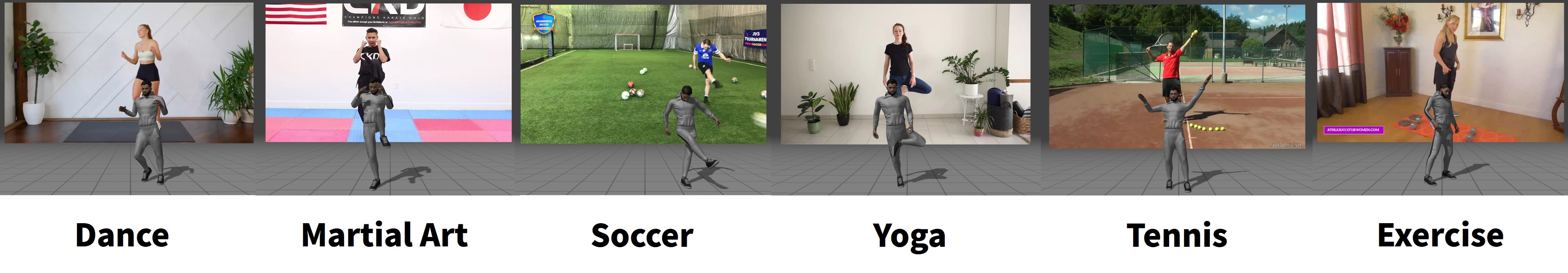}
\caption{The videos and their categories that we used in our user study.}
\label{fig:eachvideo}
\end{figure*}

Figure.~\ref{fig:studysetup} shows the setting of our study.
Hololens2 was connected to a laptop PC through the Holographic Remoting Player.
Also, the Azure Kinect camera was connected to the laptop PC though a USB-C cable.
The experimenter controlled which feature to display using Unity.
The video and avatar size and position were customized based on individual participant preferences.
We selected videos from six various categories: yoga, dance, martial arts, tennis, soccer, and exercise (Fig.~\ref{fig:eachvideo}).
Each video was one minute long.

\subsection{Study Design}
\subsubsection{Procedure}
First, we asked the participants to provide their consent.
Then, we conducted the pre-study questionnaire to ask about their familiarity with video instructions.
The study consisted of six sessions.
In each of these sessions, participants viewed one of the six different videos, each under the three conditions.
We told the participants to follow the instructor's movement and move their bodies to learn the correct movement.
To counterbalance the order effect of the conditions, we used six different orders of the three conditions for six sessions.
Also, to counterbalance the order effect of the videos, we used six different orders of the six videos using a Latin square design. 
Therefore, one video order was used by two participants.
While we were conducting \textit{Avatar} and \textit{Avatar+Video}, we showed our system's features for each session.
After each session, they answered how useful each feature was.
After all sessions, participants answered the questionnaire about their overall experience with each condition.
Finally, we interviewed the participants to gather feedback.
The study was approximately 75 minutes.
They compensate 15 USD (The actual currency has been removed for anonymity).

\subsubsection{Measurements}
To compare the overall experience between the three conditions, we asked the following questions using a 7-point Likert scale (1: Not at all, 7: Extremely): 
1) \textbf{Co-presence}: "How much do you feel the sense of the instructor's presence?", 
2) \textbf{Engagement}: "How much do you feel engaged in the instruction?",
3) \textbf{Fun}: "How fun was it to use this system?",
4) \textbf{Easy to Follow}: "How easy was it to follow the instructor's movement?".
Additionally, we asked about the usefulness of the features for each video using a 7-point Likert scale.


\subsection{Results}
\subsubsection{Overall Experienece}
\begin{figure*}[h]
\centering
\includegraphics[width=\linewidth]{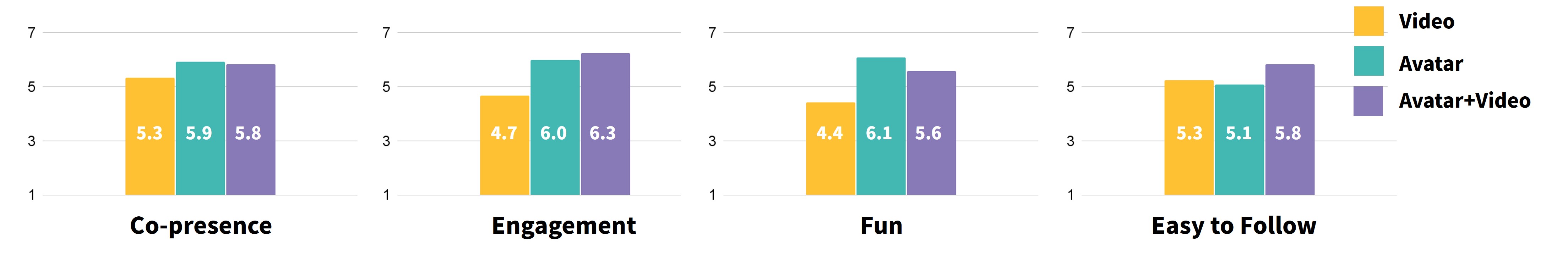}
\caption{Results for the overall experience.}
\label{fig:overall}
\end{figure*}

The results of the overall experience are shown in Fig.\ref{fig:overall}.
For \textbf{Co-presence}, participants rated \textit{Avatar} and \textit{Avatar+Video} better than \textit{Video} (\textit{Video}: 5.3, \textit{Avatar}: 5.9, \textit{Avatar+Video}: 5.8).
Showing a 3D avatar and making them move with the participants might have increased the sense of the avatar being near (P3, P10).
\textit{``It actually makes me feel that an instructor is actually standing in front of me.''}(P3).
\textit{``I can see the instructor moving with me, so I feel like they're right there''}(P10).

For \textbf{Engagement}, participants rated \textit{Avatar} and \textit{Avatar+Video} better than \textit{Video} (\textit{Video}: 4.7, \textit{Avatar}: 6.0, \textit{Avatar+Video}: 6.3).
Because the instructor is in front of the user, participants might have tried to match their postures (P3, P4, P11).
\textit{``It was engaging because I was trying to match my postures quite closely with the avatar.''}(P4).
\textit{``If I was probably just watching a video, I wasn't doing it as much. But like if it was the avatar doing it, I was trying to follow it. So I had my concentration there.''}(P11).


For \textbf{Fun}, participants rated \textit{Avatar} and \textit{Avatar+Video} better than \textit{Video} (\textit{Video}: 4.4, \textit{Avatar}: 6.1, \textit{Avatar+Video}: 5.6).
This could be because our system can give feedback to the user that cannot be done just using video.
\textit{``It's more fun when you can see that you're doing things right.''}(P10)
Also, it was because it is a different way of learning instruction (P4, P5).
\textit{``It was fun because it was like cool and exciting and new''}(P5)

For \textbf{Easy to follow}, participants rated \textit{Avatar} the lowest and \textit{Avatar+Video} the highest (\textit{Video}: 5.3, \textit{Avatar}: 5.1, \textit{Avatar+Video}: 5.8).
The reason could be because, in some scenarios, the avatar could not represent the nuanced movement (P5, P6).
\textit{``The avatar didn't capture the nuance of the human body. So stuff like for squats was actually really hard to follow.''} and \textit{``ones where the motion was slow and nuanced, the video was actually really easy to follow.''}(P6)
It was \textit{``occasionally difficult just in terms of a bit of jitter with the tracking.''}(P5).
Also, it was difficult for participants to follow when the instrument was important to follow the instructions (P2, P5, P10).
\textit{``When the instructor interacted with external objects, having just the avatar, there was very hard to follow.''}(P2)
\textit{``It was generally easy to follow the instructions except in cases where some of the context was missing things like the, the props, like the tennis rackets or the tennis balls or the soccer balls.''}(P5)



\subsubsection{Feature Evaluation}
\begin{figure*}[h]
\centering
\includegraphics[width=0.8\linewidth]{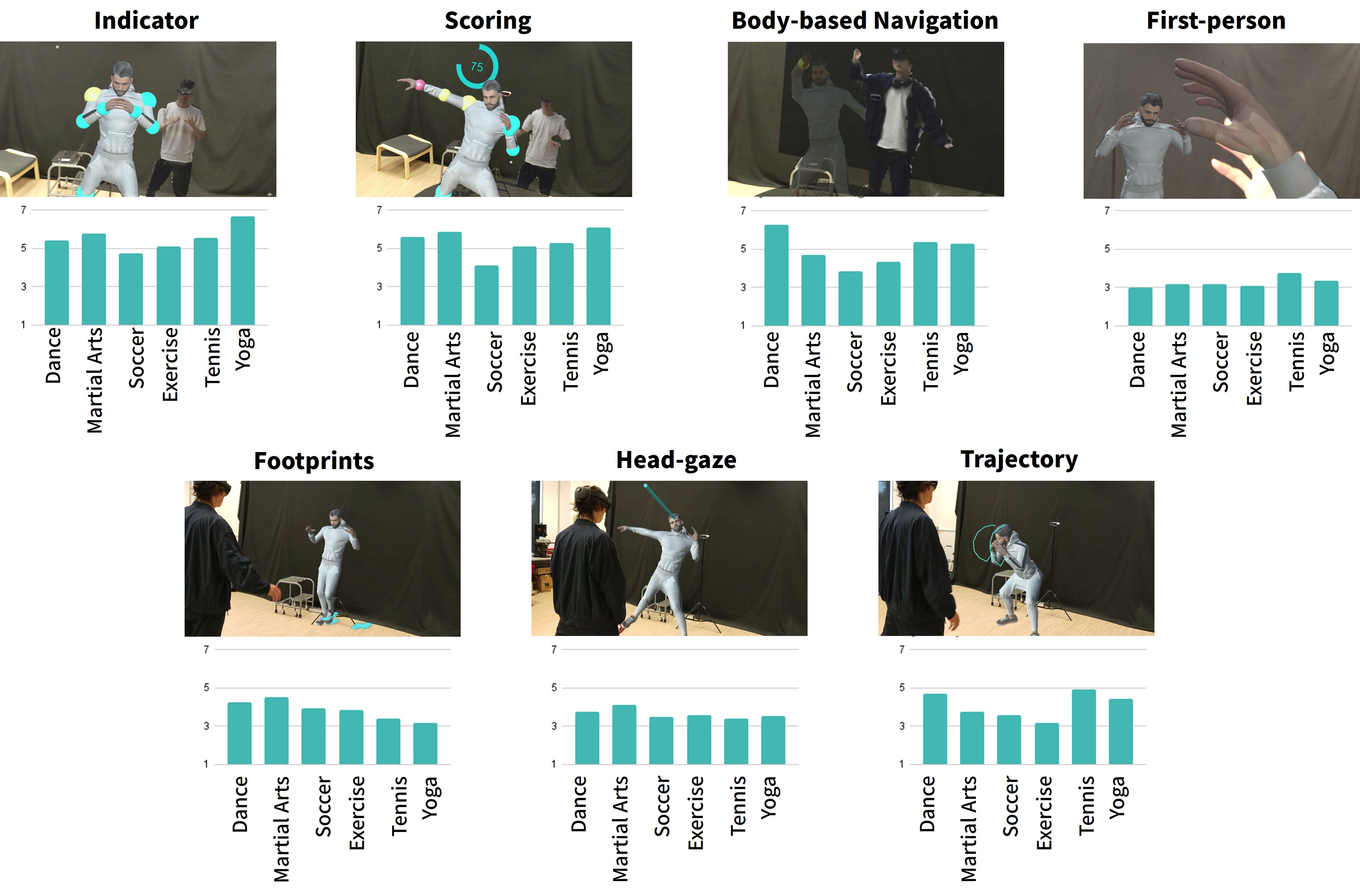}
\caption{Results of feature evaluation. }
\label{fig:eachfeature}
\end{figure*}

The results of the feature evaluation are shown in Figure\ref{fig:eachfeature}.
\textbf{Indicator} and \textbf{Scoring} were rated high compared to other features.
This helped participants to do the exact same movements with the instructor.
\textit{``The indicators were the most helpful feature because you would actually know whether your body is in the exact same position.''}(P11)
\textit{``I feel it's most useful to check whether your body part is in the right position. That's why I felt indicator and scoring the most helpful.''}(P10)
Also, some participants mentioned some issues with the indicator.
\textit{``I also kind of want the indicators on myself. It was hard to look at someone's body and understand which part it is telling.''}(P6)
\textit{``Certain joints were not as important to the instruction and so showing them wasn't as useful.''}(P4)

\textbf{Body-based Navigation} was rated high for dancing.
The movement was fast in dancing, so this feature might help them follow the instructions by stopping until the posture is correct.
For the dance scenario, \textit{``I rated the body-based navigation highly because it would be really useful to learn the dance slowly. If the instructor did difficult poses in very quick changes, body-based navigation would be useful to make sure you have the pose correct''} (P8).
Also, it helped the participants to understand the sequence.
\textit{``Body-based navigation is quite useful to make sure that you're doing everything in the right order.''}(P7)
Soccer was rated low for Body-based Navigation. 
This might be because the instructor used a huge space, and it was difficult to compare the body pose to it.
\textit{``It was more challenging to use indicators in the soccer scenario because the instructor was moving around and to follow him and the precise movement was harder.''}(P9)

The \textbf{First-person} feature could be the most affected feature by the technical issue of Hololens2.
Because of the limited field of view of Hololens2, people suffered from following the first-person instructor.
\textit{``I'm not able to always see the hands. It's not really visible except for the moment where he slides the hands across the face.''}(P1)
However, the concept of using the first-person view was appreciated by some participants.
For the tennis scenario, \textit{``Fore hands maybe first person view was a bit more helpful because I could really see like the hand is this like I'm trying to match my hand. I could understand how I have to position it, maybe the angle or how high I have to raise it.''}(P2)
For the martial arts scenario, \textit{``it was interesting because the first person view visualized the height the instructor was getting on his kick versus my own kick better.''}(P5)

For \textbf{Footprint}, participants rated Martial Arts the best.
The footprint was useful for scenarios in which the foot actively moved.
For martial arts, \textit{``Given that the stance was important for the kicking and returning to the stance was important. So I really liked the footprints this time.''}(P5)
On the other side, for exercise, \textit{``I didn't find the footprint useful for this activity because you don't really move your feet too much for the squad.''}(P5)

For \textbf{Head-gaze}, participants rated Martial Arts the best.
The head-gaze helped participants to understand where they should look.
\textit{``Head-gaze kind of helped me to know where to point my head.''}(P4)
Also, it helped how static the instructor's head is.
For the yoga scenario, \textit{``Unlike the other ones, seeing how still the instructor's gaze is really interesting. And understanding that you're not as still as them''} (P6)

For \textbf{Trajectory}, participants rated Tennis and Dance the best.
This could be because the current implementation only showed the hand trajectory, so the apps that use hands often could be more useful than apps that use lower body, such as Martial Arts, Soccer, and Exercise.
For the tennis scenario, \textit{``If I had gotten more practice, hand trajectory could be useful, particularly for tennis, where your hand movements are sort of important''}(P5).
For exercise, \textit{``I need more trajectory on the hips or something or than the knees, more than the hands when it comes to squatting''}(P7).
And the trail shape made by the hand trajectory attracted the participants.
\textit{``Seeing the arc of the hand trajectory is actually really interesting''}(P6).

\subsubsection{Sports Category-Based Evaluation}
For Dancing and Martial Arts, participants preferred \textit{Avatar} to \textit{Avatar+Video}.
The reason could be that the avatar information was enough to follow the instructions (P2, P4).
For dancing, \textit{``I don't really see the need for the video because it's way easier to just follow the avatar because it's doing the same thing with the video.''}(P2)
Another participant mentioned for dancing, \textit{``Avatar+video was just really difficult because there's a lot of information.''}(P4)

Although, for Tennis, participants prefer \textit{Avatar+Video}.
For tennis, it was difficult for participants to understand what the instructor was doing by watching the avatar without the video.
For tennis, just showing avatar, \textit{``I didn't know which hand had the tennis racket and which one was throwing the ball, for instance.''}(P4)

Martial Arts was the most preferred video.
The reason might be that in martial arts, the 3D position of the body position was important and the avatar helped them to understand it.
\textit{``It was actually quite useful to look at the avatar from all angles. It really helps particularly with the leg positioning.''}(P6)

On the other side, Exercise was the least preferred video.
The reason might be the movement was too simple.
\textit{``The squatting one, the avatar was kind of unnecessary because exactly what the instructor is doing you can just do.''}(P7)



\section{Expert Review}

\subsection{Method}
To conduct our expert review, we recruited six experts in the physical exercise domains. Our experts were between 25 - 36 years old, with 4 - 30 years of experience within their domain of expertise. Each participant was an expert in one of the following domains: (EY) yoga, (ED) dancing, (ES) soccer, (EM) martial arts, (EF) fitness and (EG) golf. To evaluate our system's suitability for their domain of expertise, We selected a 60-second clip from YouTube of a video tutorial of their domain, showed the experts the video clip on an iPad Pro's screen then had them experience \system{}. We then walked through each of our implemented features, asking them various inquisitive questions to gather qualitative feedback. Overall, watching a video clip to experience our system and the subsequent interviews took about 60 minutes to complete. Expert participants were compensated 20\$ USD (The actual currency has been removed for anonymity).

\subsection{Overall Feedback}

Since we recruited experts from a diverse array of domains, we were interested in understanding which features were most and least useful for each domain. 

The experts universally expressed interest in and found much value in the indicator the score features, albeit with concerns regarding accuracy. \textit{ES: Great idea, as a coach I would pick the most relevant positions the [user] needs to match step by step}. \textit{ED: It would be useful since it can help me in real time on if I'm performing the technique poorly. So for example, if my legs are spread too wide, the video and avatar will not progress until I perform the action correctly. It will force me to narrow them before we move to the next step}. \textit{EG: The most useful feature for golf. I like that it incorporates all the different elements of your swing. How your knees bend and move during setup and the swing. I like that it would tell me in real time if my shoulders, wrists and other parts are correctly set up for the swing in real time. I would add more joints, like hips if you can}. \textit{EY: I really like it, it's fun. If you can maybe select where you want it to be evaluating you, like only evaluating my arms, or legs, or feet, etc would be very great}. \textit{EM: Would use the feature for sure. To perform a demo of a movement then ask someone to mimic the movement would be super useful in this way}. \textit{EF: I like being able to synchronize with the instructor's avatar. Since it follows joints, it will be very useful for fitness}.

As for body-based navigation, all experts again expressed interest in the feature but found it largely geared towards beginners, as intermediate and advanced learners would likely need to practice more fluidly. \textit{ED: this is probably better for beginners to learn each step, because more experienced dancers would want to perform more fluidly}. \textit{ES: Could be difficult to use as you level up since the techniques should be done as one movement, not broken up into steps}. \textit{EG: This would allow you to see how your body is moving compared to the avatar in a slow motion fashion, which is probably a better way to learn a golf swing if you haven't learned yet. Most people have trouble with golf by overswinging, they hit the ball as hard as they can. If you go step by step and in slow motion, your body will learn the motions and positions more accurately, i.e. not smashing the ball, and you'll learn through muscle memory}. \textit{EF: Being able to show students that they're doing the proper technique step by step is very useful and especially through different positions}. \textit{EM: I don't want to see how students are positioned at the start and end of exercise, I want to see and teach them step by step where they should be located and angled through the whole exercise which this helps a lot for}.

\subsection{Domain Specific Feedback}

\subsubsection{\textbf{Yoga}}

While greatly interested in the indicator, score, and body-based navigation features, the yoga expert found little value for trajectory and footprint since yoga requires little movement. \textit{EY: "I liked that [footprints] were easy to synchronize. I could just look down and plant my feet the same as the avatar. But we don't move much in Yoga so I'd only do it once (...) since we hold the poses for an amount of time, it's not really important to see the trajectory of the body. Often there isn't even a trajectory to see. I would rather just see the outline of the pose"}. However, they explain situations in which the head-gaze feature would be useful. \textit{EY: sometimes seeing the head-gaze would be useful for poses where it's important to keep your head still or facing a certain direction. For example the Warrior 2 Pose in Yoga needs you to face forwards while your body twists, which for beginners is difficult to do. But seeing the avatar's head-gaze not moving would make it easier to understand just how little the head should move}.

\subsubsection{\textbf{Dancing}}

As we expected, the dancing expert found the footprints and trajectory features the most useful. 1st person footprints more than 3rd person, 3rd and 1st person trajectory, though 1st person trajectory had it's value diminished due to the limited Hololens FOV. \textit{ED: I can't really see it well [due to Hololens FOV]. Footprint is probably the only good one right now for first person, especially if you're learning footwork heavy dancing. 3rd person is much better for individual learning and group learning}. The expert also appreciated the head-gaze feature, albeit in combination with trajectory. \textit{ED: "During group dancing, we should all be synchronized. Viewing the trajectory and head-gaze movements together could help us all while practicing to make sure we stay synchronized, which is critical to certain kinds of dances"}.

\subsubsection{\textbf{Soccer}}

The soccer expert found the trajectory very useful for visualizing shooting the ball. \textit{ES: "It all aligns for how it should look like. You want the trajectory, especially for soccer, of the body the head and the foot and where the ball ends up. If i was teaching, i would want to draw a line from my foot starting to swing, to hitting the ball to where the ball goes. I would tell a kid to draw a line [with their dominant leg] from where the ball is to where they wanted it to go. I'd want to be able to select which limbs to track rather than the whole body at once and i want to select when to start drawing the trajectory and when to not, like when I'm explaining something"}. They also enjoyed head-gaze for understanding and visualizing different methods of heading a ball. \textit{ES: "[Head-gaze is] useful for visualizing headers, since there's different kinds of headers (ex: straight, glance, etc). You could visualize the head rotation and position for each of those as well as how to sync to them [with 1st person]"}.

\subsubsection{\textbf{Golf}}

The golf expert found our system best for learning how to set up a golf swing, both by using the first person footprints for seeing where to stand \textit{EG: "[Footprints] are useful during setup to put your feet in the right positions"} and head-gaze to remind users to stare at the ball during the entire swing. \textit{EG: "The key thing in golf is to have your head looking at the ball throughout the whole swing which [head-gaze] certainly does"}. The expert also liked the third person trajectory for visualizing the overall circular swing. \textit{EG: "I liked how it shows the natural arc of a golf swing and shows what it should look like as an example. I wish it would consider wrist positions, body positions, for the swing as well"}.

\subsubsection{\textbf{Fitness}}

The fitness expert enjoyed first and third person POVs of footprints and trajectory to prepare for and performing exercises. \textit{EF: "Seeing the path of the hands, for movements and exercises like dead-lifts would be great. Could show the movement of the exercise [and therefore the equipment] without [the avatar] actually showing the equipment"}. \textit{EF: "It's useful to see the path of footprints. For example for walking lunges, zumba, etc."}

\subsubsection{\textbf{Martial Arts}}

Similar to the fitness expert, the martial arts expert enjoyed first and third person POVs of trajectory and footprints the most. \textit{EM: "I really liked [trajectory], it's the most worthwhile and [the feature I'm most] likely to use. It's best for watching someone do an exercise and even better [combined with footprints] to see the path of the technique"}. Unlike the fitness expert they also found head-gaze useful context for learners. \textit{EM: "it would be great to offer the student the ability to see where I"m looking and match head motions"}.


\subsection{Expert Insights}

\subsubsection{\textbf{Embodiment of Professionals}}

The golfing, dancing, and soccer experts each also expressed an interest in moving beyond YouTube tutorials and instead embodying professionals and celebrities. The soccer expert expressed excitement at the ability to embody professional soccer players \textit{S1: "One of the biggest things for soccer is you want to be the professional, you want to know what the professional sees and what they do based on the information at hand. The best way to relay that to someone is to let them be the athlete. It would also be very insightful to see how quickly they dribble the ball and their footwork especially for learning tricks"}. The dancing expert echoed this sentiment \textit{ED: "If I had videos of Beyoncé on stage I'd love to learn her dances then perform it like she does on stage. I'm really interested in seeing her 1st person perspective"}. Finally, the golfing expert also communicated interest in the embodiment of professional golfers \textit{EG: "The swing might be hard to see in first person but it'd be useful to see how Tiger Woods sets up [to swing]"}.

\subsubsection{\textbf{More Scalable Education}}
Although our system was designed for 1-1 education and can only accommodate one user at a time, our experts universally expressed great interest and excitement for using our system during group lessons. Most experts explain that, financial concerns aside, they would equip a group of students with a Hololens each and have everyone watch the 3rd person avatar's movements as a group. \textit{ES: "I would use 3rd person as a group example for the kids (...) in general 3rd person seems to me to be useful for group explanations but 1st person is better for practicing the technique and individual training"}. They imagine that students would walk around and observe the movements from various angles in 3D while the expert explains what's happening or what to pay attention to. Experts would be in control of which features are active at what time (footprint, trajectory, etc) and would especially use them as additional visual context for their oral explanation. With the demonstration complete, they would then break the students off to practice individually. They believe that each student would be able to practice, learn and improve more independently due to our system's ability to 1) enable students to replay the 3D lesson (eg: for details they may have missed), 2) to embody the instructor's avatar (for a first-person perspective of the lesson) and 3) most importantly of all, constantly outputting real-time feedback tailored for each student individually (which helps them improve without the need for intervention or feedback from the expert). \textit{ES: "you could practice that mechanic over and over again building the muscle memory and body motions. If accurate, this could enable automatic training for soccer techniques because it gives you real-time feedback"}. Furthermore, they expressed interest in monitoring student scores as a whole, for an overall picture of class performance and to tailor interventions. \textit{ED: "If it could give me feedback on the scores of how the kids are doing, that would be amazing because it would let me check which techniques each kid is having difficulty with or group kids with similar difficulties to teach together in smaller groups"}. Finally, some experts also mentioned using our system as a method for supplemental or remote learning. \textit{"ES: "it would be great to give to kids as practice/homework to do while away from the instructor or learning remotely"}. \textit{EF: "Would be great to use for teaching people how to use their own home gyms for fitness training"}. \textit{EM: If we were practicing a technique and i was teaching remotely, it would be great to offer the student the ability to see where I"m looking and match head motions}.

\subsubsection{\textbf{Comparison to Traditional Instruction}}
Our golfing expert pointed out how indoor golfing simulations have become ubiquitous, capable of providing accurate, useful and immediate feedback, enabling golfers to practice their craft more effectively, though the feedback is limited to the \textbf{outcome} of a golfer's actions and not  the action itself. \textit{EG: "The tech for indoor driving ranges are really good, they can tell you the angle of the ball, speed, wind direction, spin on the ball, etc. They're very accurate, but they don't tell you anything about your actual swing, your body positions or your motions"}. They then explain how our system can combine with indoor golfing simulations to provide a more holistic and complete evaluation of a user's golfing performance. \textit{G1: "Your system on the other hand helps tremendously for learning how to setup for a golf swing and gives you feedback on the swing itself, which is something currently missing in the market. We have accurate info for what happens to the ball after you swing, but nothing for what happens prior. Combining them is the best of both worlds"}.


\subsection{\textbf{Suggestions for Feature Expansion}}

Our experts suggested a few different improvements for our features but for the most part they requested additional contextual cues. \textit{ES: "Would also be nice to have different color trajectories for each stage of the technique like before the leg hits the ball, after leg hits the ball, etc"}. \textit{ED: "An arrow or something to show the direction in which the trajectory is moving would be super helpful in quickly identifying the actual trajectory of the limb if the trajectory gets messy or I pause it"}. \textit{ED: "being able to see the whole trajectory of the whole movement at once instead of the past few second would be useful to prepare the rest of our body for the upcoming motions based on the trajectory that we can see the limbs needs to move in"}. \textit{EY: "Could you show the degree of the angle of the joints on the 3D avatar and my own [Kinect] avatar? Like if I lean to the side but keep my legs still, can you show how many degrees I've leaned or angled? I'd like that info"} \textit{EF: "Would be useful to be able to click the trajectory and see where they were at that time, especially for viewing what their other limbs were doing in that moment and even click to rewind and fast forward"}. \textit{ED: "Numbering the footprints or even highlighting the them with the tempo/lyrics, etc so I can visualize the order in which i need to step much better"}. The dancing instructor also requested the ability to modify the tempo of the input music and have the 3D avatar's animation synchronize automatically. \textit{ED: "Sometimes we synchronize [our dances] with either the lyrics, basses or instruments. I'd like it if we could visualise those alongside the avatar. Like showing which instruments are playing at what part of the dance or the music tempo"}. Expanding on synchronized dancing, the expert requests the ability modify the music and have the avatar's dance change accordingly.\textit{ED: "It would be useful if I could change the tempo of the music from the video and see the avatar dancing the same moves but automatically adjusted to the different tempo I set. Like I could quickly see a preview of the dance to the new tempo through the avatar"}.

\subsection{\textbf{Technical Limitations}}

While each domain expert had varying opinions on our system overall, as well as each of the features, some opinions were expressed by all experts. Most commonly, experts found the limited FOV of the Hololens 2 challenging to deal with, especially for the first-person features. For example, while our martial arts expert enjoyed the trajectory feature to \textit{EM: "visualize a punching or kicking"} they expressed concern and frustration with following more than one movement at a time using the first person perspective. \textit{EM: "I can only sync one hand at a time since I can't really view them both at once"}. Experts from domains that required the use of equipment, like golf and soccer, expressed disappointment from the lack of equipped equipment by the 3D avatar. \textit{ES: "I wish be you could use a real ball in with the headset to give that physical feedback to the students"}. On the other hand, the fitness expert expressed less disappointment at the lack of equipment, claiming that trajectory is still useful to see how the exercise is performed, even without the avatar equipping equipment. \textit{EF: "Seeing the path of the hands, for exercises like dead-lifts would be great. This would show the movement of the exercise and the equipment"}. Finally, all the participants expressed concern over inaccuracy of Kinect body tracking for the indicator, score and body-based navigation features causing difficulty to synchronize their body to the 3D avatar.

%% file: 6-future-work.tex
\section{Future Work}

\subsubsection*{\textbf{Object Detection}}
In our study, some participants mentioned that it was difficult to follow the avatar instructions for videos that use equipment such as tennis and soccer. 
To address this, we could attach 3D object models of equipment to the avatar.
For example, we can attach a virtual tennis racket to the avatar's hand like \textit{RealityCanvas}~\cite{xia2023realitycanvas} and rotate the racket based on the hand rotation.
Also, we could detect the objects in the video using object detection techniques.
For example, if we could detect the 3D position of the soccer ball, users can watch where the instructor kicks the ball and how the ball moves.

Also, while we learn sports that use instruments, we often use those instruments as well, such as tennis rackets and soccer balls.
If we can track the real object that the user is using, we can show the user a more immersive instructional experience.
For example, for the soccer instruction scenario, by synchronizing the avatar's animation with the position of a soccer ball, user can see the exact point where an instructor kicks the ball.

\subsubsection*{\textbf{Verbal Information}}
Verbal information is important to understand the detailed information.
Inspired by previous works, such as \textit{Reality Talk}~\cite{liao2022realitytalk}, we envision enhancing the avatar representation based on what the instructor is talking about.
We can leverage speech recognition techniques such as \textit{Google Cloud Speech-to-Text}.
By using the verbal information, we could highlight the specific body position or show additional information about the avatar based on what they are talking about.
For example, when the instructor is talking about specific body parts, we can highlight those parts.
When the instructor explains domain-specific words, we can display some related figures and text to explain them.

\subsubsection*{\textbf{2D Visualized Information}}
Some 2D videos incorporate on-screen visualizations to enhance clarity and encouragement.
For example, they use time and numbers to encourage the user and use highlighting and arrows to emphasize body parts. 
We could extract that information from the video by using optical character recognition (OCR) techniques, such as \textit{Google Cloud Vision API} and convert it into 3D could help the user understand the instructions and be encouraged.
Also, some videos refer to other videos, such as professional videos or their own previous videos.
Combining extracting body motion from several videos and showing several avatars could display more detailed information.

\subsubsection*{\textbf{Avatar Representation}}
In this paper, as we utilized a 3D avatar sourced from Mixamo, the avatar didn't resemble the video's instructor.
However, by utilizing 3D reconstruction technologies, we could create an avatar's texture from the video and make the avatar closely resemble the instructor in the video.
This could make users feel like the instructor is really in front of them and make them more engaged.
Also, a more accurate facial representation could allow to convey the instructor's emotions properly.
This could help enhance the user's comprehension of instructions.

%% file: 7-conclusion.tex
\section{Conclusion}
In this paper, we developed \system{}, which enhances and automatically generates mixed reality instructions utilizing extracted human motion from 2D instructional videos.
\system{} has four design elements: 1) Comparison, 2) Visualization, 3) Navigation, 4) Reposition, and we implemented several features based on them.
We conducted two evaluations to evaluate the usefulness of \system{} compared with just using 2D videos.
We conducted a user study with 12 participants and confirmed that our system can enhance co-presence, engagement, and fun.
Also, we found using video in the background of the avatar could help users follow the instructions.
Through the expert reviews with six participants, we confirmed that our implemented features are useful for a variety of physical exercise domains, gathered unique qualitative insights like how experts wish to embody celebrities and professionals, learned that there is a desire to enhance current instructional methods with the concepts and features presented by our system and finally that \system{} enables more scalable instruction by empowering beginners and novices to learn more independently.
Additionally, we mentioned how we can enhance the experience through object detection, verbal detection, and improving avatar representation as a future work.


%% file: main.bbl

\begin{thebibliography}{58}


\ifx \showCODEN    \undefined \def \showCODEN     #1{\unskip}     \fi
\ifx \showDOI      \undefined \def \showDOI       #1{#1}\fi
\ifx \showISBNx    \undefined \def \showISBNx     #1{\unskip}     \fi
\ifx \showISBNxiii \undefined \def \showISBNxiii  #1{\unskip}     \fi
\ifx \showISSN     \undefined \def \showISSN      #1{\unskip}     \fi
\ifx \showLCCN     \undefined \def \showLCCN      #1{\unskip}     \fi
\ifx \shownote     \undefined \def \shownote      #1{#1}          \fi
\ifx \showarticletitle \undefined \def \showarticletitle #1{#1}   \fi
\ifx \showURL      \undefined \def \showURL       {\relax}        \fi
\providecommand\bibfield[2]{#2}
\providecommand\bibinfo[2]{#2}
\providecommand\natexlab[1]{#1}
\providecommand\showeprint[2][]{arXiv:#2}

\bibitem[\protect\citeauthoryear{Adolf, K{\'a}n, Outram, Kaufmann,
  Dole{\v{z}}al, and Lhotsk{\'a}}{Adolf et~al\mbox{.}}{2019}]%
        {adolf2019juggling}
\bibfield{author}{\bibinfo{person}{Jind{\v{r}}ich Adolf},
  \bibinfo{person}{Peter K{\'a}n}, \bibinfo{person}{Benjamin Outram},
  \bibinfo{person}{Hannes Kaufmann}, \bibinfo{person}{Jarom{\'\i}r
  Dole{\v{z}}al}, {and} \bibinfo{person}{Lenka Lhotsk{\'a}}.}
  \bibinfo{year}{2019}\natexlab{}.
\newblock \showarticletitle{Juggling in vr: Advantages of immersive virtual
  reality in juggling learning}. In \bibinfo{booktitle}{\emph{Proceedings of
  the 25th ACM Symposium on Virtual Reality Software and Technology}}.
  \bibinfo{pages}{1--5}.
\newblock


\bibitem[\protect\citeauthoryear{Amores, Benavides, and Maes}{Amores
  et~al\mbox{.}}{2015}]%
        {amores2015showme}
\bibfield{author}{\bibinfo{person}{Judith Amores}, \bibinfo{person}{Xavier
  Benavides}, {and} \bibinfo{person}{Pattie Maes}.}
  \bibinfo{year}{2015}\natexlab{}.
\newblock \showarticletitle{ShowMe: A Remote Collaboration System That Supports
  Immersive Gestural Communication}. In \bibinfo{booktitle}{\emph{Proceedings
  of the 33rd Annual ACM Conference Extended Abstracts on Human Factors in
  Computing Systems}}. \bibinfo{pages}{1343–1348}.
\newblock


\bibitem[\protect\citeauthoryear{Bai, Sasikumar, Yang, and Billinghurst}{Bai
  et~al\mbox{.}}{2020}]%
        {bai2020user}
\bibfield{author}{\bibinfo{person}{Huidong Bai}, \bibinfo{person}{Prasanth
  Sasikumar}, \bibinfo{person}{Jing Yang}, {and} \bibinfo{person}{Mark
  Billinghurst}.} \bibinfo{year}{2020}\natexlab{}.
\newblock \showarticletitle{A user study on mixed reality remote collaboration
  with eye gaze and hand gesture sharing}. In
  \bibinfo{booktitle}{\emph{Proceedings of the 2020 CHI conference on human
  factors in computing systems}}. \bibinfo{pages}{1--13}.
\newblock


\bibitem[\protect\citeauthoryear{Buchenau and Suri}{Buchenau and Suri}{2000}]%
        {buchenau2000experience}
\bibfield{author}{\bibinfo{person}{Marion Buchenau} {and}
  \bibinfo{person}{Jane~Fulton Suri}.} \bibinfo{year}{2000}\natexlab{}.
\newblock \showarticletitle{Experience prototyping}. In
  \bibinfo{booktitle}{\emph{Proceedings of the 3rd conference on Designing
  interactive systems: processes, practices, methods, and techniques}}.
  \bibinfo{pages}{424--433}.
\newblock


\bibitem[\protect\citeauthoryear{Clarke, Cavdir, Chiu, Denoue, and
  Kimber}{Clarke et~al\mbox{.}}{2020}]%
        {clarke2020reactive}
\bibfield{author}{\bibinfo{person}{Christopher Clarke}, \bibinfo{person}{Doga
  Cavdir}, \bibinfo{person}{Patrick Chiu}, \bibinfo{person}{Laurent Denoue},
  {and} \bibinfo{person}{Don Kimber}.} \bibinfo{year}{2020}\natexlab{}.
\newblock \showarticletitle{Reactive video: adaptive video playback based on
  user motion for supporting physical activity}. In
  \bibinfo{booktitle}{\emph{Proceedings of the 33rd Annual ACM Symposium on
  User Interface Software and Technology}}. \bibinfo{pages}{196--208}.
\newblock


\bibitem[\protect\citeauthoryear{Conner and Poor}{Conner and Poor}{2016}]%
        {conner2016correcting}
\bibfield{author}{\bibinfo{person}{Caleb Conner} {and}
  \bibinfo{person}{Gene~Michael Poor}.} \bibinfo{year}{2016}\natexlab{}.
\newblock \showarticletitle{Correcting exercise form using body tracking}. In
  \bibinfo{booktitle}{\emph{Proceedings of the 2016 CHI Conference Extended
  Abstracts on Human Factors in Computing Systems}}.
  \bibinfo{pages}{3028--3034}.
\newblock


\bibitem[\protect\citeauthoryear{DeepMotion}{DeepMotion}{2023}]%
        {noauthor_deepmotion_nodate}
\bibfield{author}{\bibinfo{person}{DeepMotion}.}
  \bibinfo{year}{2023}\natexlab{}.
\newblock \bibinfo{title}{DeepMotion}.
\newblock
\newblock
\urldef\tempurl%
\url{https://www.deepmotion.com/}
\showURL{%
\tempurl}


\bibitem[\protect\citeauthoryear{Dittakavi, Bavikadi, Desai, Chakraborty,
  Reddy, Balasubramanian, Callepalli, and Sharma}{Dittakavi
  et~al\mbox{.}}{2022}]%
        {dittakavi2022pose}
\bibfield{author}{\bibinfo{person}{Bhat Dittakavi}, \bibinfo{person}{Divyagna
  Bavikadi}, \bibinfo{person}{Sai~Vikas Desai}, \bibinfo{person}{Soumi
  Chakraborty}, \bibinfo{person}{Nishant Reddy}, \bibinfo{person}{Vineeth~N
  Balasubramanian}, \bibinfo{person}{Bharathi Callepalli}, {and}
  \bibinfo{person}{Ayon Sharma}.} \bibinfo{year}{2022}\natexlab{}.
\newblock \showarticletitle{Pose tutor: an explainable system for pose
  correction in the wild}. In \bibinfo{booktitle}{\emph{Proceedings of the
  IEEE/CVF Conference on Computer Vision and Pattern Recognition}}.
  \bibinfo{pages}{3540--3549}.
\newblock


\bibitem[\protect\citeauthoryear{Eckhoff, Sandor, Lins, Eck, Kalkofen, and
  Hein}{Eckhoff et~al\mbox{.}}{2018}]%
        {eckhoff2018tutar}
\bibfield{author}{\bibinfo{person}{Daniel Eckhoff}, \bibinfo{person}{Christian
  Sandor}, \bibinfo{person}{Christian Lins}, \bibinfo{person}{Ulrich Eck},
  \bibinfo{person}{Denis Kalkofen}, {and} \bibinfo{person}{Andreas Hein}.}
  \bibinfo{year}{2018}\natexlab{}.
\newblock \showarticletitle{TutAR: augmented reality tutorials for hands-only
  procedures}. In \bibinfo{booktitle}{\emph{Proceedings of the 16th ACM
  SIGGRAPH International Conference on Virtual-Reality Continuum and its
  Applications in Industry}}. \bibinfo{pages}{1--3}.
\newblock


\bibitem[\protect\citeauthoryear{Elsayed, Kartono, Sch{\"o}n, Schmitz,
  M{\"u}hlh{\"a}user, and Weigel}{Elsayed et~al\mbox{.}}{2022}]%
        {elsayed2022understanding}
\bibfield{author}{\bibinfo{person}{Hesham Elsayed}, \bibinfo{person}{Kenneth
  Kartono}, \bibinfo{person}{Dominik Sch{\"o}n}, \bibinfo{person}{Martin
  Schmitz}, \bibinfo{person}{Max M{\"u}hlh{\"a}user}, {and}
  \bibinfo{person}{Martin Weigel}.} \bibinfo{year}{2022}\natexlab{}.
\newblock \showarticletitle{Understanding Perspectives for Single-and
  Multi-Limb Movement Guidance in Virtual 3D Environments}. In
  \bibinfo{booktitle}{\emph{Proceedings of the 28th ACM Symposium on Virtual
  Reality Software and Technology}}. \bibinfo{pages}{1--10}.
\newblock


\bibitem[\protect\citeauthoryear{Fang, Xie, Tai, and Lu}{Fang
  et~al\mbox{.}}{2017}]%
        {fang2017rmpe}
\bibfield{author}{\bibinfo{person}{Hao-Shu Fang}, \bibinfo{person}{Shuqin Xie},
  \bibinfo{person}{Yu-Wing Tai}, {and} \bibinfo{person}{Cewu Lu}.}
  \bibinfo{year}{2017}\natexlab{}.
\newblock \showarticletitle{Rmpe: Regional multi-person pose estimation}. In
  \bibinfo{booktitle}{\emph{Proceedings of the IEEE international conference on
  computer vision}}. \bibinfo{pages}{2334--2343}.
\newblock


\bibitem[\protect\citeauthoryear{Faridan, Kumari, and Suzuki}{Faridan
  et~al\mbox{.}}{2023}]%
        {faridan2023chameleoncontrol}
\bibfield{author}{\bibinfo{person}{Mehrad Faridan}, \bibinfo{person}{Bheesha
  Kumari}, {and} \bibinfo{person}{Ryo Suzuki}.}
  \bibinfo{year}{2023}\natexlab{}.
\newblock \showarticletitle{ChameleonControl: Teleoperating Real Human
  Surrogates through Mixed Reality Gestural Guidance for Remote Hands-on
  Classrooms}. In \bibinfo{booktitle}{\emph{Proceedings of the 2023 CHI
  Conference on Human Factors in Computing Systems}}.
  \bibinfo{numpages}{13}~pages.
\newblock


\bibitem[\protect\citeauthoryear{Fieraru, Zanfir, Pirlea, Olaru, and
  Sminchisescu}{Fieraru et~al\mbox{.}}{2021}]%
        {fieraru2021aifit}
\bibfield{author}{\bibinfo{person}{Mihai Fieraru}, \bibinfo{person}{Mihai
  Zanfir}, \bibinfo{person}{Silviu~Cristian Pirlea}, \bibinfo{person}{Vlad
  Olaru}, {and} \bibinfo{person}{Cristian Sminchisescu}.}
  \bibinfo{year}{2021}\natexlab{}.
\newblock \showarticletitle{Aifit: Automatic 3d human-interpretable feedback
  models for fitness training}. In \bibinfo{booktitle}{\emph{Proceedings of the
  IEEE/CVF conference on computer vision and pattern recognition}}.
  \bibinfo{pages}{9919--9928}.
\newblock


\bibitem[\protect\citeauthoryear{Hamanishi, Miyaki, and Rekimoto}{Hamanishi
  et~al\mbox{.}}{2019}]%
        {hamanishi2019assisting}
\bibfield{author}{\bibinfo{person}{Natsuki Hamanishi}, \bibinfo{person}{Takashi
  Miyaki}, {and} \bibinfo{person}{Jun Rekimoto}.}
  \bibinfo{year}{2019}\natexlab{}.
\newblock \showarticletitle{Assisting viewpoint to understand own posture as an
  avatar in-situation}. In \bibinfo{booktitle}{\emph{Proceedings of the 5th
  International ACM In-Cooperation HCI and UX Conference}}.
  \bibinfo{pages}{1--8}.
\newblock


\bibitem[\protect\citeauthoryear{Hamanishi and Rekimoto}{Hamanishi and
  Rekimoto}{2020}]%
        {hamanishi2020poseasquery}
\bibfield{author}{\bibinfo{person}{Natsuki Hamanishi} {and}
  \bibinfo{person}{Jun Rekimoto}.} \bibinfo{year}{2020}\natexlab{}.
\newblock \showarticletitle{Poseasquery: Full-body interface for repeated
  observation of a person in a video with ambiguous pose indexes and performed
  poses}. In \bibinfo{booktitle}{\emph{Proceedings of the Augmented Humans
  International Conference}}. \bibinfo{pages}{1--11}.
\newblock


\bibitem[\protect\citeauthoryear{Hamanishi and Rekimoto}{Hamanishi and
  Rekimoto}{2021}]%
        {hamanishi2021motion}
\bibfield{author}{\bibinfo{person}{Natsuki Hamanishi} {and}
  \bibinfo{person}{Jun Rekimoto}.} \bibinfo{year}{2021}\natexlab{}.
\newblock \showarticletitle{Motion-specific browsing method by mapping to a
  circle for personal video Observation with Head-Mounted Displays}. In
  \bibinfo{booktitle}{\emph{Proceedings of the Augmented Humans International
  Conference 2021}}. \bibinfo{pages}{240--250}.
\newblock


\bibitem[\protect\citeauthoryear{Han, Chen, Hsieh, Huang, and Hung}{Han
  et~al\mbox{.}}{2016}]%
        {han2016ar}
\bibfield{author}{\bibinfo{person}{Ping-Hsuan Han}, \bibinfo{person}{Kuan-Wen
  Chen}, \bibinfo{person}{Chen-Hsin Hsieh}, \bibinfo{person}{Yu-Jie Huang},
  {and} \bibinfo{person}{Yi-Ping Hung}.} \bibinfo{year}{2016}\natexlab{}.
\newblock \showarticletitle{Ar-arm: Augmented visualization for guiding arm
  movement in the first-person perspective}. In
  \bibinfo{booktitle}{\emph{Proceedings of the 7th Augmented Human
  International Conference 2016}}. \bibinfo{pages}{1--4}.
\newblock


\bibitem[\protect\citeauthoryear{Han, Chen, Zhong, Wang, and Hung}{Han
  et~al\mbox{.}}{2017}]%
        {han2017my}
\bibfield{author}{\bibinfo{person}{Ping-Hsuan Han}, \bibinfo{person}{Yang-Sheng
  Chen}, \bibinfo{person}{Yilun Zhong}, \bibinfo{person}{Han-Lei Wang}, {and}
  \bibinfo{person}{Yi-Ping Hung}.} \bibinfo{year}{2017}\natexlab{}.
\newblock \showarticletitle{My Tai-Chi coaches: an augmented-learning tool for
  practicing Tai-Chi Chuan}. In \bibinfo{booktitle}{\emph{Proceedings of the
  8th Augmented Human International Conference}}. \bibinfo{pages}{1--4}.
\newblock


\bibitem[\protect\citeauthoryear{Hoang, Reinoso, Vetere, and Tanin}{Hoang
  et~al\mbox{.}}{2016}]%
        {hoang2016onebody}
\bibfield{author}{\bibinfo{person}{Thuong~N Hoang}, \bibinfo{person}{Martin
  Reinoso}, \bibinfo{person}{Frank Vetere}, {and} \bibinfo{person}{Egemen
  Tanin}.} \bibinfo{year}{2016}\natexlab{}.
\newblock \showarticletitle{Onebody: remote posture guidance system using first
  person view in virtual environment}. In \bibinfo{booktitle}{\emph{Proceedings
  of the 9th Nordic Conference on Human-Computer Interaction}}.
  \bibinfo{pages}{1--10}.
\newblock


\bibitem[\protect\citeauthoryear{Huang, Qian, Wang, Patel, Sreeram, Cao,
  Ramani, and Quinn}{Huang et~al\mbox{.}}{2021}]%
        {huang2021adaptutar}
\bibfield{author}{\bibinfo{person}{Gaoping Huang}, \bibinfo{person}{Xun Qian},
  \bibinfo{person}{Tianyi Wang}, \bibinfo{person}{Fagun Patel},
  \bibinfo{person}{Maitreya Sreeram}, \bibinfo{person}{Yuanzhi Cao},
  \bibinfo{person}{Karthik Ramani}, {and} \bibinfo{person}{Alexander~J Quinn}.}
  \bibinfo{year}{2021}\natexlab{}.
\newblock \showarticletitle{Adaptutar: An adaptive tutoring system for machine
  tasks in augmented reality}. In \bibinfo{booktitle}{\emph{Proceedings of the
  2021 CHI Conference on Human Factors in Computing Systems}}.
  \bibinfo{pages}{1--15}.
\newblock


\bibitem[\protect\citeauthoryear{Iannucci, Chen, Armeni, Pollefeys, Pfister,
  and Beyer}{Iannucci et~al\mbox{.}}{2023}]%
        {iannucci2023arrow}
\bibfield{author}{\bibinfo{person}{Elena Iannucci}, \bibinfo{person}{Zhutian
  Chen}, \bibinfo{person}{Iro Armeni}, \bibinfo{person}{Marc Pollefeys},
  \bibinfo{person}{Hanspeter Pfister}, {and} \bibinfo{person}{Johanna Beyer}.}
  \bibinfo{year}{2023}\natexlab{}.
\newblock \showarticletitle{ARrow: A Real-Time AR Rowing Coach}.
\newblock  (\bibinfo{year}{2023}).
\newblock


\bibitem[\protect\citeauthoryear{Ihara, Faridan, Ichikawa, Kawaguchi, and
  Suzuki}{Ihara et~al\mbox{.}}{2023}]%
        {ihara2023holobots}
\bibfield{author}{\bibinfo{person}{Keiichi Ihara}, \bibinfo{person}{Mehrad
  Faridan}, \bibinfo{person}{Ayumi Ichikawa}, \bibinfo{person}{Ikkaku
  Kawaguchi}, {and} \bibinfo{person}{Ryo Suzuki}.}
  \bibinfo{year}{2023}\natexlab{}.
\newblock \showarticletitle{HoloBots: Augmenting Holographic Telepresence with
  Mobile Robots for Tangible Remote Collaboration in Mixed Reality}. In
  \bibinfo{booktitle}{\emph{Proceedings of the 36th Annual ACM Symposium on
  User Interface Software and Technology}}. \bibinfo{pages}{1--12}.
\newblock


\bibitem[\protect\citeauthoryear{Ikeda, Hwang, Koike, Bruder, Yoshimoto, and
  Cobb}{Ikeda et~al\mbox{.}}{2018}]%
        {ikeda2018ar}
\bibfield{author}{\bibinfo{person}{Atsuki Ikeda}, \bibinfo{person}{Dong-Hyun
  Hwang}, \bibinfo{person}{Hideki Koike}, \bibinfo{person}{Gerd Bruder},
  \bibinfo{person}{Shunsuke Yoshimoto}, {and} \bibinfo{person}{Sue Cobb}.}
  \bibinfo{year}{2018}\natexlab{}.
\newblock \showarticletitle{AR based Self-sports Learning System using Decayed
  Dynamic TimeWarping Algorithm.}. In \bibinfo{booktitle}{\emph{ICAT-EGVE}}.
  \bibinfo{pages}{171--174}.
\newblock


\bibitem[\protect\citeauthoryear{Jan, Tseng, Kao, and Hung}{Jan
  et~al\mbox{.}}{2021}]%
        {jan2021augmented}
\bibfield{author}{\bibinfo{person}{Yao-Fu Jan}, \bibinfo{person}{Kuan-Wei
  Tseng}, \bibinfo{person}{Peng-Yuan Kao}, {and} \bibinfo{person}{Yi-Ping
  Hung}.} \bibinfo{year}{2021}\natexlab{}.
\newblock \showarticletitle{Augmented Tai-Chi Chuan Practice Tool with Pose
  Evaluation}. In \bibinfo{booktitle}{\emph{2021 IEEE 4th International
  Conference on Multimedia Information Processing and Retrieval (MIPR)}}. IEEE,
  \bibinfo{pages}{35--41}.
\newblock


\bibitem[\protect\citeauthoryear{Katzakis, Tong, Ariza, Chen, Klinker,
  R{\"o}der, and Steinicke}{Katzakis et~al\mbox{.}}{2017}]%
        {katzakis2017stylo}
\bibfield{author}{\bibinfo{person}{Nicholas Katzakis},
  \bibinfo{person}{Jonathan Tong}, \bibinfo{person}{Oscar Ariza},
  \bibinfo{person}{Lihan Chen}, \bibinfo{person}{Gudrun Klinker},
  \bibinfo{person}{Brigitte R{\"o}der}, {and} \bibinfo{person}{Frank
  Steinicke}.} \bibinfo{year}{2017}\natexlab{}.
\newblock \showarticletitle{Stylo and handifact: Modulating haptic perception
  through visualizations for posture training in augmented reality}. In
  \bibinfo{booktitle}{\emph{Proceedings of the 5th Symposium on Spatial User
  Interaction}}. \bibinfo{pages}{58--67}.
\newblock


\bibitem[\protect\citeauthoryear{Kosmalla, Hupperich, Hirsch, Daiber, and
  Kr{\"u}ger}{Kosmalla et~al\mbox{.}}{2021}]%
        {kosmalla2021virtualladder}
\bibfield{author}{\bibinfo{person}{Felix Kosmalla}, \bibinfo{person}{Fabian
  Hupperich}, \bibinfo{person}{Anke Hirsch}, \bibinfo{person}{Florian Daiber},
  {and} \bibinfo{person}{Antonio Kr{\"u}ger}.} \bibinfo{year}{2021}\natexlab{}.
\newblock \showarticletitle{VirtualLadder: Using Interactive Projections for
  Agility Ladder Training}. In \bibinfo{booktitle}{\emph{Extended Abstracts of
  the 2021 CHI Conference on Human Factors in Computing Systems}}.
  \bibinfo{pages}{1--7}.
\newblock


\bibitem[\protect\citeauthoryear{Liao, Karim, Jadon, Kazi, and Suzuki}{Liao
  et~al\mbox{.}}{2022}]%
        {liao2022realitytalk}
\bibfield{author}{\bibinfo{person}{Jian Liao}, \bibinfo{person}{Adnan Karim},
  \bibinfo{person}{Shivesh~Singh Jadon}, \bibinfo{person}{Rubaiat~Habib Kazi},
  {and} \bibinfo{person}{Ryo Suzuki}.} \bibinfo{year}{2022}\natexlab{}.
\newblock \showarticletitle{RealityTalk: Real-Time Speech-Driven Augmented
  Presentation for AR Live Storytelling}. In
  \bibinfo{booktitle}{\emph{Proceedings of the 35th Annual ACM Symposium on
  User Interface Software and Technology}}. \bibinfo{pages}{1--12}.
\newblock


\bibitem[\protect\citeauthoryear{Lin, Singh, Yang, Nobre, Beyer, Smith, and
  Pfister}{Lin et~al\mbox{.}}{2021}]%
        {lin2021towards}
\bibfield{author}{\bibinfo{person}{Tica Lin}, \bibinfo{person}{Rishi Singh},
  \bibinfo{person}{Yalong Yang}, \bibinfo{person}{Carolina Nobre},
  \bibinfo{person}{Johanna Beyer}, \bibinfo{person}{Maurice~A Smith}, {and}
  \bibinfo{person}{Hanspeter Pfister}.} \bibinfo{year}{2021}\natexlab{}.
\newblock \showarticletitle{Towards an understanding of situated ar
  visualization for basketball free-throw training}. In
  \bibinfo{booktitle}{\emph{Proceedings of the 2021 CHI Conference on Human
  Factors in Computing Systems}}. \bibinfo{pages}{1--13}.
\newblock


\bibitem[\protect\citeauthoryear{Liu, Saquib, Chen, Kazi, Wei, Fu, and Tai}{Liu
  et~al\mbox{.}}{2022}]%
        {liu2022posecoach}
\bibfield{author}{\bibinfo{person}{Jingyuan Liu}, \bibinfo{person}{Nazmus
  Saquib}, \bibinfo{person}{Zhutian Chen}, \bibinfo{person}{Rubaiat~Habib
  Kazi}, \bibinfo{person}{Li-Yi Wei}, \bibinfo{person}{Hongbo Fu}, {and}
  \bibinfo{person}{Chiew-Lan Tai}.} \bibinfo{year}{2022}\natexlab{}.
\newblock \showarticletitle{PoseCoach: A Customizable Analysis and
  Visualization System for Video-based Running Coaching}.
\newblock \bibinfo{journal}{\emph{IEEE Transactions on Visualization and
  Computer Graphics}} (\bibinfo{year}{2022}).
\newblock


\bibitem[\protect\citeauthoryear{Liu, Wu, Liao, Nishioka, Furuya, and
  Koike}{Liu et~al\mbox{.}}{2023}]%
        {liu2023pianosyncar}
\bibfield{author}{\bibinfo{person}{Ruofan Liu}, \bibinfo{person}{Erwin Wu},
  \bibinfo{person}{Chen-Chieh Liao}, \bibinfo{person}{Hayato Nishioka},
  \bibinfo{person}{Shinichi Furuya}, {and} \bibinfo{person}{Hideki Koike}.}
  \bibinfo{year}{2023}\natexlab{}.
\newblock \showarticletitle{PianoSyncAR: Enhancing Piano Learning through
  Visualizing Synchronized Hand Pose Discrepancies in Augmented Reality}. In
  \bibinfo{booktitle}{\emph{2023 IEEE International Symposium on Mixed and
  Augmented Reality (ISMAR)}}. IEEE, \bibinfo{pages}{859--868}.
\newblock


\bibitem[\protect\citeauthoryear{Marquardt, Beira, Em, Paiva, and
  Kox}{Marquardt et~al\mbox{.}}{2012}]%
        {marquardt2012super}
\bibfield{author}{\bibinfo{person}{Zoe Marquardt}, \bibinfo{person}{Jo{\~a}o
  Beira}, \bibinfo{person}{Natalia Em}, \bibinfo{person}{Isabel Paiva}, {and}
  \bibinfo{person}{Sebastian Kox}.} \bibinfo{year}{2012}\natexlab{}.
\newblock \showarticletitle{Super Mirror: a kinect interface for ballet
  dancers}.
\newblock In \bibinfo{booktitle}{\emph{CHI'12 Extended Abstracts on Human
  Factors in Computing Systems}}. \bibinfo{pages}{1619--1624}.
\newblock


\bibitem[\protect\citeauthoryear{Matsumoto, Wu, and Koike}{Matsumoto
  et~al\mbox{.}}{2022}]%
        {matsumoto2022skiing}
\bibfield{author}{\bibinfo{person}{Takashi Matsumoto}, \bibinfo{person}{Erwin
  Wu}, {and} \bibinfo{person}{Hideki Koike}.} \bibinfo{year}{2022}\natexlab{}.
\newblock \showarticletitle{Skiing, Fast and Slow: Evaluation of Time
  Distortion for VR Ski Training}. In \bibinfo{booktitle}{\emph{Proceedings of
  the Augmented Humans International Conference 2022}}.
  \bibinfo{pages}{142--151}.
\newblock


\bibitem[\protect\citeauthoryear{Mostajeran, Steinicke, Ariza~Nunez, Gatsios,
  and Fotiadis}{Mostajeran et~al\mbox{.}}{2020}]%
        {mostajeran2020augmented}
\bibfield{author}{\bibinfo{person}{Fariba Mostajeran}, \bibinfo{person}{Frank
  Steinicke}, \bibinfo{person}{Oscar~Javier Ariza~Nunez},
  \bibinfo{person}{Dimitrios Gatsios}, {and} \bibinfo{person}{Dimitrios
  Fotiadis}.} \bibinfo{year}{2020}\natexlab{}.
\newblock \showarticletitle{Augmented reality for older adults: exploring
  acceptability of virtual coaches for home-based balance training in an aging
  population}. In \bibinfo{booktitle}{\emph{Proceedings of the 2020 CHI
  Conference on Human Factors in Computing Systems}}. \bibinfo{pages}{1--12}.
\newblock


\bibitem[\protect\citeauthoryear{Piumsomboon, Lee, Hart, Ens, Lindeman, Thomas,
  and Billinghurst}{Piumsomboon et~al\mbox{.}}{2018}]%
        {piumsomboon2018mini}
\bibfield{author}{\bibinfo{person}{Thammathip Piumsomboon},
  \bibinfo{person}{Gun~A Lee}, \bibinfo{person}{Jonathon~D Hart},
  \bibinfo{person}{Barrett Ens}, \bibinfo{person}{Robert~W Lindeman},
  \bibinfo{person}{Bruce~H Thomas}, {and} \bibinfo{person}{Mark Billinghurst}.}
  \bibinfo{year}{2018}\natexlab{}.
\newblock \showarticletitle{Mini-me: An adaptive avatar for mixed reality
  remote collaboration}. In \bibinfo{booktitle}{\emph{Proceedings of the 2018
  CHI conference on human factors in computing systems}}.
  \bibinfo{pages}{1--13}.
\newblock


\bibitem[\protect\citeauthoryear{Piumsomboon, Lee, Irlitti, Ens, Thomas, and
  Billinghurst}{Piumsomboon et~al\mbox{.}}{2019}]%
        {piumsomboon2019shoulder}
\bibfield{author}{\bibinfo{person}{Thammathip Piumsomboon},
  \bibinfo{person}{Gun~A Lee}, \bibinfo{person}{Andrew Irlitti},
  \bibinfo{person}{Barrett Ens}, \bibinfo{person}{Bruce~H Thomas}, {and}
  \bibinfo{person}{Mark Billinghurst}.} \bibinfo{year}{2019}\natexlab{}.
\newblock \showarticletitle{On the shoulder of the giant: A multi-scale mixed
  reality collaboration with 360 video sharing and tangible interaction}. In
  \bibinfo{booktitle}{\emph{Proceedings of the 2019 CHI conference on human
  factors in computing systems}}. \bibinfo{pages}{1--17}.
\newblock


\bibitem[\protect\citeauthoryear{Saenz-de Urturi and
  Garcia-Zapirain~Soto}{Saenz-de Urturi and Garcia-Zapirain~Soto}{2016}]%
        {saenz2016kinect}
\bibfield{author}{\bibinfo{person}{Zelai Saenz-de Urturi} {and}
  \bibinfo{person}{Begonya Garcia-Zapirain~Soto}.}
  \bibinfo{year}{2016}\natexlab{}.
\newblock \showarticletitle{Kinect-based virtual game for the elderly that
  detects incorrect body postures in real time}.
\newblock \bibinfo{journal}{\emph{Sensors}} \bibinfo{volume}{16},
  \bibinfo{number}{5} (\bibinfo{year}{2016}), \bibinfo{pages}{704}.
\newblock


\bibitem[\protect\citeauthoryear{Sch{\"o}nauer, Fukushi, Olwal, Kaufmann, and
  Raskar}{Sch{\"o}nauer et~al\mbox{.}}{2012}]%
        {schonauer2012multimodal}
\bibfield{author}{\bibinfo{person}{Christian Sch{\"o}nauer},
  \bibinfo{person}{Kenichiro Fukushi}, \bibinfo{person}{Alex Olwal},
  \bibinfo{person}{Hannes Kaufmann}, {and} \bibinfo{person}{Ramesh Raskar}.}
  \bibinfo{year}{2012}\natexlab{}.
\newblock \showarticletitle{Multimodal motion guidance: techniques for adaptive
  and dynamic feedback}. In \bibinfo{booktitle}{\emph{Proceedings of the 14th
  ACM international conference on Multimodal interaction}}.
  \bibinfo{pages}{133--140}.
\newblock


\bibitem[\protect\citeauthoryear{Sekhavat and Namani}{Sekhavat and
  Namani}{2018}]%
        {sekhavat2018projection}
\bibfield{author}{\bibinfo{person}{Yoones~A Sekhavat} {and}
  \bibinfo{person}{Mohammad~S Namani}.} \bibinfo{year}{2018}\natexlab{}.
\newblock \showarticletitle{Projection-based AR: Effective visual feedback in
  gait rehabilitation}.
\newblock \bibinfo{journal}{\emph{IEEE Transactions on Human-Machine Systems}}
  \bibinfo{volume}{48}, \bibinfo{number}{6} (\bibinfo{year}{2018}),
  \bibinfo{pages}{626--636}.
\newblock


\bibitem[\protect\citeauthoryear{Skreinig, Stanescu, Mori, Heyen, Mohr,
  Sedlmair, Schmalstieg, and Kalkofen}{Skreinig et~al\mbox{.}}{2022}]%
        {skreinig2022ar}
\bibfield{author}{\bibinfo{person}{Lucchas~Ribeiro Skreinig},
  \bibinfo{person}{Ana Stanescu}, \bibinfo{person}{Shohei Mori},
  \bibinfo{person}{Frank Heyen}, \bibinfo{person}{Peter Mohr},
  \bibinfo{person}{Michael Sedlmair}, \bibinfo{person}{Dieter Schmalstieg},
  {and} \bibinfo{person}{Denis Kalkofen}.} \bibinfo{year}{2022}\natexlab{}.
\newblock \showarticletitle{AR Hero: Generating interactive augmented reality
  guitar tutorials}. In \bibinfo{booktitle}{\emph{2022 IEEE Conference on
  Virtual Reality and 3D User Interfaces Abstracts and Workshops (VRW)}}. IEEE,
  \bibinfo{pages}{395--401}.
\newblock


\bibitem[\protect\citeauthoryear{Sodhi, Benko, and Wilson}{Sodhi
  et~al\mbox{.}}{2012}]%
        {sodhi2012lightguide}
\bibfield{author}{\bibinfo{person}{Rajinder Sodhi}, \bibinfo{person}{Hrvoje
  Benko}, {and} \bibinfo{person}{Andrew Wilson}.}
  \bibinfo{year}{2012}\natexlab{}.
\newblock \showarticletitle{LightGuide: projected visualizations for hand
  movement guidance}. In \bibinfo{booktitle}{\emph{Proceedings of the SIGCHI
  Conference on Human Factors in Computing Systems}}.
  \bibinfo{pages}{179--188}.
\newblock


\bibitem[\protect\citeauthoryear{Suzuki, Kazi, Wei, DiVerdi, Li, and
  Leithinger}{Suzuki et~al\mbox{.}}{2020}]%
        {suzuki2020realitysketch}
\bibfield{author}{\bibinfo{person}{Ryo Suzuki}, \bibinfo{person}{Rubaiat~Habib
  Kazi}, \bibinfo{person}{Li-Yi Wei}, \bibinfo{person}{Stephen DiVerdi},
  \bibinfo{person}{Wilmot Li}, {and} \bibinfo{person}{Daniel Leithinger}.}
  \bibinfo{year}{2020}\natexlab{}.
\newblock \showarticletitle{Realitysketch: Embedding responsive graphics and
  visualizations in AR through dynamic sketching}. In
  \bibinfo{booktitle}{\emph{Proceedings of the 33rd Annual ACM Symposium on
  User Interface Software and Technology}}. \bibinfo{pages}{166--181}.
\newblock


\bibitem[\protect\citeauthoryear{Suzuki, Sakamoto, and Ono}{Suzuki
  et~al\mbox{.}}{2023}]%
        {suzuki2023gino}
\bibfield{author}{\bibinfo{person}{Yuto Suzuki}, \bibinfo{person}{Daisuke
  Sakamoto}, {and} \bibinfo{person}{Tetsuo Ono}.}
  \bibinfo{year}{2023}\natexlab{}.
\newblock \showarticletitle{Gino. Aiki: Mixed Reality-based Physical Motor
  Skill Training in Aikido}. In \bibinfo{booktitle}{\emph{2023 IEEE
  International Symposium on Mixed and Augmented Reality Adjunct
  (ISMAR-Adjunct)}}. IEEE, \bibinfo{pages}{519--524}.
\newblock


\bibitem[\protect\citeauthoryear{Thanyadit, Punpongsanon, and Pong}{Thanyadit
  et~al\mbox{.}}{2019}]%
        {thanyadit2019observar}
\bibfield{author}{\bibinfo{person}{Santawat Thanyadit},
  \bibinfo{person}{Parinya Punpongsanon}, {and} \bibinfo{person}{Ting-Chuen
  Pong}.} \bibinfo{year}{2019}\natexlab{}.
\newblock \showarticletitle{ObserVAR: Visualization system for observing
  virtual reality users using augmented reality}. In
  \bibinfo{booktitle}{\emph{2019 IEEE International Symposium on Mixed and
  Augmented Reality (ISMAR)}}. IEEE, \bibinfo{pages}{258--268}.
\newblock


\bibitem[\protect\citeauthoryear{Tharatipyakul, Choo, and
  Perrault}{Tharatipyakul et~al\mbox{.}}{2020}]%
        {tharatipyakul2020pose}
\bibfield{author}{\bibinfo{person}{Atima Tharatipyakul},
  \bibinfo{person}{Kenny~TW Choo}, {and} \bibinfo{person}{Simon~T Perrault}.}
  \bibinfo{year}{2020}\natexlab{}.
\newblock \showarticletitle{Pose estimation for facilitating movement learning
  from online videos}. In \bibinfo{booktitle}{\emph{Proceedings of the
  International Conference on Advanced Visual Interfaces}}.
  \bibinfo{pages}{1--5}.
\newblock


\bibitem[\protect\citeauthoryear{Thoravi~Kumaravel, Anderson, Fitzmaurice,
  Hartmann, and Grossman}{Thoravi~Kumaravel et~al\mbox{.}}{2019}]%
        {thoravi2019loki}
\bibfield{author}{\bibinfo{person}{Balasaravanan Thoravi~Kumaravel},
  \bibinfo{person}{Fraser Anderson}, \bibinfo{person}{George Fitzmaurice},
  \bibinfo{person}{Bjoern Hartmann}, {and} \bibinfo{person}{Tovi Grossman}.}
  \bibinfo{year}{2019}\natexlab{}.
\newblock \showarticletitle{Loki: Facilitating remote instruction of physical
  tasks using bi-directional mixed-reality telepresence}. In
  \bibinfo{booktitle}{\emph{Proceedings of the 32nd Annual ACM Symposium on
  User Interface Software and Technology}}. \bibinfo{pages}{161--174}.
\newblock


\bibitem[\protect\citeauthoryear{Turmo~Vidal, M{\'a}rquez~Segura, Boyer, and
  Waern}{Turmo~Vidal et~al\mbox{.}}{2019}]%
        {turmo2019enlightened}
\bibfield{author}{\bibinfo{person}{Laia Turmo~Vidal}, \bibinfo{person}{Elena
  M{\'a}rquez~Segura}, \bibinfo{person}{Christopher Boyer}, {and}
  \bibinfo{person}{Annika Waern}.} \bibinfo{year}{2019}\natexlab{}.
\newblock \showarticletitle{Enlightened yoga: Designing an augmented class with
  wearable lights to support instruction}. In
  \bibinfo{booktitle}{\emph{Proceedings of the 2019 on Designing Interactive
  Systems Conference}}. \bibinfo{pages}{1017--1031}.
\newblock


\bibitem[\protect\citeauthoryear{Turmo~Vidal, Zhu, and
  Riego-Delgado}{Turmo~Vidal et~al\mbox{.}}{2020}]%
        {turmo2020bodylights}
\bibfield{author}{\bibinfo{person}{Laia Turmo~Vidal}, \bibinfo{person}{Hui
  Zhu}, {and} \bibinfo{person}{Abraham Riego-Delgado}.}
  \bibinfo{year}{2020}\natexlab{}.
\newblock \showarticletitle{Bodylights: Open-ended augmented feedback to
  support training towards a correct exercise execution}. In
  \bibinfo{booktitle}{\emph{Proceedings of the 2020 CHI Conference on Human
  Factors in Computing Systems}}. \bibinfo{pages}{1--14}.
\newblock


\bibitem[\protect\citeauthoryear{Velloso, Bulling, and Gellersen}{Velloso
  et~al\mbox{.}}{2013}]%
        {velloso2013motionma}
\bibfield{author}{\bibinfo{person}{Eduardo Velloso}, \bibinfo{person}{Andreas
  Bulling}, {and} \bibinfo{person}{Hans Gellersen}.}
  \bibinfo{year}{2013}\natexlab{}.
\newblock \showarticletitle{Motionma: motion modelling and analysis by
  demonstration}. In \bibinfo{booktitle}{\emph{Proceedings of the SIGCHI
  Conference on Human Factors in Computing Systems}}.
  \bibinfo{pages}{1309--1318}.
\newblock


\bibitem[\protect\citeauthoryear{Wang, Zhou, Fitzmaurice, and Anderson}{Wang
  et~al\mbox{.}}{2022}]%
        {wang2022videoposevr}
\bibfield{author}{\bibinfo{person}{Cheng~Yao Wang}, \bibinfo{person}{Qian
  Zhou}, \bibinfo{person}{George Fitzmaurice}, {and} \bibinfo{person}{Fraser
  Anderson}.} \bibinfo{year}{2022}\natexlab{}.
\newblock \showarticletitle{VideoPoseVR: Authoring Virtual Reality Character
  Animations with Online Videos}.
\newblock \bibinfo{journal}{\emph{Proceedings of the ACM on Human-Computer
  Interaction}} \bibinfo{volume}{6}, \bibinfo{number}{ISS}
  (\bibinfo{year}{2022}), \bibinfo{pages}{448--467}.
\newblock


\bibitem[\protect\citeauthoryear{Wu, Yu, Xu, Deng, Wang, Xie, Zhang, and Wu}{Wu
  et~al\mbox{.}}{2023}]%
        {wu2023ar}
\bibfield{author}{\bibinfo{person}{Yihong Wu}, \bibinfo{person}{Lingyun Yu},
  \bibinfo{person}{Jie Xu}, \bibinfo{person}{Dazhen Deng},
  \bibinfo{person}{Jiachen Wang}, \bibinfo{person}{Xiao Xie},
  \bibinfo{person}{Hui Zhang}, {and} \bibinfo{person}{Yingcai Wu}.}
  \bibinfo{year}{2023}\natexlab{}.
\newblock \showarticletitle{AR-Enhanced Workouts: Exploring Visual Cues for
  At-Home Workout Videos in AR Environment}. In
  \bibinfo{booktitle}{\emph{Proceedings of the 36th Annual ACM Symposium on
  User Interface Software and Technology}}. \bibinfo{pages}{1--15}.
\newblock


\bibitem[\protect\citeauthoryear{Xia, Fang, Arakawa, and Sugiura}{Xia
  et~al\mbox{.}}{2022}]%
        {xia2022volearn}
\bibfield{author}{\bibinfo{person}{Chengshuo Xia}, \bibinfo{person}{Xinrui
  Fang}, \bibinfo{person}{Riku Arakawa}, {and} \bibinfo{person}{Yuta Sugiura}.}
  \bibinfo{year}{2022}\natexlab{}.
\newblock \showarticletitle{VoLearn: A Cross-Modal Operable Motion-Learning
  System Combined with Virtual Avatar and Auditory Feedback}.
\newblock \bibinfo{journal}{\emph{Proc. ACM Interact. Mob. Wearable Ubiquitous
  Technol.}} (\bibinfo{year}{2022}), \bibinfo{numpages}{26}~pages.
\newblock


\bibitem[\protect\citeauthoryear{Xia, Monteiro, Van, and Suzuki}{Xia
  et~al\mbox{.}}{2023}]%
        {xia2023realitycanvas}
\bibfield{author}{\bibinfo{person}{Zhijie Xia}, \bibinfo{person}{Kyzyl
  Monteiro}, \bibinfo{person}{Kevin Van}, {and} \bibinfo{person}{Ryo Suzuki}.}
  \bibinfo{year}{2023}\natexlab{}.
\newblock \showarticletitle{RealityCanvas: Augmented Reality Sketching for
  Embedded and Responsive Scribble Animation Effects}. In
  \bibinfo{booktitle}{\emph{Proceedings of the 36th Annual ACM Symposium on
  User Interface Software and Technology}}. \bibinfo{pages}{1--14}.
\newblock


\bibitem[\protect\citeauthoryear{Xiu, Li, Wang, Fang, and Lu}{Xiu
  et~al\mbox{.}}{2018}]%
        {xiu2018pose}
\bibfield{author}{\bibinfo{person}{Yuliang Xiu}, \bibinfo{person}{Jiefeng Li},
  \bibinfo{person}{Haoyu Wang}, \bibinfo{person}{Yinghong Fang}, {and}
  \bibinfo{person}{Cewu Lu}.} \bibinfo{year}{2018}\natexlab{}.
\newblock \showarticletitle{Pose Flow: Efficient online pose tracking}.
\newblock \bibinfo{journal}{\emph{arXiv preprint arXiv:1802.00977}}
  (\bibinfo{year}{2018}).
\newblock


\bibitem[\protect\citeauthoryear{Yan, Ding, Guan, Sun, Li, and Zhang}{Yan
  et~al\mbox{.}}{2015}]%
        {yan2015outsideme}
\bibfield{author}{\bibinfo{person}{Shuo Yan}, \bibinfo{person}{Gangyi Ding},
  \bibinfo{person}{Zheng Guan}, \bibinfo{person}{Ningxiao Sun},
  \bibinfo{person}{Hongsong Li}, {and} \bibinfo{person}{Longfei Zhang}.}
  \bibinfo{year}{2015}\natexlab{}.
\newblock \showarticletitle{Outsideme: Augmenting dancer's external self-image
  by using a mixed reality system}. In \bibinfo{booktitle}{\emph{Proceedings of
  the 33rd Annual ACM Conference Extended Abstracts on Human Factors in
  Computing Systems}}. \bibinfo{pages}{965--970}.
\newblock


\bibitem[\protect\citeauthoryear{Ye, Chen, Chu, Wang, Fu, Shen, Zhou, and
  Wu}{Ye et~al\mbox{.}}{2020}]%
        {ye2020shuttlespace}
\bibfield{author}{\bibinfo{person}{Shuainan Ye}, \bibinfo{person}{Zhutian
  Chen}, \bibinfo{person}{Xiangtong Chu}, \bibinfo{person}{Yifan Wang},
  \bibinfo{person}{Siwei Fu}, \bibinfo{person}{Lejun Shen},
  \bibinfo{person}{Kun Zhou}, {and} \bibinfo{person}{Yingcai Wu}.}
  \bibinfo{year}{2020}\natexlab{}.
\newblock \showarticletitle{Shuttlespace: Exploring and analyzing movement
  trajectory in immersive visualization}.
\newblock \bibinfo{journal}{\emph{IEEE transactions on visualization and
  computer graphics}} \bibinfo{volume}{27}, \bibinfo{number}{2}
  (\bibinfo{year}{2020}), \bibinfo{pages}{860--869}.
\newblock


\bibitem[\protect\citeauthoryear{Yu, Angerbauer, Mohr, Kalkofen, and
  Sedlmair}{Yu et~al\mbox{.}}{2020}]%
        {yu2020perspective}
\bibfield{author}{\bibinfo{person}{Xingyao Yu}, \bibinfo{person}{Katrin
  Angerbauer}, \bibinfo{person}{Peter Mohr}, \bibinfo{person}{Denis Kalkofen},
  {and} \bibinfo{person}{Michael Sedlmair}.} \bibinfo{year}{2020}\natexlab{}.
\newblock \showarticletitle{Perspective matters: Design implications for motion
  guidance in mixed reality}. In \bibinfo{booktitle}{\emph{2020 IEEE
  International Symposium on Mixed and Augmented Reality (ISMAR)}}. IEEE,
  \bibinfo{pages}{577--587}.
\newblock


\bibitem[\protect\citeauthoryear{Zhao, Kiciroglu, Vinzant, Cheng, Katircioglu,
  Salzmann, and Fua}{Zhao et~al\mbox{.}}{2022}]%
        {zhao20223d}
\bibfield{author}{\bibinfo{person}{Ziyi Zhao}, \bibinfo{person}{Sena
  Kiciroglu}, \bibinfo{person}{Hugues Vinzant}, \bibinfo{person}{Yuan Cheng},
  \bibinfo{person}{Isinsu Katircioglu}, \bibinfo{person}{Mathieu Salzmann},
  {and} \bibinfo{person}{Pascal Fua}.} \bibinfo{year}{2022}\natexlab{}.
\newblock \showarticletitle{3d pose based feedback for physical exercises}. In
  \bibinfo{booktitle}{\emph{Proceedings of the Asian Conference on Computer
  Vision}}. \bibinfo{pages}{1316--1332}.
\newblock


\bibitem[\protect\citeauthoryear{Zhou, Xu, and Yatani}{Zhou
  et~al\mbox{.}}{2021}]%
        {zhou2021syncup}
\bibfield{author}{\bibinfo{person}{Zhongyi Zhou}, \bibinfo{person}{Anran Xu},
  {and} \bibinfo{person}{Koji Yatani}.} \bibinfo{year}{2021}\natexlab{}.
\newblock \showarticletitle{Syncup: Vision-based practice support for
  synchronized dancing}.
\newblock \bibinfo{journal}{\emph{Proceedings of the ACM on Interactive,
  Mobile, Wearable and Ubiquitous Technologies}} \bibinfo{volume}{5},
  \bibinfo{number}{3} (\bibinfo{year}{2021}), \bibinfo{pages}{1--25}.
\newblock


\end{thebibliography}
